\documentclass[11pt]{article}
\usepackage[letterpaper, left=.8in, top=0.9in, right=.8in, bottom=0.70in,nohead,includefoot, verbose, ignoremp]{geometry}
\usepackage{charter} 
\usepackage[round]{natbib}
\usepackage{enumerate} 
\usepackage{latexsym,amssymb,amsmath,amsfonts,color,enumerate}
\usepackage{graphicx}

\usepackage[svgnames,dvipsnames,x11names]{xcolor}
\usepackage{theorem}

\def\eq#1{equation (\ref{#1})}

\def\beginmat{ \left( \begin{array} }
\def\endmat{ \end{array} \right) }

\def\log{{\rm log}}

\def\cM{\mathcal{M}}

\newtheorem{cor}{Corollary}[section]
\newtheorem{prp}{Proposition}[section]
\theorembodyfont{\rmfamily}

\usepackage{subcaption}

\usepackage{booktabs}
\usepackage{multirow}

\def\qed{\hfill $\square$}

\begin{document}
\vspace{-1in}
\title{\bf Bayesian Dynamic Fused LASSO}
\author{\Large Kaoru Irie\thanks{
Assistant Professor of Economics, The University of Tokyo. 
The author thanks Christopher Glynn, Tsuyoshi Kunihama, Jouchi Nakajima, Hedibert Lopes, Mike West and Meng Xie for their discussion and suggestion on the draft of this paper. He also thanks the anonymous referees for their valuable comments. 
This research was supported in part by Grant-in-Aid for Scientific Research from Japan Society for the Promotion of Science  (17K17659).
}}
\maketitle 
\begin{abstract}
	The new class of Markov processes is proposed to realize the flexible shrinkage effects for the dynamic models. The transition density of the new process consists of two penalty functions, similarly to Bayesian fused LASSO in its functional form, that shrink the current state variable to its previous value and zero. The normalizing constant of the density, which is not ignorable in the posterior computation, is shown to be essentially the log-geometric mixture of double-exponential densities. This process comprises the state equation of the dynamic regression models, which is shown to be conditionally Gaussian and linear in state variables and utilize the forward filtering and backward sampling in posterior computation by Gibbs sampler. The problem of overshrinkage that is inherent in lasso is moderated by considering the hierarchical extension, which can even realize the shrinkage of horseshoe priors marginally. The new prior is compared with the standard double-exponential prior in the estimation of and prediction by the dynamic linear models for illustration. It is also applied to the time-varying vector autoregressive models for the US macroeconomic data, where we examine the (dis)similarity of the additional shrinkage effect to dynamic variable selection or, specifically, the latent threshold models. 
	\par\vspace{4mm}
\end{abstract}

\section{Introduction}

The univariate dynamic linear models (DLMs) in practice are frequently over-parametrized because of the massive amount of predictors, most of which are believed to be noises. For example, the time-varying vector autoregressive models are typically decomposed into the multiple univariate sub-models for the computational feasibility (e.g., \citealt{zhao2016dynamic} and \citealt{gruber2016gpu}), but this approach results in the excess amount of predictors in those sub-models even for the moderate dimensional observations. This research contributes to the appropriate modeling of sparsity in the univariate DLMs with many predictors by defining a new shrinkage prior on the dynamic coefficients, and its application to the modeling of multivariate time series. 

Specifically, for the univariate state variable $x_t$, which is the time-varying regression coefficient in the context of DLMs, we consider the new Markov process defined by its transition density, 
\begin{equation} \label{eq:original}
p(x_t|x_{t-1}) \propto \exp \left\{ \ - \alpha |x_t| - \beta |x_t-x_{t-1}| \ \right\} ,
\end{equation} 
where weights $\alpha$ and $\beta$ are positive. The two penalty functions in the exponential realize the shrinkage effects conditional on the latest state $x_{t-1}$. By using this process as the prior in DLMs, we shrink the state variable at time $t$ toward zero by the first penalty function, while shrinking it to the previous state variable at $t-1$ as well to penalize the excess dynamics by the second penalty. The technical difficulty in using this prior in statistical analysis is the unknown normalizing constant abbreviated in \eq{eq:original} that involves state variable $x_{t-1}$ and is not ignorable in the posterior analysis. The objective of this research is to compute this normalizing constant explicitly and to provide the computational methodology for the efficient posterior analysis by Markov chain Monte Carlo methods. 

The prior in (\ref{eq:original}) is named {\it dynamic fused LASSO} (DFL) prior for its similarity to Bayesian fused LASSO models (e.g., \citealt{kyung2010penalized} and \citealt{betancourt2017bayesian}). The Bayesian fused LASSO has rarely been applied to the time series analysis and, consequently, the problem of unknown normalizing constant has not been discussed. This is because, in Bayesian fused LASSO, the state variables are modeled {\it jointly}, not conditionally as in (\ref{eq:original}), by exponentiating the various penalty functions to define the joint density of all the state variables. 
By modeling the joint density directly, the normalizing constant becomes free from the state variables, which simplifies the Bayesian inference for the fused LASSO models and enables the scale mixture representation of the double exponential priors, as in the standard Bayesian LASSO models \citep{park2008bayesian}. From this viewpoint, our research is clearly different from the existing fused LASSO models in modeling the {\it conditional} distribution of state variables to realize our prior belief in the dynamic modeling, which instead poses the problem of computing the normalizing constant that can be ignored (or not required to compute) in the study of the Bayesian fused LASSO. The modeling of the conditional distribution is crucial in predictive analysis; the direct application of the existing fused LASSO to the joint distribution of time-varying parameters does not define the conditional evolution of state variables coherently and, as a result, cannot be used for sequential posterior updating and forecasting. 

Although the normalizing constant complicates the prior and posterior distributions of the DLMs with the DFL process, we prove the augmented model representation of the DFL prior as the conditionally dynamic linear models (CDLMs), for which the efficient posterior sampling of state variables by the forward filtering and backward sampling (FFBS) is available. Facilitating the posterior computation by FFBS with the help of the CDLM representation is the standard strategy in the literature of econometrics and forecasting, where the dynamic sparsity has been realized by the hierarchical DLMs (e.g., \citealt{fruhwirth2010stochastic}, \citealt{belmonte2014hierarchical} and \citealt{bitto2019achieving}). This hierarchical version of dynamic linear models is obtained by the natural extension of DLMs with another prior on its scale parameters in the state equation. While the posterior of state variables is easily computed by FFBS, these priors penalize only the distance between the two consecutive state variables, $x_t$ and $x_{t-1}$, which is understood as the special case of the prior of this study with $\alpha =0$ in \eq{eq:original}. Our approach, in contrast, integrates the additional penalty for the shrinkage toward zero explicitly into the conditional transition density of the prior process. This modeling approach reflects our prior belief that the state variable is likely to be either zero or unchanged from its previous value. The additional shrinkage effect to zero in the DFL prior can also address the problem of the shrinkage effect restricted to be uniform over time as ``horizontal shrinkage'' \citep{uribe2017dynamic,rockova2020dynamic} by localizing the shrinkage effect at each time by customized latent parameters as practiced in, but in a different way to,  \cite{kalli2014time} and \cite{kowal2019dynamic}. 

The conditional normality and linearity of the DFL prior is based on the fact that the prior process is decomposed into two parts: the synthetic likelihood and synthetic prior. These terminologies literally mean that the prior consists of two components, one of which is treated as (part of) likelihood and the other of which serves as the prior in the computation of the full conditional posterior of state variables. The synthetic prior is just the well-known scale mixture of Gaussian random walks, hence normal and linear in state variables. The synthetic likelihood part is equivalent to observing the artificial data $z_t=0$ with mean $x_t$ that provides the additional information to shrink the state variable to zero. The posterior of this CDLM is proportional to the full conditional of state variables up to constant, which justifies the use of FFBS for this synthetic model. The idea of synthetic likelihood and prior approach has been utilized in the studies of optimal portfolios that are sparse and less switching (e.g., \citealt{KolmRitter2015} and \citealt{irie2019bayesian}), and this research consider the same idea in the context of statistical modeling. 

The rest of the paper is structured as follows. 
Section~\ref{sec:DFL} focuses on the DFL prior in (\ref{eq:original}) and proves its CDLM representation, followed by the comparison with the existing approaches in Section~\ref{sec:inter}. Section~3 considers the effect of weights parameters $(\alpha ,\beta )$ and the extension of the DFL models with the prior on weights.  Section~\ref{sec:SSM} introduces the DLMs with the DFL prior and provides the MCMC algorithm by using the properties proven in the previous sections, in addition to discussing the estimation of observational variances. In Section~\ref{sec:data}, the proposed model is applied to the simulation data for illustration (Section~\ref{sec:sim}) and to the US macroeconomic time series for the comparison with the model of the variable selection type (Section~\ref{sec:LTM}). The paper is concluded in Section~\ref{sec:conclusion} with the list of potential future research. All the proofs are given in the Supplementary Materials. 

\

\textbf{Notations:} The density of the univariate normal distribution with mean $\mu$ and variance $\sigma ^2$ evaluated at $x$ is denoted by $N(x|\mu ,\sigma ^2)$. The double-exponential density with parameter $a$ is denoted by $DE(x|a)=(a/2) e^{-a|x|}$. The gamma distribution with shape $a$ and rate $b$ is written as $Ga(a,b)$ with mean $a/b$. The exponential distribution with rate $b$ is $\mathrm{Ex} (b) = Ga(1,b)$. The beta distribution with positive shapes $a$ and $b$ with mean $a/(a+b)$ is $Be(a,b)$. The generalized inverse Gaussian distribution is denoted by $GIG(p,a,b)$, the density of which is proportional to $x^{p-1}\exp \{ -(a x + b /x)/2 \}$. In our study, $p$ is either $1/2$ or $3/2$, so it is the inverse Gaussian distribution or its reciprocal.

\section{Dynamic fused LASSO} \label{sec:DFL}

\subsection{Definitions} \label{sec:def}

A new class of univariate, stationary Markov processes $\{ x_t \} _{t=1,2,\dots}$ is defined by its conditional density of transition $p(x_t|x_{t-1})$ given in (\ref{eq:original}). The two conflicting $\ell^1$-penalty functions, $\alpha |x_t|$ and $\beta |x_t - x_{t-1}|$, represent the shrinkage toward zero and the latest state $x_{t-1}$, respectively. As discussed in the introduction, this process is expected to reflect our prior belief on the dynamic sparsity in coefficients of DLMs, for which we assume $\alpha < \beta$ throughout the paper. With this restriction, we intend to have weight $\alpha$ sufficiently small, because large $\alpha$ might result in the excess shrinkage toward zero and lead to poor predictive accuracy. In this section, we study the property of this process as the prior for time-varying regression coefficients, while the likelihood of DLMs is introduced in Section~\ref{sec:SSM}. 

We first write the two penalties as the densities of double-exponential distributions,
\begin{equation*}
\begin{split}
&f(x) \equiv \frac{\alpha }{2}e^{-\alpha |x|} = \int _0^{\infty} N(x|0,\tau _1) p_1(\tau_1) d\tau _1, \\
&g(x,x') \equiv \frac{\beta }{2} e^{-\beta |x-x'|} = \int _0^{\infty} N(x|x',\tau _2 ) p_2(\tau_2) d\tau _2,
\end{split}
\end{equation*}
where $\tau_1 \sim \mathrm{Ex}(\alpha ^2/2)$ and $\tau_2\sim \mathrm{Ex}(\beta ^2/2)$ for non-negative weights $\alpha$ and $\beta$ \citep{andrews1974scale,west1987scale,park2008bayesian}. Then, the transition density of the DFL process from $x'$ to $x$ is defined by 
\begin{equation} \label{eq:fgh}
p(x|x') \equiv \frac{f(x) g(x,x')}{h(x)}, \ \ \ \mathrm{where} \ \ \ h(x') \equiv \int _{-\infty}^{\infty} f(x) g(x,x')dx.
\end{equation}
The analytical expression of $h(x')$ is discussed later in Proposition~\ref{prp:1}. 

Denote the marginal distribution of $x$ by $\pi (x)$. The condition for this process to be stationary is
\begin{equation*} 
\pi (x) = \int _{-\infty}^{\infty} \frac{f(x) g(x,x')}{h(x')} \pi (x') dx',
\end{equation*}
and one solution of this functional equation is $\pi (x) = h(x) f(x)$, with which the process also becomes reversible. In the following, we assume the marginal density of this form, hence the DFL process in this paper is stationary. 

The representation by the scale mixture of normals is the key to the computational feasibility, as used in Bayesian LASSO (\citealt{park2008bayesian}). Using the latent scale parameters, $\tau_1$ and $\tau _2$, we have 
\begin{equation*}
f(x)g(x,x') = \int _0^{\infty} \!\!\! \int _0^{\infty} N\!\left( x \left| \frac{\tau _1}{\tau _1 + \tau _2 } x' , \ \frac{\tau_1\tau_2}{\tau_1+\tau_2} \right. \right) N\!\left( x' \left| 0, \tau _1+\tau _2   \right. \right) p_1(\tau _1 ) \ p_2(\tau _2 ) \ d\tau _1 d\tau _2,
\end{equation*}
hence the normalizing function $h(x')$ is
\begin{equation} \label{eq:hmix}
h(x') = \int _0^{\infty} \!\!\! \int _0^{\infty}  N\!\left( x' \left| 0, \tau _1+\tau _2   \right. \right) p_1(\tau _1 ) \ p_2(\tau _2 ) \ d\tau _1 d\tau _2,
\end{equation}
i.e., the scale mixture of normals with the convolution of $p_1(\tau _1)$ and $p_2(\tau _2)$. The conditional density also has the following mixture representation, 
\begin{equation} \label{eq:process2}
p(x|x') = \int _0^{\infty} \!\!\! \int _0^{\infty} N\!\left( x \left| \frac{\tau _1}{\tau _1 + \tau _2 } x' , \ \frac{\tau_1\tau_2}{\tau_1+\tau_2} \right. \right) \frac{ N\!\left( x' \left| 0, \tau _1+\tau _2   \right. \right) p_1(\tau _1 ) \ p_2(\tau _2 ) }{h(x')} \ d\tau _1 d\tau _2,
\end{equation}
i.e., the location-scale mixture of normals. The integrand is the regression of $x$ on $x'$ with coefficient $\tau _1 / (\tau _1+\tau _2)$. Conditioned by the latest state $x'$, the current state $x$ is shrunk toward $x'$ in this form. The amount of this conditional shrinkage is determined by the balance of two conflicting loss functions, or weights $\alpha$ and $\beta$.

\subsection{Normalizing constant} \label{sec:nc}

\begin{figure}[!htbp]
	\centering
	\includegraphics[width=4.5in]{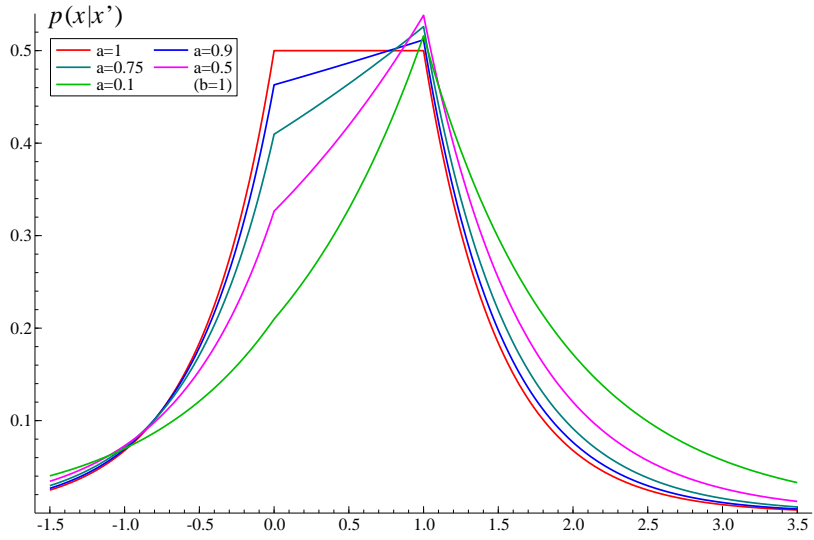}
	\caption{Conditional prior density of state $p(x|x')$ for $b=1$, $x'=1$ and various $a$. If $b=a$, the plateau appears between two shrinkage points, which means that and the values between $x'$ and zero are indifferent in the prior. As $a$ becomes small, the shrinkage effect to $x'$ dominates the functional form of the density, while some probability mass still remains around zero.}
	\label{fig:cond}
\end{figure}

\begin{figure}[!htbp]
	\centering
	\includegraphics[width=4.5in]{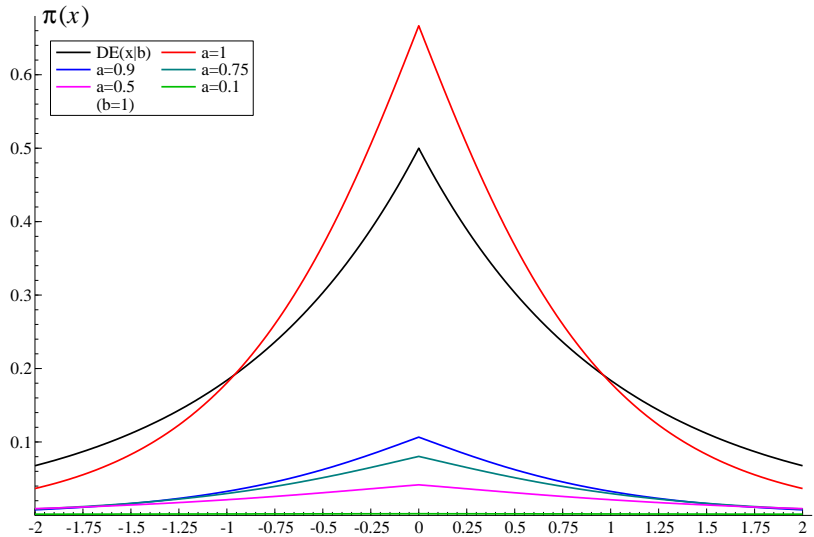}
	\caption{Marginal prior density of $(w|x')$ for $b=10$ and various $a$. The smaller $a$ is, the fatter tails the prior density becomes. The double-exponential prior density is also shown in the same figure. The density of the stationary distribution of the DFL process is clearly different from the double-exponential density for any $a$. }
	\label{fig:marg}
\end{figure}

The computation of normalizing function $h$ is straightforward. 
\begin{prp}
	\label{prp:1}
	For $\alpha < \beta$, the normalizing function $h(x)$ is, 
	\begin{equation} \label{eq:h}
	h(x) = \frac{\alpha \beta^2}{2(\beta^2-\alpha^2)} e^{-\alpha|x|} \left\{ 1 - \left( \frac{\alpha}{\beta} \right) e^{-(\beta-\alpha)|x|}  \right\}.
	\end{equation}
\end{prp}
The proof is given in Section~S1. This proposition enables the evaluation of densities involving $h(x)$, such as the conditional density $p(x|x')$ and the stationary density $\pi (x)$. In Figure~\ref{fig:cond}, the conditional density $p(x|x')$ with $x'=1$ and $\beta =1$ is plotted for the different choices of $\alpha$. As weight $\alpha$ increases, the shrinkage effect to zero becomes visually clear. If $\alpha =\beta$, then the two shrinkage effects are completely balanced, which is expressed as the plateau of the density between $x'$ and zero. With this density as the prior, one does not discriminate any state between these two points a priori. In practice, however, we do not want the extreme amount of shrinkage to zero when transitioning the last state $x'$ to the next $x$ and, for this reason, we assume $\alpha < \beta$ in application. Figure~\ref{fig:marg} shows the marginal, stationary density $\pi (x)$. In general, the smaller $\alpha$ leads to the fatter tails, avoiding the shrinkage effect on the outliers. The sharp increase of density around the origin is seen even with small $\alpha$, so the marginal shrinkage effect to zero is preserved. For large $\alpha$, the density is more spiky, but clearly different from the double-exponential density. These characteristics of priors are revisited from the different viewpoints later in Section~\ref{sec:param} to study the effect of priors on weights $(\alpha ,\beta)$.

\subsection{Log-geometric augmentation} \label{sec:count}

In the expression of the second line of \eq{eq:h}, note that $(\alpha /\beta )e^{-(\beta -\alpha )|x|} < 1$ because $\alpha < \beta$ by assumption. It guarantees the absolute convergence of the series expression of the reciprocal normalizing function as
\begin{equation*}
\frac{1}{h(x)} = \frac{2(\beta ^2-\alpha ^2)}{\alpha \beta ^2} e^{\alpha |x|} \left\{ 1 - \left( \frac{\alpha}{\beta} \right) e^{-(\beta -\alpha )|x|}  \right\} ^{-1} = \frac{2(\beta ^2-\alpha ^2)}{\alpha \beta ^2} e^{\alpha |x|} \sum _{n=0}^{\infty} \left( \frac{\alpha}{\beta} \right) ^n e^{-n(\beta -\alpha )|x|} .
\end{equation*}
We can further rewrite this expression by using probability densities. First, the exponential function $e^{\alpha |x|}$ is the reciprocal of the double exponential density with parameter $\alpha$ with the appropriate adjustment of constants. Second, in the geometric series, we can read off the mixture of double-exponential densities with the running index of series, $n$, as the latent non-negative integer. 
This mixture consists of two components: (i) constant, for $n=0$, which has no contribution to the posterior, and (ii) the discrete mixture of double exponential distributions with parameter $n(\beta -\alpha )$ for $n>0$. The mixture weight is proportional to the probability function of log-geometric distribution with parameter $\alpha /\beta$, defined by the series representation of $-\log (1-\alpha/\beta)$. 
We summarize this observation as a proposition. 
\begin{prp}
	\label{prp:2}
	For $\alpha < \beta$, the reciprocal of normalizing function $h$ is written as 
	\begin{equation*}
	\frac{1}{h(x)} = \frac{2(\alpha +\beta)}{\beta^2} \frac{(C_0+C_+)}{f(x)} q(x|\alpha,\beta)
	\end{equation*}
	where $q(x|\alpha,\beta )$ is the mixture of a constant and the log-geometric mixture of double-exponential distributions, 
	\begin{equation} \label{eq:qmix}
	q(x|\alpha ,\beta) \equiv \frac{C_0}{C_0+C_+} + \frac{C_+}{C_0+C_+} \sum _{n=1}^{\infty} w_n DE(x|n(\beta -\alpha))
	\end{equation}
	where $C_0 = (\beta -\alpha )/2$, $C_+ =\log (\beta ) - \log (\beta -\alpha )$ and $w_n = (\alpha /\beta )^n / nC_+$.
\end{prp}

\subsection{Joint distribution} \label{sec:ex}

Using the expression of normalizing constant $h(\cdot)$ in Proposition~\ref{prp:2}, we can compute the joint distribution of $x_{1:T} = (x_1,\dots , x_T)$ as follows: 
\begin{equation*}
\begin{split}
p(x_{1:T}) = \pi (x_1) \prod _{t=2}^T  p(x_t|x_{t-1}) &= \left\{ \frac{\pi (x_1)}{f(x_1)} \prod _{t=2}^T \frac{f(x_t)}{h(x_{t-1})} \right\} \left\{ f(x_1) \prod _{t=2}^T g(x_t,x_{t-1}) \right\} \\
&\propto \underbrace{  \left\{ f(x_T) \prod _{t=2}^{T-1} q(x_t|\alpha ,\beta ) \right\} }_{``likelihood"} \underbrace{\left\{ f(x_1) \prod _{t=2}^T g(x_t,x_{t-1}) \right\} }_{``prior"}.
\end{split}
\end{equation*}
The reciprocal of the normalizing constant $1/h(x_{t-1})$ involves the double-exponential density $f(x_{t-1})$, which is canceled out with another $f(x_{t-1})$ in the transition from $t{-}1$ to $t$, simplifying the joint prior to the product of two components named ``prior'' and ``likelihood.'' We can, literally, treat these components as the likelihood and prior in order to define the synthetic model whose posterior distribution is equivalent to the original joint density $p(x_{1:T})$. This redundant expression of the prior is the key to the CDLM representation for the efficient posterior computation with the DFL prior. 

The ``prior'' part is the Markov process defined by the initial distribution $f(x_1)$ and the transition $g(x_t,x_{t-1})$, where the explicit shrinkage of $x_t$ is set only toward a single point $x_{t-1}$, not toward zero. This Markov process has widely been used in the state space modeling, for $g(x,x')$ is the scale mixture of normals hence simplifies the posterior computation. 

The rest of the density, phrased ``likelihood'' in the expression above, completes the synthetic model. The density $f(x_T)$ is equivalent to $DE(z_T-x_T|\alpha )$ with $z_T=0$ as the function of $x_T$, where we introduce the ``observation'' $z_T$ in the synthetic model. For function $q(x_t|\alpha ,\beta )$, we revisit equation (\ref{eq:qmix}) and relabel the latent integer as 
\begin{equation*}
q(x_t | \alpha ,\beta ) = \frac{C_0}{C_0+C_+} 1[ n_t = 0] +  \frac{C_+}{C_0+C_+} \sum _{n_t=1}^{\infty} w_{n_t} DE(z_t-x_t|n_t(\beta - \alpha )), 
\end{equation*}
where we define $z_t = 0$ for $t=2,\dots ,T{-}1$, and $C_0$, $C_+$ and $w_{n_t}$ are given in Proposition~\ref{prp:2}. The synthetic model obtained in this way is linear in state variables, and also conditionally Gaussian by writing the double-exponential distributions as the scale mixture of normals. 

\begin{prp}
	\label{prp:3}
	For fixed $\alpha$ and $\beta$ ($\alpha < \beta$), the joint density of $x_{1:T}$ of the DFL prior is proportional to the joint posterior density of the following conditional dynamic linear model; the state evolution and the prior for the associated latent parameter are 
	\begin{equation} \label{eq:evol}
	\begin{split}
	x_t|x_{t-1},\lambda _{\beta ,t} &= x_{t-1} + N(0,\lambda _{\beta ,t}), \\
	\lambda _{\beta ,t} &\sim \mathrm{Ex}(\beta ^2/2), 
	\end{split}
	\end{equation}
	with the initial distribution at $t=1$,
	\begin{equation} \label{eq:ini}
	\begin{split}
	x_1|\lambda _{\alpha ,1} &\sim N(0,\lambda _{\alpha ,1}), \\
	\lambda _{\alpha ,1} &\sim \mathrm{Ex}(\alpha ^2/2). 
	\end{split}
	\end{equation}
	The synthetic observation is defined as, at $t=T$, 
	\begin{equation} \label{eq:synzT}
	\begin{split}
	z_T|x_T,\lambda _{\alpha ,T} &\sim N(x_T,\lambda _{\alpha ,T}), \ \ \ \ \ \ \ \ z_T = 0, \\
	\lambda _{\alpha ,T} &\sim \mathrm{Ex}(\alpha ^2/2). 
	\end{split}
	\end{equation}
	For $t=2,\dots ,T{-}1$, the latent count $n_t$ follows the discrete distribution, 
	\begin{equation} \label{eq:synn}
	n_t \sim \frac{C_0}{C_0+C_+} 1[ n_t=0 ] + \frac{C_+}{C_0+C_+} \mathrm{log\mathchar`-Geo}(\alpha /\beta ), 
	\end{equation}
	where $1[n_t=0]$ here is the point mass on $n_t=0$ and $\mathrm{log\mathchar`-Geo}(\alpha /\beta )$ is the discrete distribution on positive integers $\{ 1,2,\dots \}$ whose probability function is $Pr[n_t=n] = w_n$. The quantities $w_n$, $C_0$ and $C_+$ are defined in Proposition~\ref{prp:2}. Conditional on $n_t$, and if $n_t>0$, we additionally have observational equations defined by, 
	\begin{equation} \label{eq:synz}
	\begin{split}
	z_t | x_t, \lambda _{n,t},n_t &\sim N(x_t,\lambda _{n,t}), \ \ \ \ \ \ \ \ z_t = 0, \\
	\lambda _{n,t} | n_t &\sim \mathrm{Ex}(n_t^2(\beta -\alpha )^2/2). 
	\end{split}	
	\end{equation}
	If $n_t=0$, we have no additional observation equation at time $t$.
\end{prp}
See Section~S3 for the proof. In this synthetic model, the shrinkage effects of the new Markov process gains new interpretation. If $n_t=0$ is sampled at time $t$, then the model has no synthetic observation at $t$, or $z_t$ is ``missing.'' If $n_t>0$ is sampled, then $z_t=0$ is observed, providing the additional information that encourages the shrinkage to zero at time $t$. This is exactly the local (vertical) shrinkage effect, that is different from the global (horizontal) shrinkage that is uniformly applied to the state variables at all the time points, as pointed out by \cite{uribe2017dynamic}. The amount of this shrinkage is indirectly controlled by weights $\alpha$ and $\beta$; the larger $\beta$ is, the more likely it is that $n_t = 0$, having less shrinkage effect to zero, which is consistent with the interpretation of weights in the loss function. 

For the use of this prior in DLMs where we have the ``real'' likelihood, we name the ``likelihood'' part of the prior, or (\ref{eq:synzT}) and (\ref{eq:synz}), as the {\it synthetic likelihood}. Likewise, the ``prior'' part, or (\ref{eq:evol}), (\ref{eq:ini}) and (\ref{eq:synn}), is named the {\it synthetic prior}.

\subsection{Other possible approach to dynamic shrinkage} \label{sec:inter}

The DFL prior is characterized by the two conflicting shrinkage effects that are not seen in the other continuous shrinkage priors used in the time series analysis. Proposition~\ref{prp:3} gives the new interpretation to treat these shrinkage effects separately; the shrinkage to $x_{t-1}$ is based on the state equation (\ref{eq:evol}), while the shrinkage to zero is achieved by the synthetic observations in (\ref{eq:synz}). The former has been seen in the literature of state space modeling with sparsity, where the state transition is defined by
\begin{equation*}
p( x_t | x _{t-1}, \tau _{2t} ) = N(x_t | x_{t-1},\tau _{2t}).
\end{equation*}
If scale $\tau _{2t}$ follows the exponential distribution, then this is the special case of DFL prior with $\alpha =0$. Another example is the case where $\tau _{2t} = \tau _2$ for all $t$, and $\tau _2$ follows the half-Cauchy or scaled-beta priors (\citealt{fruhwirth2010stochastic}, \citealt{belmonte2014hierarchical}, \citealt{feldkircher2017sophisticated} and \citealt{bitto2019achieving}). As discussed already, the shrinkage effect of this prior is limited for its constant scale $\tau _2$; the dynamic scale $\tau_{2t}$ (or the equivalent concept) is discussed in \cite{kalli2014time} and \cite{kowal2019dynamic}. The DFL prior is different from these approaches in adding the new shrinkage effect to zero. 

The concept of simultaneous shrinkage to two points has been discussed as variable selection, or the finite mixture of point mass on zero and (conditionally) Gaussian AR(1) process (dynamic spike-and-slab priors, \citealt{uribe2017dynamic} and \citealt{rockova2020dynamic}). The exact shrinkage to zero achieved by this approach comes at the cost of the other shrinkage directed not to the previous state $x_{t-1}$ but its discounted value $\phi x_t$ with $|\phi|<1$, in addition to the slow convergence of Markov chains that is inevitable in variable selection. Alternatively, the point mass distribution on zero can be replaced by the thresholding the latent state variables. \cite{NakajimaWest2013JBES} considers the thresholding the state variables to zero, while \cite{eisenstat2016stochastic} and \cite{huber2019should} model the thresholding of dynamics. The similarity and difference of the former approach and the DFL model is discussed in Section~\ref{sec:LTM}.

From the viewpoint of the use of multiple penalty functions, another alternative that could achieve the same objective is the existing fused LASSO prior in \cite{kyung2010penalized} that directly models the joint distribution of state variables (the joint fused LASSO prior, or JFL). The joint distribution of all the states variables is defined as 
\begin{equation} \label{eq:JFL}
p_T( x_{1:T}|\alpha ,\beta ) \propto \exp \left\{ \ -\alpha \sum _{t=1}^T |x_t| - \beta \sum _{t=1}^T | x_t-x_{t-1}| \ \right\} .
\end{equation}
This density has the similar, but simpler, synthetic model representation as 
\begin{equation*}
p_T(x_{1:T}|\alpha ,\beta ) \propto \underbrace{  \left\{ \prod _{t=1}^T DE(z_t-x_t|\alpha ) \right\} }_{``likelihood"} \underbrace{  \left\{ \prod _{t=1}^T DE(x_t-x_{t-1}|\beta ) \right\} }_{``prior"}, 
\end{equation*}
where $z_t=0$ for all $t$. The CDLM representation is available for this model with the scale mixture augmentation of the double-exponential distributions. However, the JFL model lacks several desired properties that the proposed DFL possesses. First, the flexibility of the JFL prior in shrinkage effect is limited. This is indicated in the difference of the DFL/JFL priors in the synthetic likelihood. The DFL prior allows for the possibility of missing the synthetic observation and the shrinkage effect to zero can vary across time based on the value of $n_t$, but the latent $z_t$ in the JFL model is always observed, and its shrinkage effects are equally controlled by parameter $\alpha$ at any time point. In this sense, the flexibility of the local shrinkage is limited in the JFL models. Second, the normalizing constant of the JFL prior in (\ref{eq:JFL}) is unknown. This involves none of the state variables, but weight parameters $(\alpha ,\beta )$, so the the posterior analysis on weights is extremely difficult. At last, but most importantly, predictive analysis is not available with the JFL prior. The joint density of the JFL prior does not specify the evolution of state variables, and does not cover the existence of state variables after $T$ in the model. If one defines the joint distributions of $x_{1:T}$ and $x_{1:T+1}$ by (\ref{eq:JFL}) individually, then
\begin{equation*}
\int p_{T+1}( x_{1:(T+1)}|a,b) dx_{T+1} \not= p_T( x_{1:T} |a,b),
\end{equation*} 
and the conditional density is not coherently defined after $T$. This concludes that the JFL prior is not suitable for the formal sequential and predictive analysis.

\section{Estimation of weight parameters} \label{sec:param}

\subsection{Effect on marginal distribution of $x_t$} 

The weights $\alpha$ and $\beta$ determine the structure of sparsity in the prior. As the hyperparameters, these weights must be carefully chosen so that the prior appropriately reflects one's belief on the dynamics and sparsity of the state variables. For the aim of prior elicitation, one should consider the value of the conditional mean of the state transition in (\ref{eq:process2}), $E[x|x']=E[w|x']x'$, and the density of shrinkage effect $w=\tau _1/(\tau_1+\tau_2)$. They are analytically available and provide further information on the prior belief we structure by using the DFL prior. For their functional forms and graphical examples, see Section~S4, S5 and S6. 

\subsection{Conjugate priors on weights} \label{sec:prior}

In practice, tuning the weight parameters of all the state variables is not realistic. It is desirable if the automated adjustment of those hyperparameters is fully or partially available. The formal approach to this goal is the Bayesian posterior analysis by placing the prior distribution on the hyperparameters. Consider the re-parametrization by $\alpha = \rho \beta$, where $\rho \in (0,1)$, to allow that the inequality, $\alpha < \beta$, always holds. Then, we apply the gamma prior $\beta \sim Ga(r^b_0,c^b_0)$ that is conditionally conjugate. For $\rho$, we consider the beta distribution $\rho\sim Be(a^r_0,b^r_0)$; the conditional posterior density of $\rho$ is analytically available, from which we can easily sample by discretizing the prior on interval $(0,1)$. For the details including the derivation of the full conditionals, see Section~S7.

\subsection{Horseshoe as hierarchical Bayesian LASSO} \label{sec:hs}

The use of $\ell^1$-penalty is often criticized for its undesirable pattern of shrinkage, as clearly demonstrated by \cite{carvalho2010horseshoe}. Partly for this reason, it is common, as in the literature mentioned in Section~\ref{sec:inter}, to consider the scale mixture of normals by gamma distributions whose shape is not necessarily unity, but frequently $0.5$ to induce the strong shrinkage effect as the horseshoe prior. Naturally, the extension of the DFL to the dynamic fused horseshoe prior is desired, but the normalizing constant of such prior, and its mixture representation for computational feasibility, have been unknown. 

In fact, we can bypass such mathematical difficulty to the extension by considering another type of hyperprior on weights $(\alpha ,\beta )$ under the DFL prior. As proven in Section~S8, if $\beta ^2|\delta \sim Ga(0.5,\delta )$ and $\delta \sim Be(0.5,0.5)$, then the marginal of $x_t$ with $\rho = 0$ is the horseshoe prior, i.e., the scale mixture of normals by the half-Cauchy distribution. We name the hierarchical DFL prior with this set of priors for the weight parameters as {\it dynamic fused horseshoe prior}, or DFHS prior. 

\begin{figure}[!htbp]
	\centering
	\includegraphics[width=4.5in]{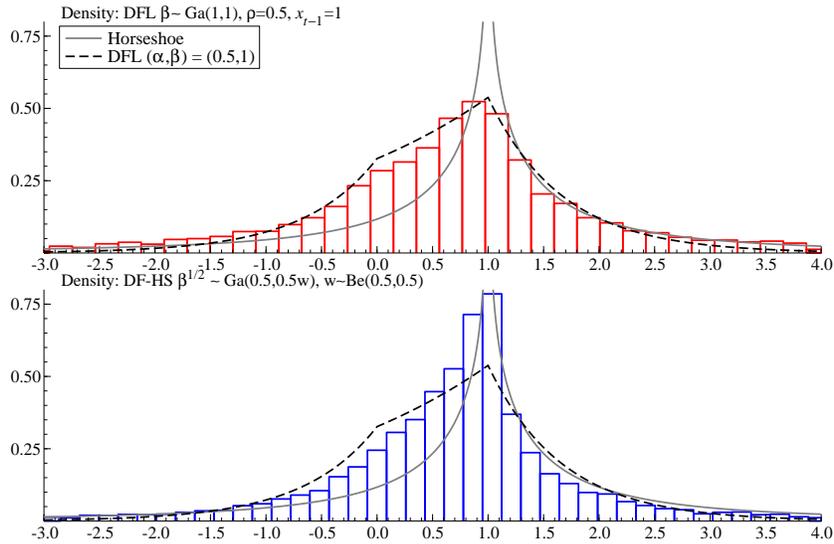}
	\caption{The prior density of $(x|x', \rho)$ with $x' = 1$, $\alpha = \rho \beta$ and $\rho = 0.5$  when $\beta$ is marginalized out. The histograms are drawn based on 10,000 random samples. The dashed line is the density of the DFL prior with $x'=1$, $\alpha = 0.5$ and $\beta=1$, as given in Proposition~\ref{prp:1}. The gray, solid lines shows the density function of horseshoe prior centered at $x'=1$. Top: $\beta \sim Ga(1,1)$. Bottom: $\beta ^2 \sim Ga(0.5,0.5\delta )$ and $\delta \sim Be(0.5,0.5)$. The DFL prior in the top panel with the conjugate prior for weight $\beta$ has the heavier tails than the DFL prior with fixed weights, but the spike of density around $x'=1$ is not found. The DFHS prior in the bottom panel shows the similarity to the horseshoe prior both in density tails and the spike around $x'=1$. }
	\label{fig:mdfl}
\end{figure}

Figure~\ref{fig:mdfl} shows the conditional distribution of $x_t|x_{t-1}=1$, where the weight parameters follow the priors introduced here and the conjugate priors in Section~\ref{sec:prior}, with $\rho = 0.5$. The DFL prior with the fixed weights has the thinner tails and overshrink the significant change of coefficients, while the conjugate priors in Section~\ref{sec:prior} has the heavier tails than the horseshoe prior to integrate such observations into posteriors. The spike around $x_{t-1}=1$, which reflects the strength in shrinking the dynamics of states, is not seen in the case of the hierarchical DFL model with the conjugate priors. In contrast, as the theory predicts, the DFHS prior exhibits the same spike of the density at $x_{t-1}=1$, and also the same heaviness of its tails as that of the horseshoe prior.

The conditional posteriors of $\beta$ and $\delta$ for Gibbs sampler are the known class of distributions. The full conditional of $\beta$ is, after the appropriate re-parametrization, a special case of extended gamma distributions. We customize the rejection sampling from this distribution, based on the discussions by \cite{finegold2011robust} and \cite{liu2012rejection}. 
The full conditional of $\delta$ belongs to the class of Kummer-beta distribution \citep{gordy1998generalization}, for which the method of random number generation is not trivial. We simply approximate the prior by the discrete distribution on the girds of $(0,1)$, whose probabilities are propositional to the densities of $Be(0.5,0.5)$, as we treat parameter $\rho$ in Section~\ref{sec:prior}. The approximation bias, or the deviation from the original prior by this discretization, is negligible.  See Section~S8 for the derivation and more discussions.

\subsection{Baseline of prior process} \label{sec:base}

The marginal prior mean of states is set to be zero in order that the additional shrinkage is directed to zero. The direction of shrinkage can be changed from zero to any value (baseline) and, in fact, estimated with some prior. Denote this baseline by $\mu$, and modify the transition density of the DFL process as 
\begin{equation*}
p(x_t|x_{t-1},\mu ) = \frac{f( x_t-\mu ) g(x_t,x_{t-1})}{h(x_{t-1}-\mu)}.
\end{equation*}
The parameter augmentation proved in Section~\ref{sec:DFL} is still valid for this prior. The only difference from the original DFL prior is that the value of the synthetic observation is now the baseline, i.e., $z_t = \mu$. The baseline is also the location parameter in the synthetic prior at $t=1$, $x_1\sim N(\mu , \lambda _{\alpha,1} )$. We use the normal prior for the baseline that is conditionally conjugate in the synthetic model. Alternatively, one can choose the scale mixture of normals to introduce the shrinkage effect on the baseline. Furthermore, although not pursued in our study, the modeling of baseline as the function of time, $\mu (t)$, could be considered under the specific applied context, where the concept of shrinkage is extended and directed to the pre-specified ``function,'' or $\mu (t)$.

\section{Application to state-space modeling} \label{sec:SSM}

\subsection{Estimation by Markov chain Monte Carlo} \label{sec:mcmc}

Consider the Gaussian and linear observational equation given by 
\begin{equation} \label{obs}
p(y_t|\theta _t) = N(y_t | F_t' \theta _t,V_t) ,
\end{equation}
where $F_t$ is $p{\times}1$ vector of predictors known at time $t$, $\theta _t = (\theta _{1t},\dots , \theta _{pt})'$ is the vector of state variables, and $V_t$ is the observational variance parameter and modeled later in Section~\ref{sec:sv}. Each state variable independently follows the DFL process, 
\begin{equation} \label{state}
p(\theta _{it}|\theta _{i,t-1}) \propto \exp \left\{ -\alpha_{i}|\theta _{it} - \mu _i| -\beta _{i}|\theta _{it}-\theta _{i,t-1}| \right\}
\end{equation}
where the baseline and weights are customized for each predictor $i$ and denoted by $\mu _i$ and $(\alpha _{i},\beta _{i})$. The CDLM representation of the prior given in Proposition~\ref{prp:3} is now combined with the ``real'' likelihood in (\ref{obs}). Following the notation in Proposition~\ref{prp:3}, the set of all the latent variables introduced for this state variable for the $i$-th state $\{ \theta _{it} \}$ is denoted by $\Lambda _{i,T} = \{ \lambda _{\alpha ,i,1}, \lambda _{\alpha ,i,T}, \lambda_{\beta ,i,2:T}, \lambda _{n,i,2:(T-1)} \}$ and $n_{i,2:(T-1)}$. The algorithm of Gibbs sampler for the posterior inference can be derived easily from the CDLM representation, where the forward filtering and backward sampling (FFBS) is utilized in sampling the state variables from the full conditionals \citep{carter1994gibbs,fruhwirth1994data}. 
 
\clearpage 
\noindent \hrulefill \\Gibbs sampler for the Bayesian dynamic fused LASSO models
\begin{enumerate}	
	\item Sampling $\theta _{1:T}$ by forward filtering and backward sampling (FFBS). 
	
	The conditional posterior of $\theta _{1:T}$ is equivalent to the posterior of the conditionally dynamic linear model with ``real'' likelihood in (\ref{obs}) and the following ``synthetic'' likelihoods and prior. For each $t\in2:(T-1)$, if $n_{it}>0$, then the model has the synthetic likelihood,
	\begin{equation*}
	z_{it}|\theta _{it} \sim N(\theta _{it} , \lambda _{n,it}), \ \ \ \ \ z_{it} = \mu _i
	\end{equation*}
	At $t=T$, the model always has the synthetic likelihood,
	\begin{equation*}
	z_{iT}|\theta _{iT} \sim N(\theta _{iT} , \lambda _{\alpha ,iT}), \ \ \ \ \ z_{iT} = \mu _i
	\end{equation*}
	The state evolution of the CDLM is defined by the synthetic prior; for $t>1$, 
	\begin{equation*}
	\theta _{it}|\theta _{i,t-1} \sim N(\theta _{i,t-1}, \lambda _{\beta ,it}), 
	\end{equation*}
	and, for $t=1$, 
	\begin{equation*}
	\theta _{i1} \sim N(\mu _i, \lambda _{\alpha ,i1}), 
	\end{equation*}
	By FFBS, one can sample from the full posterior of $\theta _{1:T}$ of this CDLM (e.g., \citealt[Chap. 4.8]{West1984a}). 
	
	\item Sampling $\Lambda _{1:p,T}$. 
	
	The components of $\Lambda _{1:p,T}$ are independently sampled from the full conditionals below: for $i=1,\dots , p$, 
	\begin{itemize}
		\item Sample $\lambda _{\alpha ,it}$ from $GIG(1/2,\alpha _i^2,(\theta _{it}-\mu_i)^2)$ for $t=1$ and $t=T$. 
		\item Sample $\lambda _{n,it}$ from $GIG(1/2,\{ n_{it}(\beta _i-\alpha _i) \} ^2,(\theta _{it}-\mu_i)^2)$ for $t\in \{ 2,\dots , T{-}1 \}$ if $n_{it} > 0$. 
		\item Sample $\lambda _{\beta ,it}$ from $GIG(1/2,\beta _{it}^2,(\theta _{it}-\theta _{i,t-1})^2)$ for $t=2,\dots , T$.
	\end{itemize}
	
	\item Sampling $n_{1:p,2:(T-1)}$. 
	
	The sampling of $n_{it}$ is based on the conditional posterior with $\lambda _{n,it}$ marginalized out. For $i=1,\dots ,p$ and $t=2,\dots, T{-}1$, the latent counts, $n_{it}$'s, are independently sampled from 
	\begin{equation*}
	n_{it} \sim Geo\left( \frac{\alpha _i}{\beta _i} e^{-(\beta _i-\alpha _i)|\theta _{it}-\mu_i|} \right) ,
	\end{equation*}
	where $Geo(q)$ means the geometric distribution; the probability function of random variable $N\sim Geo (q)$ is $Pr[N=n] = (1-q)q^n$ for $n\in \{ 0,1,2,\dots \}$. 
	
	\item Sampling $(\beta _{1:p}, \rho _{1:p})$.
	
	Depending on the choice of priors given in Section~\ref{sec:prior} and \ref{sec:hs}, one can sample from the conditional posteriors. See Section~S7.1 and S7.3 for details.  
	
	\item Sampling variances $V_t$: see the next subsection. 
	
\end{enumerate}
\hrulefill 

Note that the latent scales, $\Lambda _{i,t}$, are marginalized out when sampling the latent counts $n_{it}$ and weights $(\beta _i,\rho _i)$ to work on the double-exponential densities directly. This marginalization not only simplifies the sampling procedure but also facilitates the mixing of Markov chains (partially collapsed Gibbs sampler, \citealt{van2008partially}).

\subsection{Modeling of stochastic volatility} \label{sec:sv}

The modeling of observational variance $V_t$, or stochastic volatility, can be discussed independently of the use of the DFL priors; one can import an arbitrary model and computational method for $V_t$. In our study, we consider the following two models. For details, see also Section~S9.

The first example is the constant variance, $V_t = V$ for all $t$, in the ``scale-free'' DLMs that also includes $V$ in the state equation to scale the observational and state variances simultaneously. The conjugate prior for $V$ in the DLMs of this type is the inverse gamma gamma prior, $V^{-1}\sim Ga(n_0/2,n_0S_0/2)$ (\citealt[Chap. 4.5]{West1997}). The DFL process in \eq{eq:original} is applied to the scaled state variable $(x_t-\mu) / \sqrt{V}$; this affects the variance of the synthetic likelihood and prior but not the other parts of the model. Consequently, variance parameter $V$ appears in the synthetic likelihoods and prior of the CDLM representation in Section~\ref{sec:mcmc} as their scales, for which the FFBS is available to sample from the conditional joint posterior of states and observational variance. We consider this model in the simulation study, Section~\ref{sec:DFL}. 

In contrast, if the stochastic volatility appears only in observational equation (\ref{obs}), the MCMC algorithm in Section~\ref{sec:mcmc} is easily extended by incorporating the sampling of this parameter. We consider the model of this type as the second example and, specifically, chose the log-Gaussian models for the empirical study in Section~\ref{sec:LTM}; log-volatility $h_t = \log (V_t)$ follows the Gaussian AR(1) process as,
\begin{equation*}
h_t - \mu _h = \phi _h ( h_{t-1}-\mu ) + \eta _t, \ \ \ \ \ \eta:ind \sim N(0,\sigma _h^2),
\end{equation*}
with the normal, beta and inverse-gamma priors for AR(1) parameters $(\mu _h, \phi _h, \sigma ^2_h)$. The posterior computation is a challenge in computational statistics. In Section~\ref{sec:LTM}, we used multi-move sampler \citep{shephard1997likelihood,watanabe2004multi} that is also used in the existing literature, but we would like to note that there has been continuing advancement in this area (e.g., \citealt{kastner2014ancillarity}). 

Although not discussed in this research, the conjugate inverse gamma-scaled beta process can also be considered as a simpler alternative for the computational ease (\citealt[Chap. 10.8]{West1997}).

\section{Illustration via data analysis} \label{sec:data}

The posterior analysis by the DLM with the DFL/DFHS priors is conducted for the simulated and real datasets. In the simulation study, the proposed model is compared with the standard dynamic models with double-exponential/horseshoe priors to clarify the difference in their shrinkage effects. In the analysis of real macroeconomic data, the patterns of shrinkage under the DFL priors is highlighted in comparison with the dynamic variable selection by the latent threshold models. 

All computations are implemented by Ox \citep{ox}.

\subsection{Simulated dataset} \label{sec:sim}

\begin{figure}[!ht]
	\centering
	\begin{subfigure}{.45\textwidth}
		\centering
		\includegraphics[width = 1\textwidth]{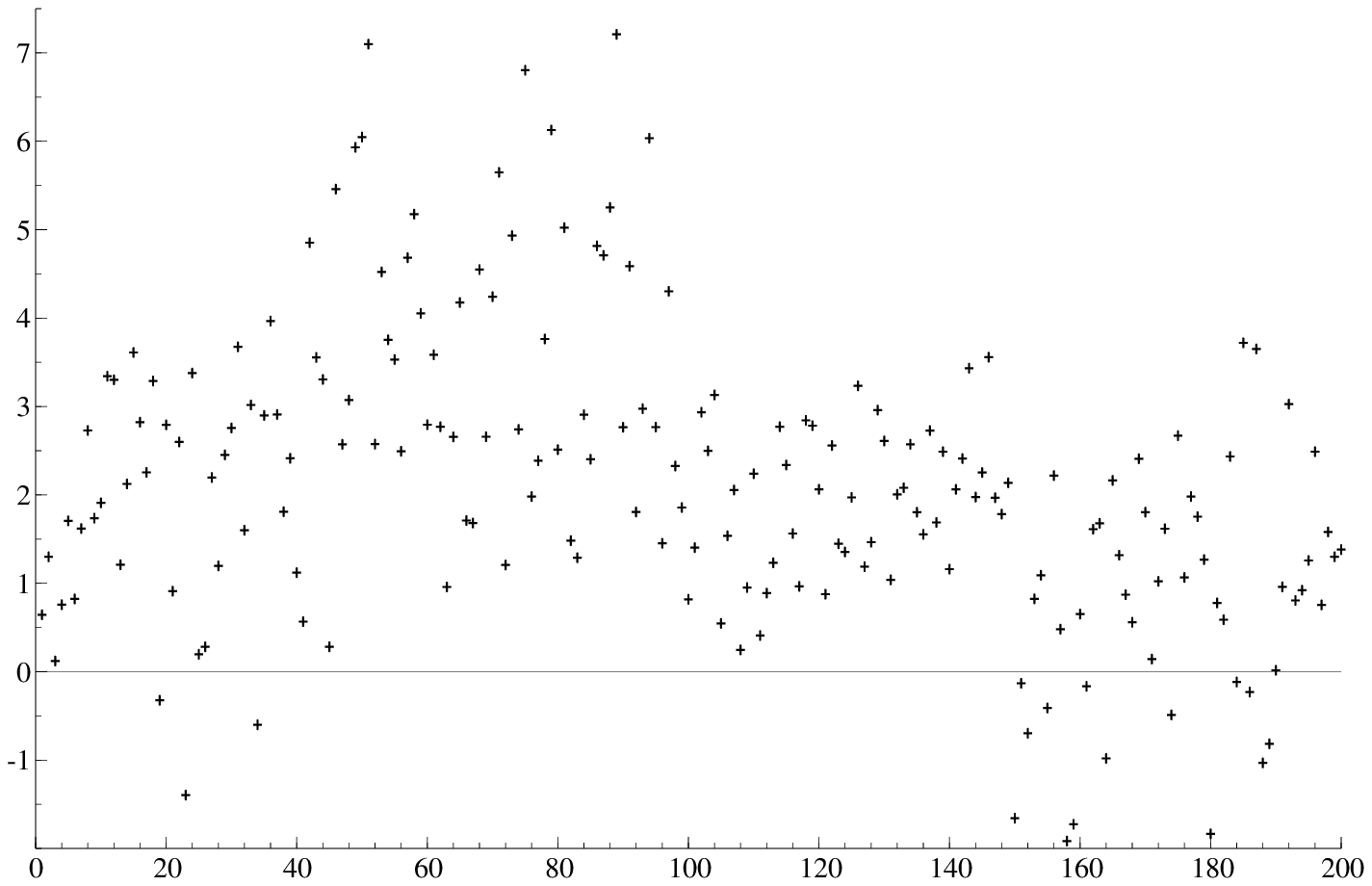}
		\caption{Simulated data $y_t$}
		\label{subfig:simdata}
	\end{subfigure}
	\begin{subfigure}{.45\textwidth}
		\centering
		\includegraphics[width = 1\textwidth]{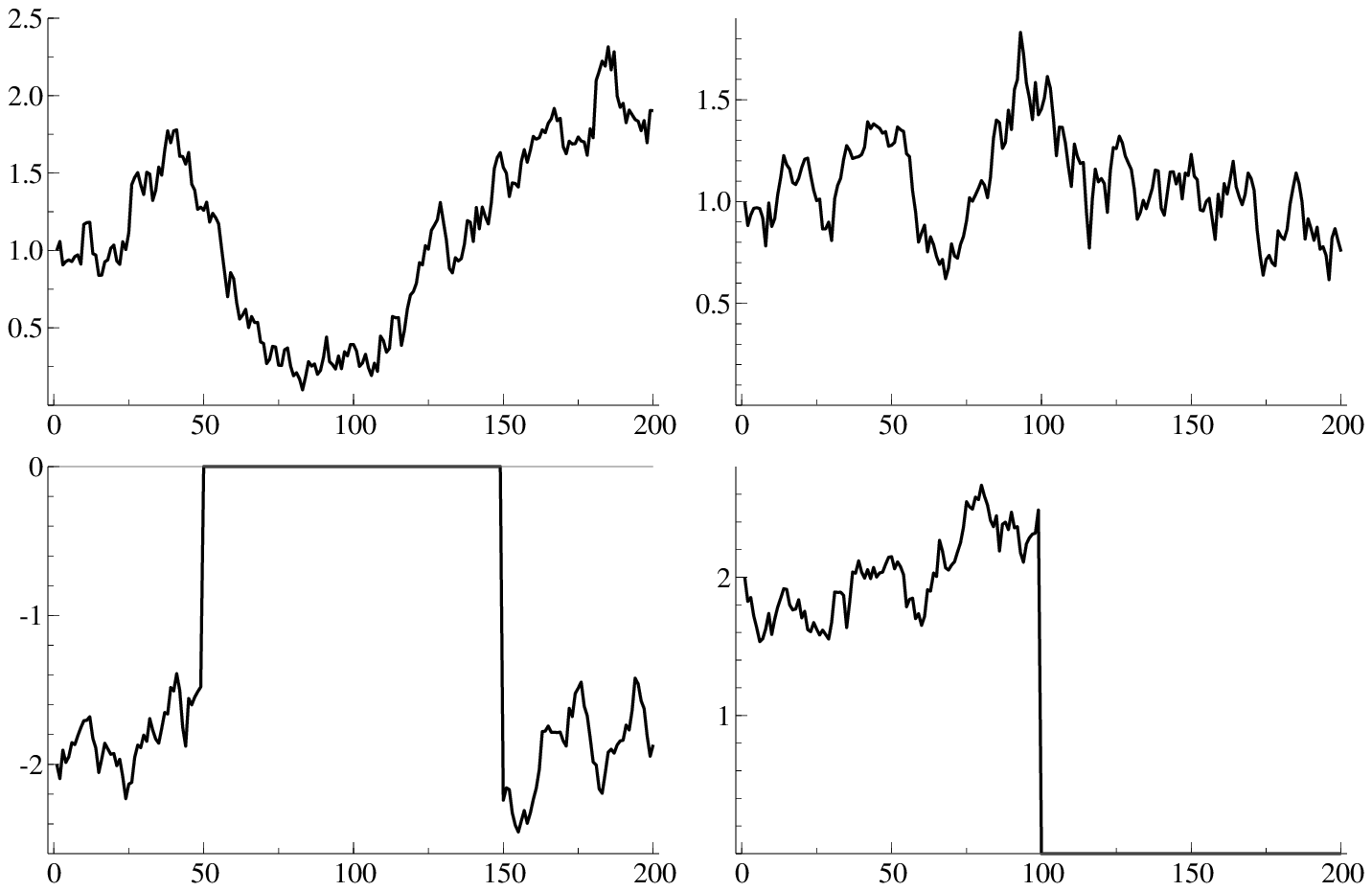}
		\caption{True values of dynamic coefficients $\theta _{1:4,t}$}
		\label{subfig:truebeta}
	\end{subfigure}
	\caption{Left: the simulated time series $\{ y_t \} _{t=1:T}$ with $T=200$. Right: the true values of dynamic coefficients $\theta _{1:4,t}$. The first predictor is constant and unity (intercept). The third and forth predictors are temporarily inactive, i.e., the coefficients become exact zero in certain periods; $\theta _{3t} = 0$ for $50\le t < 150$ and $\theta _{4t}=0$ for $t>100$. The other predictors do not contribute to the prediction of $y_t$ at all, i.e., $\theta _{it} = 0$ for all $t$ and $i\in \{ 5,6,\dots ,12\}$. }
\end{figure}

The purpose of the study in this subsection is the illustration and comparative analysis of the proposed and existing models. The univariate time series $\{ y_t \}$ is generated from (\ref{obs}), based on the predictors and the true values of the parameters specified below. Twelve predictors are generated from the uniform distribution on $(0,2)$, except for the first predictor that is constant as the intercept. The first four predictors are ``active'' and the others are ``inactive,'' in the sense that the coefficients of the former are non-zeros for all/some $t$ and the others are always zero, creating the situation suitable for the over-parametrized but sparse linear model. The data process $\{ y_t \}$ is plotted in Figure~\ref{subfig:simdata}, with the coefficients of the four active predictors, $\{ \theta _{1:4,t} \}$, in \ref{subfig:truebeta}. The true value of the observational variance is $V_t = V = 1.5$ for all $t$.

For comparison, in addition to the DFL and DFHS priors, we also consider the double-exponential (DE) and horseshoe (HS) priors, i.e., the DFL/DFHS with $\alpha =0$. For clarity, we explicitly indicate the type of prior in the name of models, as in ``DFL-DLMs.'' These dynamic models have the same likelihood in (\ref{obs}) and differ in the modeling of state evolution. The observational variance is assumed to be constant over time, $V_t = V$ for all $t$, and to follow inverse gamma prior $V^{-1} \sim Ga(n_0/2,n_0S_0/2)$; see Section~\ref{sec:sv} (and S9.1). We set $n_0 = 1$ and $S_0=1$ in all the models. For baseline $\mu _i$, we set $\mu _i=0$ or $\mu _i:iid\sim N(0,10^2)$. The weight parameters $\beta _i$'s follow the independent gamma priors as $\beta _i:iid \sim Ga(1,0.1)$ in all models. For the parameter $\rho _i$ in the DFL-DLMs and DFHS-DLMs, we consider (i) $\rho _i:iid \sim Be(1,2)$, (ii) $Be(1,5)$ and (iii) $Be(1,10)$. In the DE-DLMs and HS-DLMs, the initial state at $t=1$ is modeled by $\theta _{i1} \sim DE(\mu _i|\alpha _i)$ for the fair comparison with the DLF-DLMs. The three sets of hyperparameters are used:  (i) $\alpha _i \sim Ga(1,0.1)$, (ii) $Ga(0.5,0.5)$ and (iii) $Ga(1,10)$. The posteriors are computed based on 3,000 iterations after 300 burn-in period. In predictive analysis, the posterior computation by the MCMC method is repeated for the dataset $y_{1:s}$ for each $s$, including the estimation of the model parameters $(\beta _i , \rho _i)$, to generate $y_{s+1}$ from the one-step ahead predictive distribution during the iterations of MCMC and compute the point forecast $E[y_{s+1}|y_{1:s}]$. To sample from the predictive distributions, it is necessary to simulate $\theta _{s+1,i}$ from prior $p(\theta _{s+1,i}|\theta _{s,i})$; see Section~S6 for the technical details. 

\subsubsection*{Posterior and predictive results}

In Table~\ref{tab:summary1}, the posterior analyses of all the twelve models with various DFL/DE priors are summarized by the averages of the mean squared errors (MSEs) of the Bayes estimates (posterior means) for the dynamic coefficients and one-step ahead predictions. The computation of MSEs starts at $t=25$, because the posteriors of DE-DLM with baseline zero are strongly biased to zero due to the prior of the initial state variable, as seen later in Figure~\ref{subfig:beta1}. Among the models listed in the table, the DFL-DLMs with baseline zero perform better in almost all of the MSE measures, especially in the estimation of dynamically sparse coefficients, $\theta _{3t}$, and those of noises, $\theta _{it}$ for $i\ge5$, supported by the newly-added shrinkage. In contrast, the MSEs of the active state variables for DE-DLMs are large due to the strong penalty on the dynamics assumed in the prior. The exception is the estimation of coefficient $\theta _{2t}$, where the shrinkage to zero is not coherent with the true values of $\theta _{2t}$ and, probably for this reason, the DE-DLMs are more successful. The introduction of baselines, which is expected to improve the fitting of the models, increases the uncertainty in estimation and prediction under the DFL-DLMs, which could also explain their disproved predictive performance. This is because the baselines are redundant parameters for inactive predictors, in addition to the inefficiency for the estimation of the volatile state variables; see the discussions in Section~S10.2 and Figure~S7. In the following, we focus on the DFL-DLM with the mild shrinkage to zero, $\cM_3$, the best DE-DLM in predictions, $\cM_9$, to investigate their posterior and predictive distributions.

\begin{table}[!htbp]
	\begin{center}
		\caption{\small (DFL and DE) Mean squared errors of coefficients ($\times 10^4$) and one-step ahead forecasting ($\times 10^3$), the lengths of 95\% credible intervals of forecasting distributions (CIs) and their empirical coverage rates (CRs) for $t\ge 25$. For the DFL models, we assume (i) $\rho \sim Be(1,2)$, (ii) $\rho \sim Be(1,5)$ and (iii) $\rho \sim Be(1,10)$. For the DE models, (i) $\alpha \sim Ga(1,0.1)$, (ii) $\alpha \sim Ga(0.5,0.5)$ and (iii) $\alpha \sim Ga(1,10)$. } \label{tab:summary1}
		{\small 
			\begin{tabular}{lccccccccccc}
				\toprule
				No. & Model & $\rho$ or $\alpha$ & Baseline & $\theta _{1t}$ & $\theta _{2t}$ &  $\theta _{3t}$ &  $\theta _{4t}$ & $\theta _{5:12,t}$ &  $y_{t+1}$ & CIs & CRs \\
				\midrule
				$\cM_1$    &DFL & i   & 0       & 250.93 & 136.80 & 199.95 & 189.68 & 0.08 & 250.08 & 1.64 & 0.94 \\ 
				$\cM_2$    &    & ii  & 0       & 249.38 & 134.22 & 198.34 & 190.97 & 0.16 & 246.35 & 1.65 & 0.94 \\ 
				$\cM_3$    &    & iii & 0       & 241.12 & 130.19 & 192.59 & 186.73 & 0.53 & 247.59 & 1.64 & 0.94 \\ 
				$\cM_4$    &    & i   & $\mu_i$ & 275.11 & 154.23 & 219.91 & 212.66 & 20.45 & 344.09 & 1.75 & 0.91 \\ 
				$\cM_5$    &    & ii  & $\mu_i$ & 293.88 & 147.81 & 208.71 & 215.10 & 16.52 & 289.69 & 1.75 & 0.93 \\ 
				$\cM_6$    &    & iii & $\mu_i$ & 290.35 & 139.77 & 207.66 & 206.17 & 13.84 & 301.74 & 1.75 & 0.92 \\ \midrule
				$\cM_7$    & DE & i   & 0       & 312.76 & 129.69 & 207.31 & 226.57 & 14.61 & 278.45 & 2.38 & 0.97 \\ 
				$\cM_8$    &    & ii  & 0       & 284.38 & 134.92 & 202.55 & 224.75 & 12.64 & 277.71 & 2.43 & 0.97 \\ 
				$\cM_9$    &    & iii & 0       & 277.77 & 137.32 & 202.06 & 225.57 & 10.05 & 275.67 & 2.47 & 0.98 \\ 
				$\cM_{10}$ &    & i   & $\mu_i$ & 337.10 & 129.38 & 199.84 & 223.88 & 18.12 & 304.78 & 1.96 & 0.95 \\ 
				$\cM_{11}$ &    & ii  & $\mu_i$ & 345.93 & 122.63 & 205.24 & 214.07 & 18.07 & 304.05 & 1.96 & 0.96 \\ 
				$\cM_{12}$ &    & iii & $\mu_i$ & 333.70 & 129.71 & 200.33 & 222.79 & 17.86 & 304.85 & 1.96 & 0.95 \\ \bottomrule 
			\end{tabular} 
		}
	\end{center}
\end{table}%

\begin{table}[!htbp]
	\begin{center}
		\caption{\small (DFHS and HS) Mean squared errors of coefficients ($\times 10^4$) and one-step ahead forecasting ($\times 10^3$), the lengths of 95\% credible intervals of forecasting distributions (CIs) and their empirical coverage rates (CRs) for $t\ge 25$. For the DFHS models, we assume (i) $\rho \sim Be(1,2)$, (ii) $\rho \sim Be(1,5)$ and (iii) $\rho \sim Be(1,10)$. For the HS models, (i) $\alpha \sim Ga(1,0.1)$, (ii) $\alpha \sim Ga(0.5,0.5)$ and (iii) $\alpha \sim Ga(1,10)$.} \label{tab:summary2}
		{\small 
			\begin{tabular}{lccccccccccc}
				\toprule
				No. & Model & $\rho$ or $\alpha$ & Baseline & $\theta _{1t}$ & $\theta _{2t}$ &  $\theta _{3t}$ &  $\theta _{4t}$ & $\theta _{5:12,t}$ &  $y_{t+1}$ & CIs & CRs \\
				\midrule
				$\cM_{13}$ &DFHS& i   & 0       & 233.08 & 132.30 & 197.08 & 194.20 & 1.10 & 248.47 & 1.82 & 0.95 \\ 
				$\cM_{14}$ &    & ii  & 0       & 243.96 & 130.55 & 196.14 & 207.46 & 1.72 & 251.31 & 1.89 & 0.95 \\ 
				$\cM_{15}$ &    & iii & 0       & 259.61 & 127.47 & 191.68 & 208.54 & 3.34 & 257.21 & 1.95 & 0.97 \\ 
				$\cM_{16}$ &    & i   & $\mu_i$ & 355.53 & 142.04 & 215.38 & 219.99 & 16.58 & 320.10 & 1.93 & 0.94 \\ 
				$\cM_{17}$ &    & ii  & $\mu_i$ & 341.35 & 142.37 & 208.04 & 229.33 & 19.11 & 309.38 & 2.03 & 0.95 \\ 
				$\cM_{18}$ &    & iii & $\mu_i$ & 385.78 & 139.92 & 205.93 & 222.58 & 17.56 & 301.53 & 2.13 & 0.96 \\ \midrule
				$\cM_{19}$ & HS & i   & 0       & 260.62 & 138.17 & 199.10 & 216.92 & 11.66 & 272.37 & 2.43 & 0.98 \\ 
				$\cM_{20}$ &    & ii  & 0       & 274.58 & 136.65 & 201.25 & 213.32 & 9.93 & 274.60 & 2.43 & 0.98 \\ 
				$\cM_{21}$ &    & iii & 0       & 266.22 & 135.74 & 201.99 & 226.11 & 9.40 & 272.69 & 2.42 & 0.98 \\ 
				$\cM_{22}$ &    & i   & $\mu_i$ & 319.69 & 128.58 & 201.88 & 221.49 & 17.26 & 308.28 & 2.00 & 0.96 \\ 
				$\cM_{23}$ &    & ii  & $\mu_i$ & 329.86 & 130.62 & 198.41 & 218.08 & 18.32 & 310.41 & 1.98 & 0.96 \\ 
				$\cM_{24}$ &    & iii & $\mu_i$ & 326.85 & 129.95 & 200.13 & 216.51 & 17.57 & 302.03 & 1.97 & 0.97 \\ 
				\bottomrule 
			\end{tabular} 
		}
	\end{center}
\end{table}%

The same posterior and predictive summaries for the DFHS-DLMs and HS-DLMs are given in Table~\ref{tab:summary2}. The DFHS-DLMs are as competitive as, or slightly improved from, the DFL-DLMs in the estimation of the non-zero coefficients, while the MSEs for zero coefficients and predictions slightly increase under the DFHS-DLMs due to the strengthened shrinkage on dynamics. One notable observation here is that the best DFHS-DLM in terms of predictive performance is $\cM_{13}$ with the strongest shrinkage to zero, unlike the case of DFL-DLMs. It could be explained as that the balance of two conflicting shrinkage effects is affected by the use of horseshoe prior, which change the optimal amount of shrinkage to zero.

\begin{figure}[!htbp]
	\centering
	\includegraphics[width=6in]{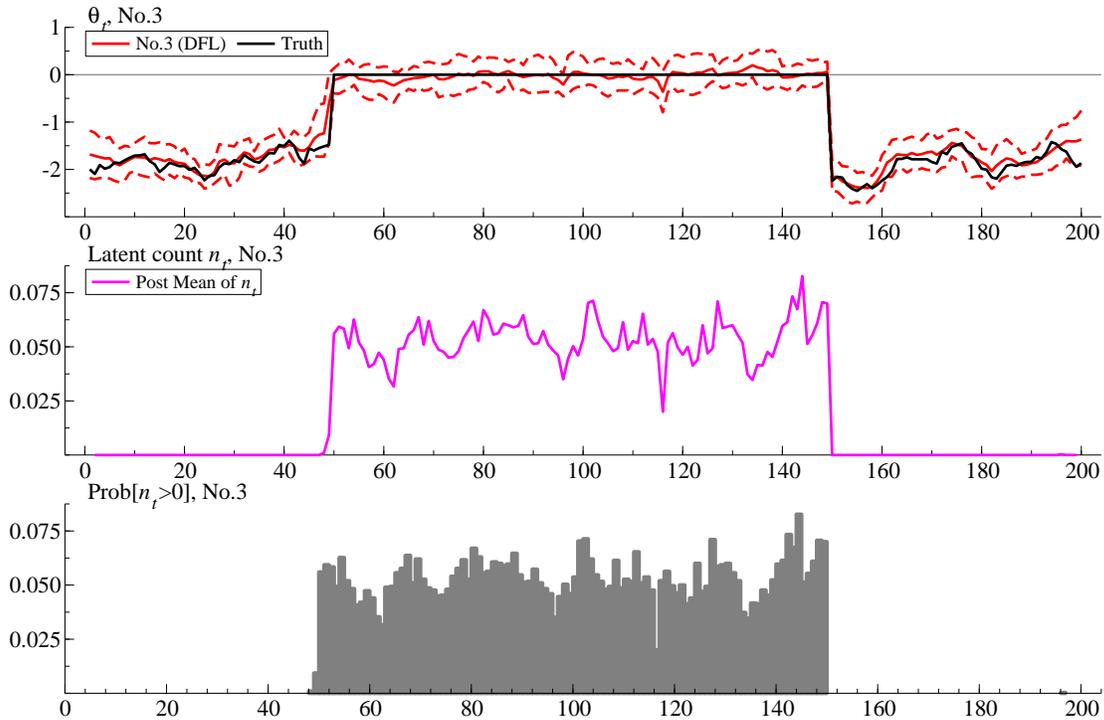}
	\caption{\small Top: the posterior means and 95\% credible intervals of $\theta _{3t}$ with Model 1 (red), compared with the true values (black). Middle: the posterior means of latent count $n_{3t}$ (pink). They spike up only in the period when this predictor becomes inactive in the data generating process, suggesting that this latent variable indicates the insignificance of the predictor. Bottom: the posterior probability of positive counts, i.e., $Pr[n_{3t}>0]$. In the period of interest, this probability becomes significantly higher than those in the other periods, but its value is at most 0.075. 
	} \label{fig:comp} 
\end{figure}

In the estimation of dynamic coefficients, the third predictor is of the special interest for its dynamic ``significance.'' The posterior of $\theta _{3t}$ computed by model $\cM_3$ is shown in Figure~\ref{fig:comp} with its true values, the posterior means of the latent count $n_{3t}$, and the posterior probabilities of positive count $Pr[n_{3t}>0|y_{1:T}]$. 
In addition to the success of the posterior of $\theta _{3t}$ in tracking the true values in the top panel of this figure, we can confirm in the middle and bottom panels that the latent count is more likely to be positive in the period when the predictor is inactive. This result is easily expected from the structure of the augmented model; if the positive count is sampled, then we have the additional observational equation of $z_{3t}=0$ in the synthetic model, whose information helps to shrink the state variable to zero in the posterior. This clear correspondence between the sparsity in state variables and the values of latent counts is not assumed a priori but the result of posterior learning (Section~S10.1), hence implies the potential interpretation of the posterior of $n_{it}$ as the indicator of ``insignificance'' of state variables. This point is further examined in the next subsection through the application to the real macroeconomic dataset.

\begin{figure}[!htbp]
	\centering
	\includegraphics[width=5in]{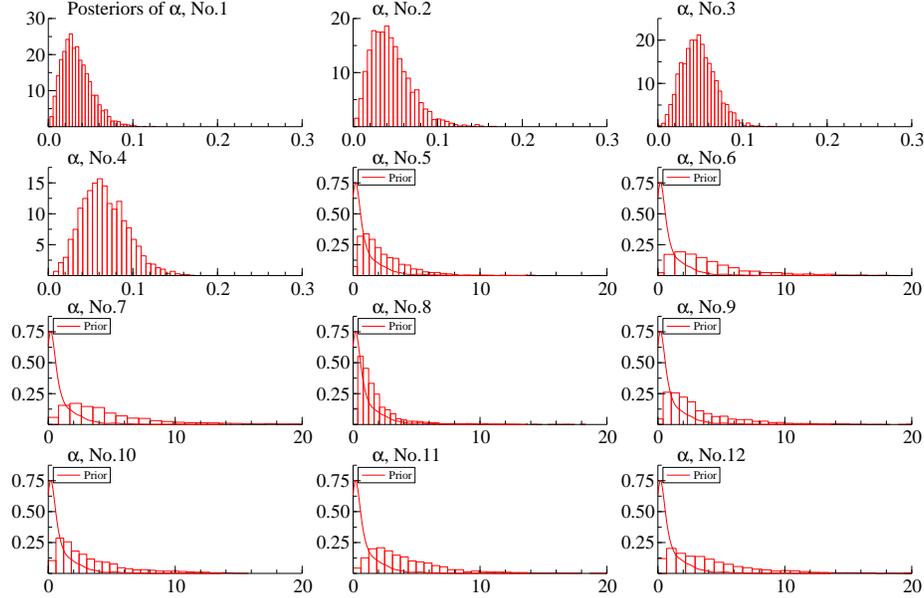}
	\caption{\small The posteriors of weight parameters $\alpha _i$ for $i\in 1:12$. For the four active predictors, the value of $\alpha _i$ is extremely small, concluding that the DFL priors for these coefficients are almost equivalent to the DE priors. For the inactive predictors, the weights can take larger values as well. In fact, the posteriors are almost the same as the priors (drawn by the solid lines). 
	} \label{fig:alpha} 
\end{figure}

\begin{figure}[!htbp]
	\centering
	\begin{subfigure}{.45\textwidth}
		\centering
		\includegraphics[width = 1\textwidth]{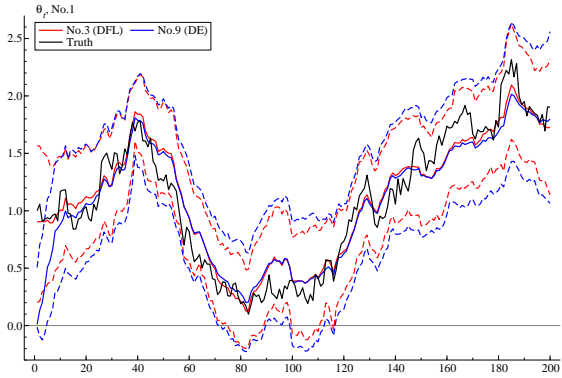}
		\caption{The posteriors of $\theta _{1t}$}
		\label{subfig:beta1}
	\end{subfigure}
	\begin{subfigure}{.45\textwidth}
		\centering
		\includegraphics[width = 1\textwidth]{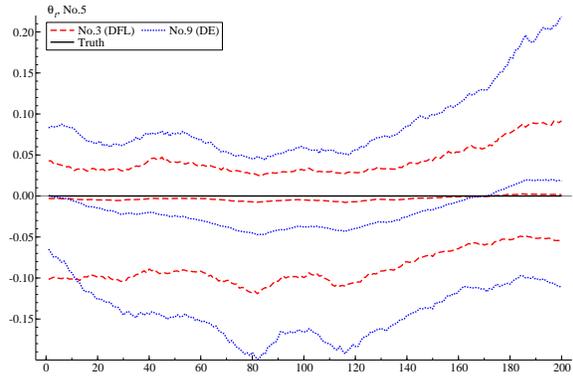}
		\caption{The posteriors of $\theta _{5t}$}
		\label{subfig:beta5}
	\end{subfigure}
	\caption{Left: the posterior means and 95\% credible intervals of $\theta _{1t}$ of the DFL-DLM $\cM_3$ (red) and the DE-DLM $\cM_9$ (blue) with true values (black). The two posteriors are almost identical after $t\ge 25$ both in the locations and uncertainty quantification. This is because of the small $\alpha _1$ observed in Figure~\ref{fig:alpha}, with which the DFL prior is approximately the DE prior. Right: the same posterior plots for $\theta _{5t}$. For the inactive predictors, including this example, the posterior of the DFL-DLM concentrates around zero over time. This is also due to $\alpha _i$, that is large now and induces the additional shrinkage to zero at each time point. In another DFL-DLM $\cM_1$, the posterior is almost degenerate at zero for its strong shrinkage effect to zero assumed in the prior.}
\end{figure}

The posteriors of $\alpha _i$'s are listed in Figure~\ref{fig:alpha}. It is clear in this figure that the values of these weights become extremely small for the active predictors, while being sufficiently large for the inactive predictors. The extremely small value of weight $\alpha _i$ indicates that the first penalty, or the shrinkage effect to zero, is almost negligible. The large value of weight $\alpha _i$ forces the model to shrink the state variable to zero more aggressively, which is the appropriate treatment for the zero coefficients.

Figure~\ref{subfig:beta1} and \ref{subfig:beta5} show the superimposed posterior trajectories of the state variables, $\theta _{1t}$ and $\theta _{5t}$, of the DFL-DLM $\cM_3$ and DE-DLM $\cM_9$. For the active predictor, as seen in the example of $\theta _{1t}$, the posterior means and 95\% credible intervals of both models almost overlap one another all the time. This result is consistent with our observation in Figure~\ref{fig:alpha} that the posterior of weight $\alpha _i$ concentrates around small values and the DFL prior reduces to the DE prior. In contrast, the difference of the two models is clear in the uncertainty of state variables for the inactive predictors. The large weight on the first penalty induces the strong shrinkage to zero, shrinking the posterior locations to zero and narrowing the credible intervals. 
Consequently, those noise predictors do not contribute to the variation of observation $y_t$ in the predictive analysis.

\begin{figure}[!htbp]
	\centering
	\begin{subfigure}{.45\textwidth}
		\centering
		\includegraphics[width = 1\textwidth]{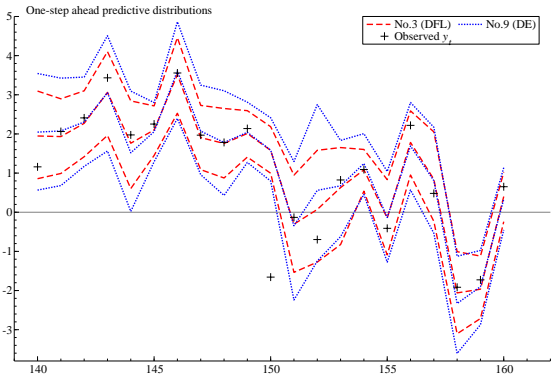}
		\caption{Forecasting by DFL-DLM $\cM_3$ and DE-DLM $\cM_3$}
		\label{subfig:pred1}
	\end{subfigure}
	\begin{subfigure}{.45\textwidth}
		\centering
		\includegraphics[width = 1\textwidth]{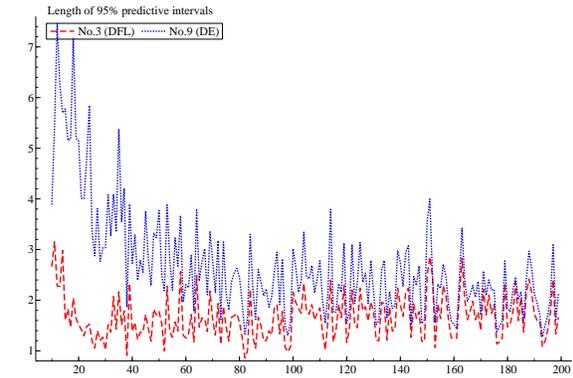}
		\caption{Length of 95\% credible intervals}
		\label{subfig:pred2}
	\end{subfigure}
	\caption{Left: The means and 95\% credible intervals of the one-step ahead forecast distributions $p(y_{t+1}|y_{1:t})$ for $t \in [140,160]$ under the DFL-DLM $\cM_3$ (red, dashed) and DE-DLM $\cM_9$ (blue, dotted), Right: the lengths of intervals shown in the left panel. Both models are able to predict the next observation reasonably, but the DFL-DLM is more successful in uncertainty quantification for its narrower credible intervals. This difference is especially evident before $t=150$, at which the true value $\theta _{3t}$ switches from zero to non-zero value. The accuracy of point forecasting in these models can be seen in Table~\ref{tab:summary1}. }
\end{figure}

The twelve models are repeatedly estimated by the MCMC method in the sequential way, starting at time $t=25$, to provide the one-step ahead forecast distributions. In addition to the summary of predictive performances in Table~\ref{tab:summary1}, we here focus on $\cM_3$ and $\cM_9$ and their predictions at each time point. 
Figure~\ref{subfig:pred1} shows the predictive means and 95\% credible intervals of $\cM_3$ and $\cM_9$ with the actual observations at $t\in [140,160]$. Notably, the predictive uncertainty under $\cM_3$ is smaller than $\cM_9$ more clearly before $t=150$. This is exactly the change-point for $\theta_{3t}$ to becomes non-zero again and, before $t=150$, model $\cM_3$ benefits the shrinkage to zero and could improve the predictive performance. The dynamics of the lengths of 95\% credible intervals, which are summarized into the averages in Table~\ref{tab:summary1}, are shown in Figure~\ref{subfig:pred2}, where we confirm that the predictive intervals of $\cM_3$ is narrower than $\cM_9$ not just on average, but at all $t$.

\subsection{Macroeconomic data and latent threshold models} \label{sec:LTM}

The posterior plot of latent counts $n_{it}$'s in Figure~\ref{fig:comp} suggests, informally but empirically, the potential use of this quantity as the indicator of ``(in)significance'' of coefficients. In this section, we further examine this aspect of the DFL prior-- the (dis)similarity of shrinkage to zero and variable selection-- through the comparison with the latent threshold models (LTMs), that are the more formal approach to the dynamic variable selection (e.g., \citealt{Nakajima2013jfe,NakajimaWest2015DSP}). The LTMs explicitly distinguish the latent and realized state variables, $\theta _{it}^{\ast}$ and $\theta _{it}$ for $i$-th coefficient, and connect them by the indicator function as, 
\begin{equation} \label{ltm}
\theta _{it} = s_{it} \theta _{it}^{\ast}, \ \ \ \ \ \ \mathrm{and} \ \ \ \ \ \ \ s_{it} = 1[ \ |\theta _{it}^{\ast} | > d_i \ ],
\end{equation}
where latent state variable $\theta _{it}^{\ast}$ follows a Gaussian AR(1) process and $d_i$ is some positive, non-dynamic threshold. The probability that this coefficient is thresholded to zero at time $t$ is parametrized by $Pr[s_t = 0]$ and can be computed explicitly in the posterior analysis. The posterior analysis of the LTMs by the MCMC method is feasible, but computationally costly both in coding and computational time, as typical in the models of variable selection. 

The example of the analysis of the US macroeconomic variables by the LTMs are taken from \cite{NakajimaWest2013JBES}. The vector of the inflation, unemployment and nominal interest rates is denoted by $y_t$ and observed quarterly between 1977 and 2007. The base model is the time-varying vector autoregressive model with order 3, the likelihood of which is 
\begin{equation*}
y_t = c_t + B_{1t} y_{t-1} + B_{2t} y_{t-2} + B_{3t} y_{t-3} + N(0,\Omega _t^{-1}), \ \ \ \ \ \ \ \Omega _t= (I-A_t)\Lambda _t (I-A_t)',
\end{equation*}
where $\Lambda _t$ is the diagonal matrix of log-Gaussian stochastic volatilities and $A_t$ is the lower-triangular matrix with diagonal zeros. Each entry of $c_t$, $B_{1:3,t}$ and $A_t$ follows the Gaussian, stationary AR(1) process with latent threshold in (\ref{ltm}). The parameters of interest are $B_{1:3,t}$ and $A_t$, that determine the lagged/simultaneous correlation between variables. The graphical, conditional dependence structure of the three macroeconomic variables captured by $A_t$, or $\Omega _t$, is of great importance in the macroeconomic studies. We followed the computational procedure and used the same hyperparameter settings in \cite{NakajimaWest2013JBES}. 

In this study, our interest is in whether the time-varying vector autoregressive models with the DFL prior produces the same patterns of sparse structure in $B_{1:3,t}$ and $A_t$ as the LTM does. We replace the prior processes for $c_t$, $B_{1:3,t}$ and $A_t$ by the DFL priors, and compute the posterior of the latent counts $n_{it}$, or $Pr[n_{it}>0|y_{1:T}]$, that are compared with the posterior exclusion probability $Pr[ s_{it} = 0 | y_{1:T} ]$ of the LTM. To make the DFL prior compatible with the prior processes used in \cite{NakajimaWest2013JBES} for state variable that are very persistent over time, we adjust the hyperparameters in the DFL prior as $\beta \sim Ga(500,10)$ and $\rho \sim Be(0.1,1)$ for all the entries of $c_t$ and $B_{1:3,t}$, and $Ga(50,1)$ and $\rho \sim Be(0.1,1)$  for $A_t$. For this choice of priors, we confirmed that the posteriors of error variances, or stochastic volatilities, become almost identical in both DFL-DLM and LTM (Section~S11.1). 
The posteriors are computed by the MCMC method of 50,000 iterations after 5,000 burn-in. The chain is set relatively long, due to the slow convergence of the regression coefficients in the LTM that are sampled by the single-move sampler, not by FFBS, as reported in \cite{NakajimaWest2013JBES}. 

\begin{figure}[!ht]
	\centering
	\begin{subfigure}{.45\textwidth}
		\centering
		\includegraphics[width = 1\textwidth]{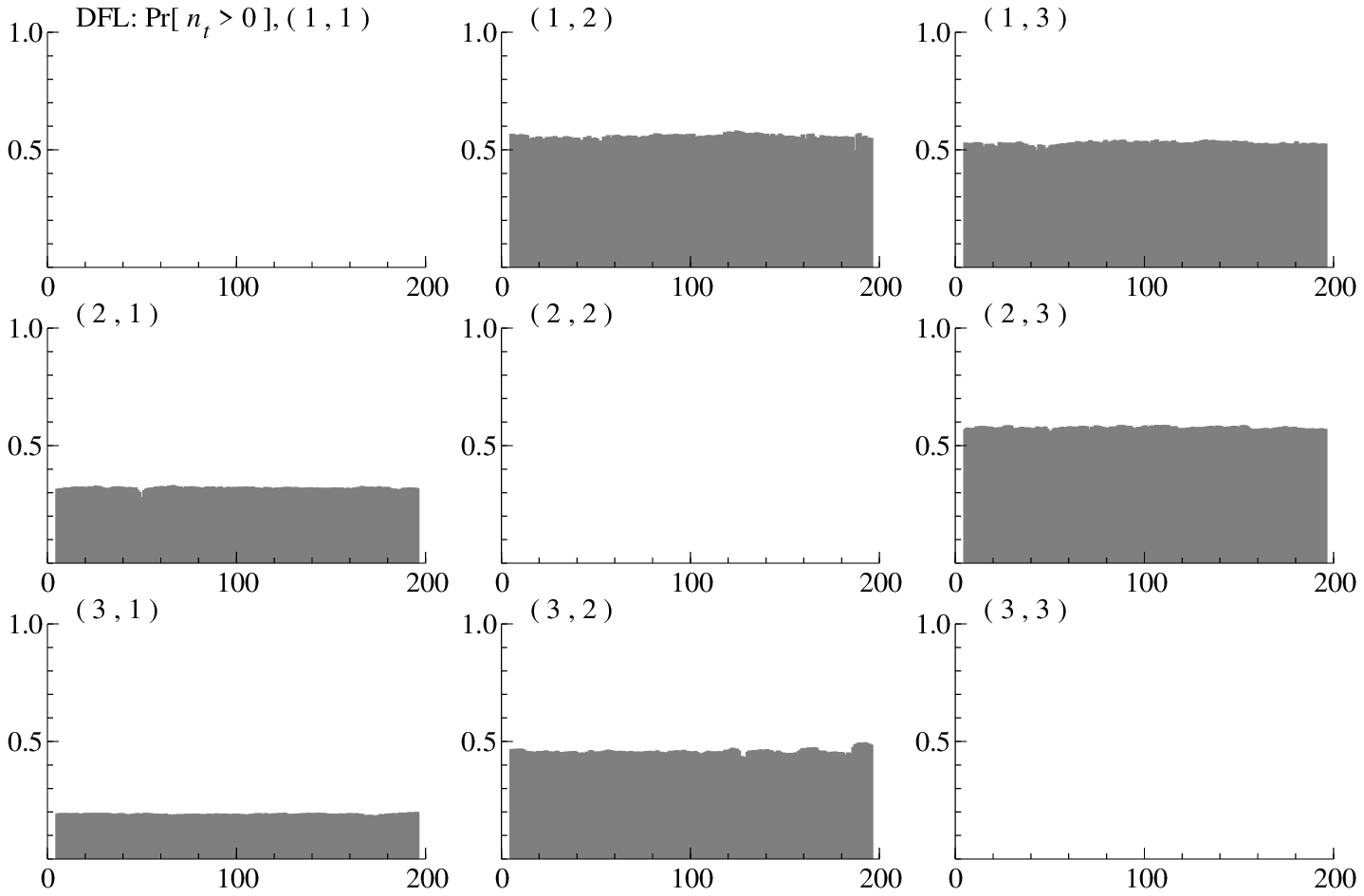}
		\caption{DFL: Posteriors of positive counts.}
		\label{subfig:DFLsb}
	\end{subfigure}
	\begin{subfigure}{.45\textwidth}
		\centering
		\includegraphics[width = 1\textwidth]{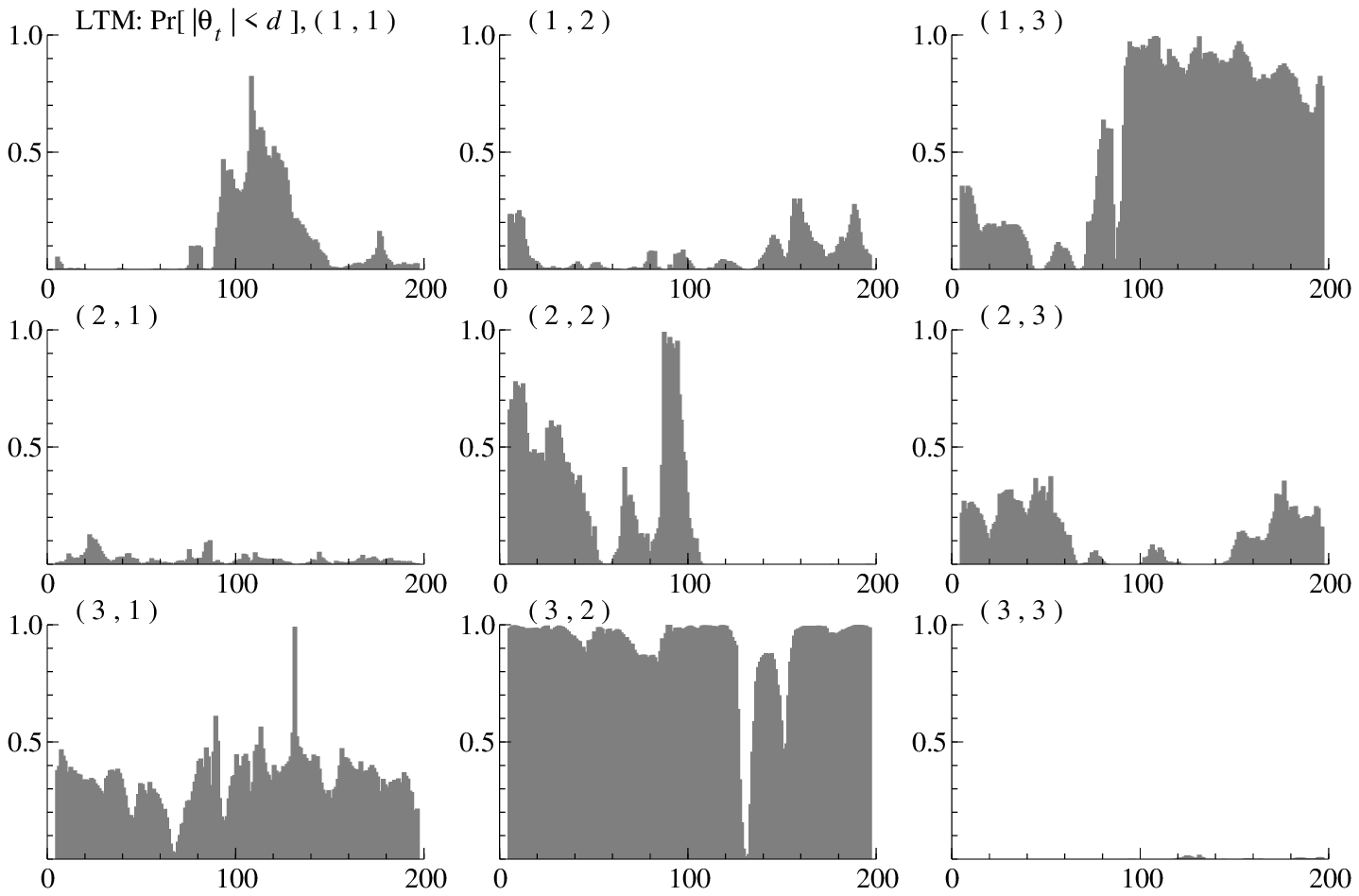}
		\caption{LTM: Posterior of thresholding $B_{ijt}$. }
		\label{subfig:LTMsb}
	\end{subfigure}
	\caption{Left: the posterior probability of observing positive latent counts for each of $B_{1t}$ in the DFL-DLM. Right: the posterior probability of the LTM that $|\theta _{ijt}|<d_{ij}$, where $\theta _{ijt}$ is the latent version of the $(i,j)$-entry of $B_{1t}$ and $d_{ij}$ is the corresponding threshold. The former is stable over time, while the latter is very volatile. The diagonal elements are the autoregressive effects, all of which are stably non-zero in the DFL-DLM. For the off-diagonal elements, the results of shrinkage/thresholding of the two models coincide in some cases.}
\end{figure}

For the nine state variables of lag-1 coefficient matrix $B_{1t}$, the posterior probability of positive counts for the DFL-DLM and the posterior exclusion probabilities for the LTM are displayed in Figure~\ref{subfig:DFLsb} and \ref{subfig:LTMsb}, respectively. The estimates of diagonal entries in the DFL-DLM are least affected by shrinkage, while most of the off-diagonal entries are subject to the strong shrink to zero. The LTM, in contrast, shows the dynamic pattern of thresholding, even in the diagonal entries (e.g., (3,1)-entry). This pattern is sensitively affected by the choice of priors, while we observed the similar posterior results of the DFL-DLMs for different choices of priors and stochastic volatilities. We also double-checked the less-dynamic shrinkage under the DFL-DLM in the posterior expectation of latent counts (Section~S11.2).

\begin{figure}[!ht]
	\centering
	\begin{subfigure}{.45\textwidth}
		\centering
		\includegraphics[width = 1\textwidth]{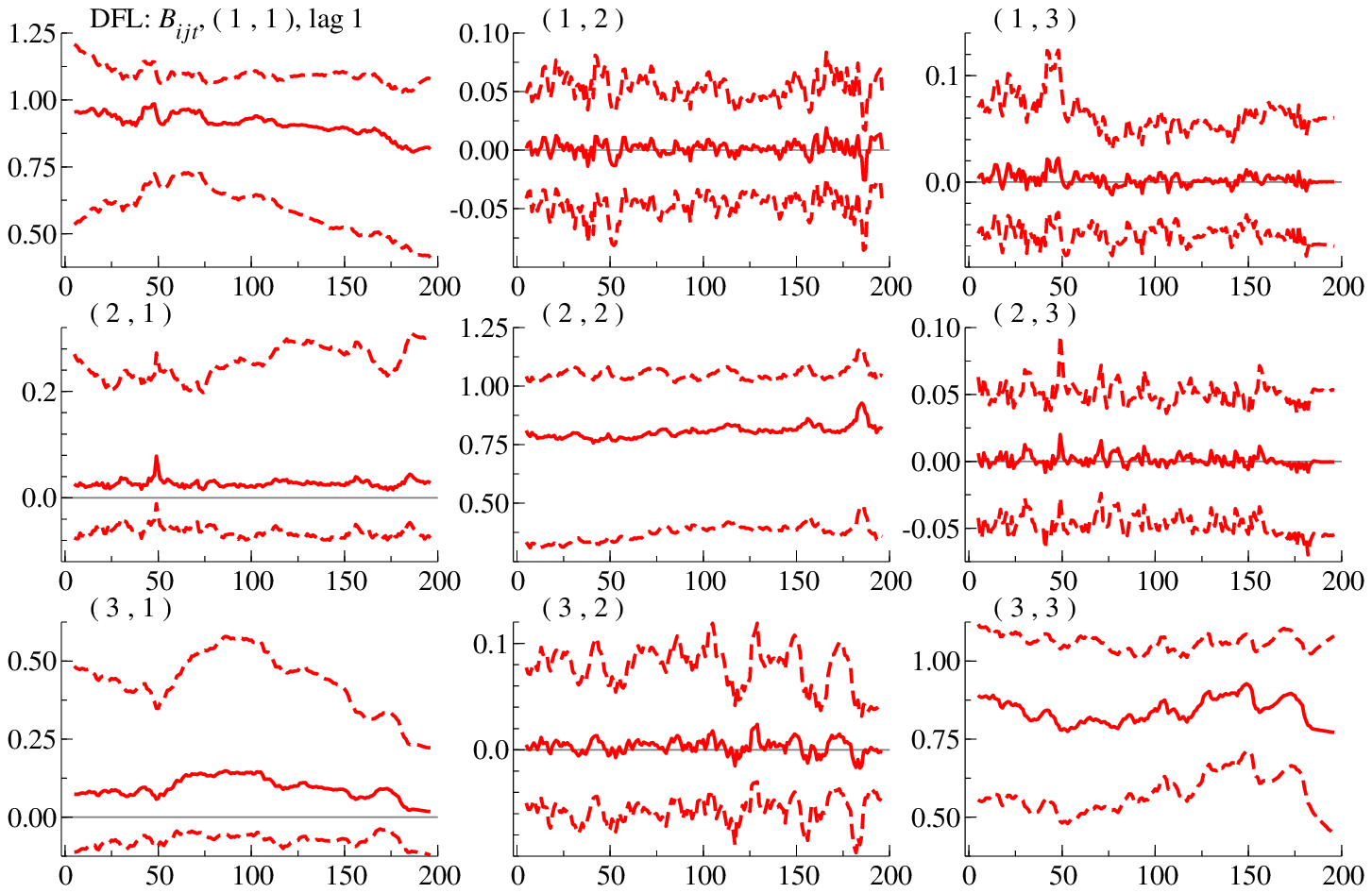}
		\caption{The posteriors of $B_{ijt}$ (DFL-DLM)}
		\label{subfig:DFLb}
	\end{subfigure}
	\begin{subfigure}{.45\textwidth}
		\centering
		\includegraphics[width = 1\textwidth]{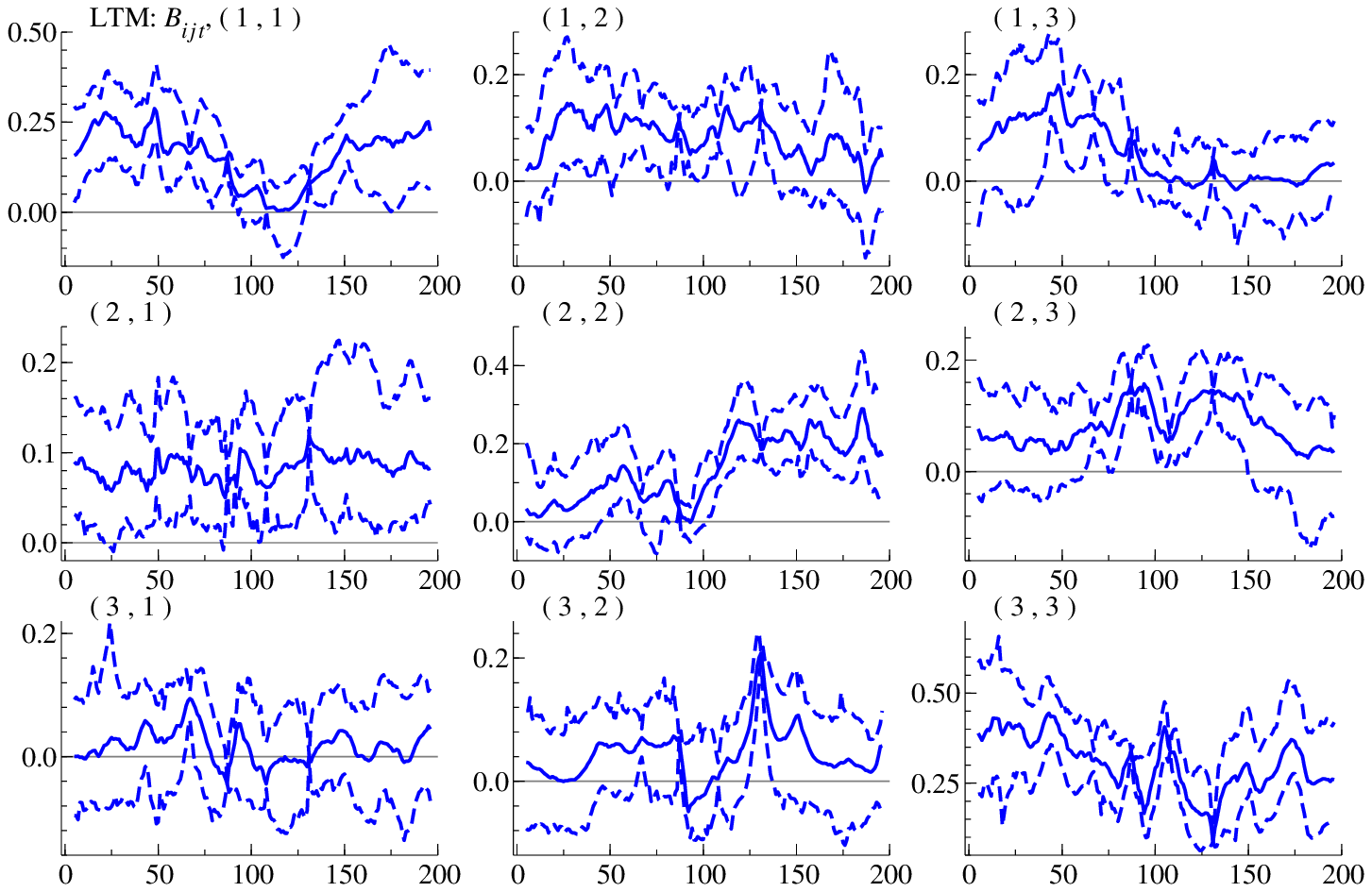}
		\caption{The posteriors of latent $B_{ijt}^{\ast}$ (LTM)}
		\label{subfig:LTMb}
	\end{subfigure}
	\caption{Left: the posteriors of the state variables in $B_{1t}$ in the DFL-DLM, Right: the posteriors of latent state variables in $B_{1t}$ for the LTM. The diagonal elements in the DFL-DLM are larger than those of the LTM, while the most of the other off-diagonal entries are shrunk to zero in the DFL-DLM. Note that the latent state variables in the LTM are not identified when they are thresholded and should be interpreted in combination with the result in Figure~\ref{subfig:DFLsb}. }
\end{figure}

\begin{figure}[!ht]
	\centering
	\begin{subfigure}{.45\textwidth}
		\centering
		\includegraphics[width = 1\textwidth]{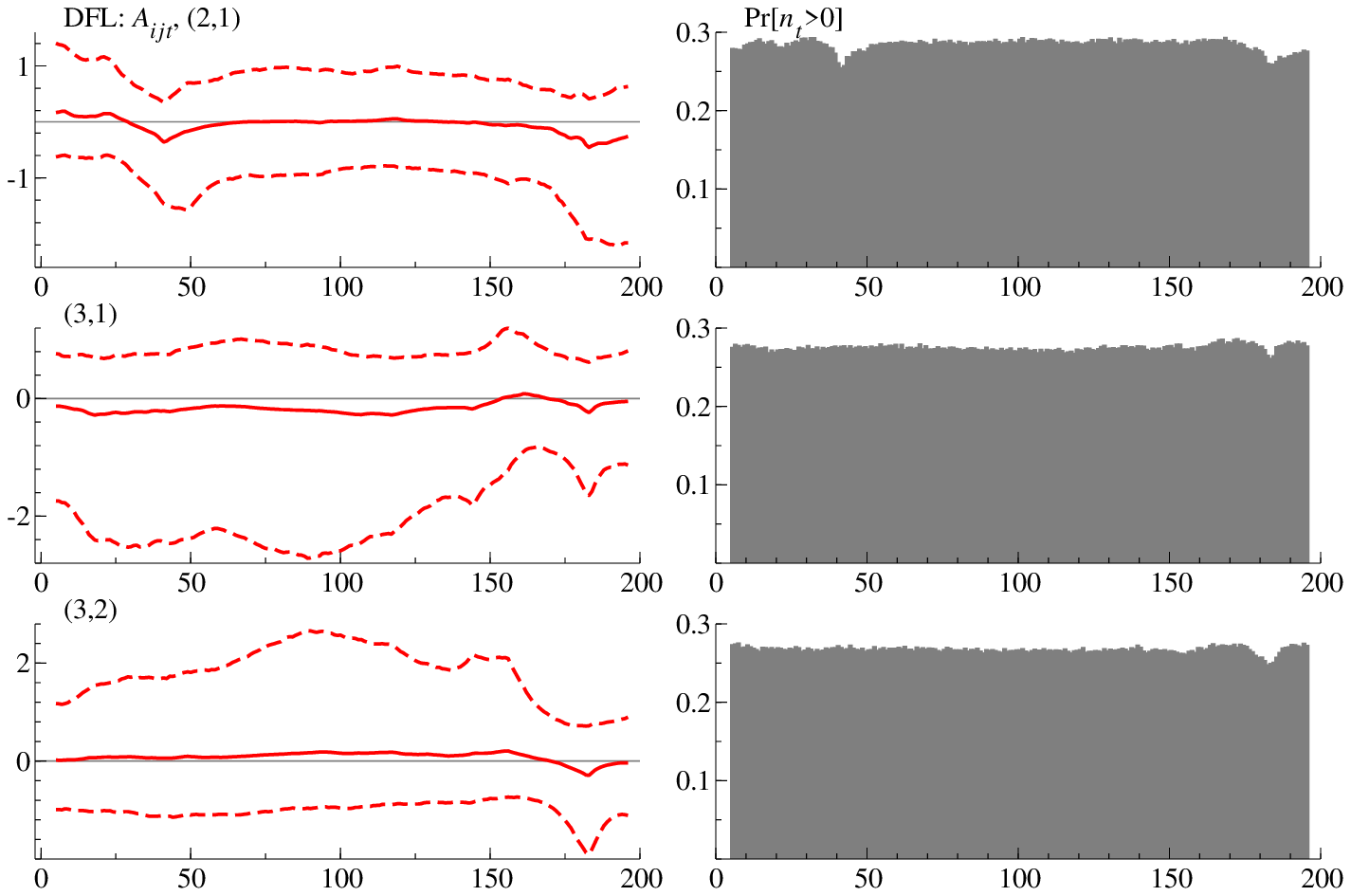}
		\caption{$A_t$ and positive latent counts (DFL-DLM)}
		\label{subfig:ma1}
	\end{subfigure}
	\begin{subfigure}{.45\textwidth}
		\centering
		\includegraphics[width = 1\textwidth]{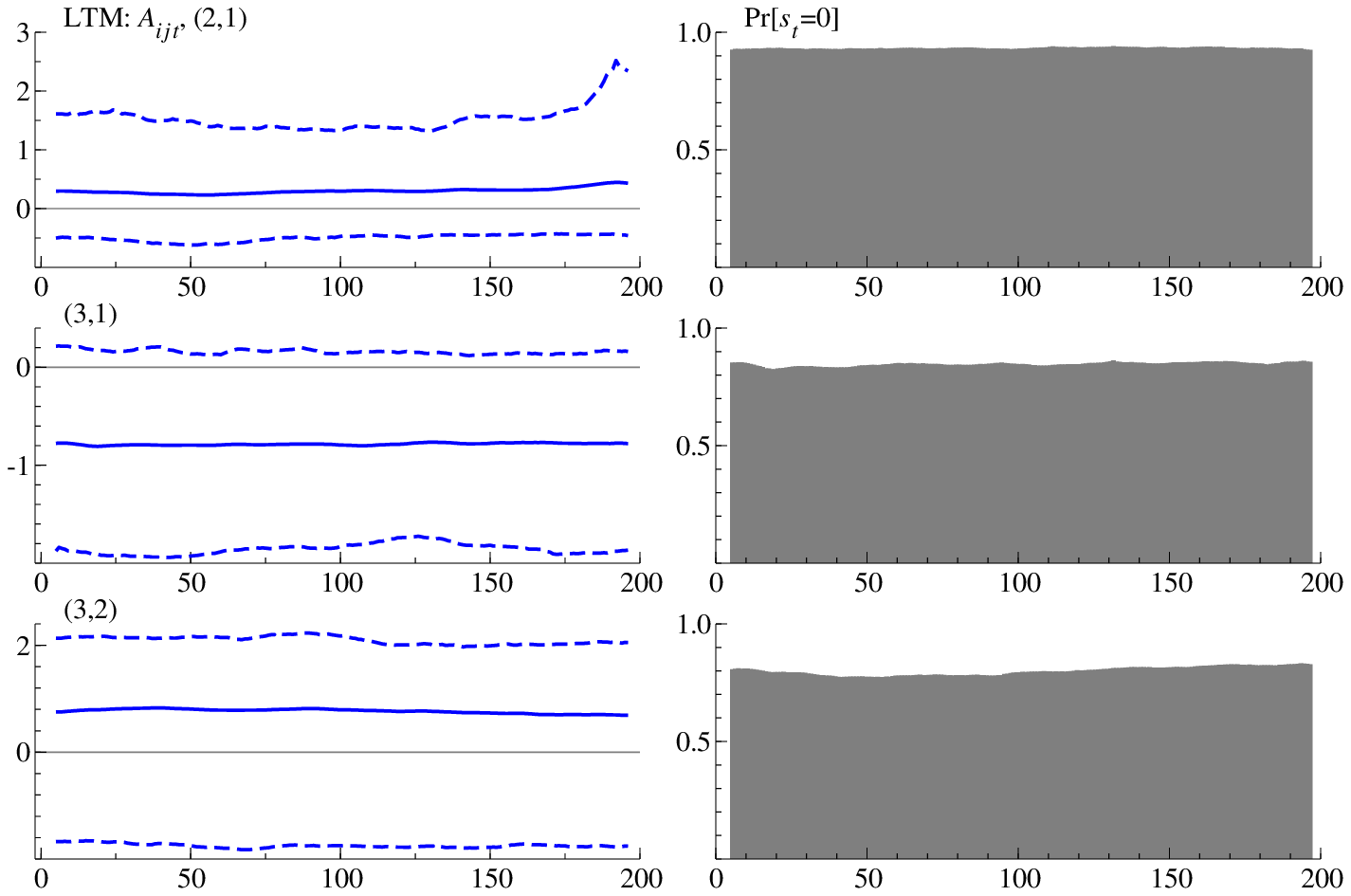}
		\caption{Latent $A_t^{\ast}$ and thresholding probabilities (LTM)}
		\label{subfig:ma2}
	\end{subfigure}
	\caption{Left: the posteriors and the associated positive counts for the lower-triangular elements of $A_t$ in the DFL-DLM, Right: the posteriors of latent state variables $A_t^{\ast}$ and their exclusion probabilities in the LTM. In both results, the simultaneous regressors are all close to zero and it is difficult to find out the graphical structure in this dataset. The DFL-DLM applied the shrinkage to zero at all time points, but their point estimates are not exactly zero, leaving some rooms for dynamic correlations. Correspondingly, the thresholding probabilities of the LTM are not exact one. Especially for the third parameter, i.e., the simultaneous regression coefficient between the unemployment and interest rates, the exclusion probability of the corresponding predictor is about 0.8. }
\end{figure}

The same outputs of the posterior analysis for simultaneous correlation $A_t$ are shown in Figure~\ref{subfig:ma1} and \ref{subfig:ma2}. Both models deny the activeness of simultaneous correlations, but not completely. For example, the LTM leaves about 20\% posterior probability of including the third parameter, i.e., the necessity of the simultaneous regression of the interest rate on the unemployment rate. 
In the DFL-DLM, the posterior probabilities of positive counts in the DFL are much smaller than those of $B_{1t}$, and the estimates of coefficients are not the exact zero at many points. 
We may conclude that both models agree that the information on the graphical structure in this dataset is limited. 

Overall, in contrast to the study of simulated data, the posterior probabilities of positive counts are less dynamic, both for $B_{1t}$ and $A_t$. One may inflate the dynamics of shrinkage in the posteriors by choosing another set of hyperparameters that makes the DLF prior more volatile, but the overall trend of posterior results, such as the significance of the diagonal elements of $B_{1t}$, remains unchanged. 

\section{Concluding Remark} \label{sec:conclusion}

This research on the DFL process is the formal Bayesian attempt at the introduction of the multiple, conflicting shrinkage effects. The additional shrinkage newly introduced in the DLF process is separated from the existing shrinkage on dynamics and comprises the synthetic likelihood, with which the model becomes the CDLM and allows the posterior computation by FFBS. Although the posterior computation is limited to the MCMC method in this study, the CDLM representation could be useful potentially in deriving other computational methodologies, such as the customized version of sequential Monte Carlo methods. 

The research of this type can be further developed in several directions. For example, the concept of fusing two penalty functions into the prior is generalized for arbitrary loss functions, $L_1(x)$ and $L_2(x-x')$, as 
\begin{equation*}
p(x|x') \propto \exp \{ - L_1(x) - L_2(x-x') \},
\end{equation*}
with the integrability condition to make $p(x|x')$ a proper density. This ``translation'' of loss functions to statistical models has been well studied in decision making problems in general \citep{mueller1999} and developed particularly in the context of shrinkage priors \citep{fahrmeir2010bayesian,polson2019bayesian}. The DFL prior belongs to this class of priors by setting $L_1(x) = \alpha |x|$ and $L_2(x-x') = \beta |x-x'|$. The important examples include; the dynamic version of Bayesian elastic net as the combination of $\ell^1$ and $\ell^2$ loss functions \citep{hans2011elastic}; the Bayesian bridge with $\ell^{\nu}$ loss functions for $\nu > 0$ \citep{polson2014bayesian}; and the loss functions induced by horseshoe density, where $\exp \{ -L_1(x) \} \propto e^{\alpha x^2/2} \Gamma (0,\alpha x^2/2)$ \citep{carvalho2009handling,carvalho2010horseshoe}. In those examples, the representation of the scale mixture of normals for $\exp \{ -L_i(x) \}$ $(i=1,2)$ are available, for which the augment in Section~\ref{sec:def} is valid even for this general class. Yet, it has not been known for those models, especially for the horseshoe ones, whether the latent augmentation of normalizing constant is available to enable the fast posterior computation by FFBS as in the DFL prior. 

The vector-autoregressive models in Section~\ref{sec:LTM} can be, as mentioned in the introduction, an important application of the proposed shrinkage priors. The problem of increased number of parameters has been raised at the early stage of development \citep{primiceri2005time,del2015time}, and the use of shrinkage prior for such problem is actively investigated \citep{kastner2019sparse,kastner2020sparse}. 
It is worth revisiting the problem of sparse estimation of these econometric models from the viewpoint of the priors of fused penalties for the time-varying parameters.

\bibliographystyle{ba}
\bibliography{DynamicShrinkage}

\clearpage

\appendix 
\renewcommand{\theequation}{S\arabic{equation}}
\setcounter{equation}{0}

\renewcommand{\thefigure}{S\arabic{figure}}
\setcounter{figure}{0}

\renewcommand{\thetable}{S\arabic{table}}
\setcounter{table}{0}

\renewcommand{\thesection}{S\arabic{section}}
\setcounter{section}{0}

\begin{center}
	{\LARGE\bf Supplemental Materials for ``Bayesian Dynamic Fused LASSO''} \\
	\vspace{6pt}Kaoru Irie \\
	Faculty of Economics, The University of Tokyo
\end{center}

\section*{Overviews}

The contents covered in this document are summarized as follows: 
\begin{itemize}
	\item Section~\ref{app:prp1} proves Proposition~2.1 in the main text and provides the closed form of normalizing constant $h(x)$. 
	
	\item Section~\ref{app:nc} also computes $h(x)$ when $\alpha = \beta$. This result is used in drawing the density function in Figure~1 of the main text. 
	
	\item Section~\ref{app:prp3} proves Proposition~2.3 and writes down the augmented joint densities of $x_{1:T}$ in which we read off the synthetic state space model. 
	
	\item Section~\ref{app:prior} gives the explicit form of the expectation and distribution of conditional shrinkage effect, $E[w|x']$ and $q(w|x')$. The examples of such expectations and densities are drawn and discussed from the viewpoint of subjective elicitation of the prior, while the proof will be given in the later sections. 
	
	\item Section~\ref{app:prp4} computes $E[w|x']$. 
	
	\item Section~\ref{app:sim} derives $q(w|x')$, and also explains the procedure of simulating from $p(x|x')$. This is an essential step in forecasting. 
	
	\item Section~\ref{app:weight} is for the discussion and details of estimation of weight parameters $(\alpha ,\beta)$ and baseline $\mu$. In each subsection, we discuss, 
	\begin{itemize}
		\item the conjugate priors for $(\beta ,\rho )$ and the simulation from the full conditionals. 
		\item the prior of half-Cauchy type for $(\beta ,\rho )$ and the simulation from the full conditionals. 
		\item the prior for baseline $\mu$ and its full conditional. 
	\end{itemize}
	
	\item Section~\ref{app:hs} proves that the horseshoe prior can be expressed as the hierarchical Bayesian LASSO. 
	
	\item Section~\ref{app:sv} provides the details of variance model $V_t$: the scaled version of DFL-DLMs with variance $V$ and the log-Gaussian stochastic volatility $V_t$. 
	
	\item Section~\ref{app:simstudy} stores the additional results in the simulation study in Section~5.1. 
	
	\item Section~\ref{app:app} records some posterior plots in the application to the macroeconomic study in Section~5.2. 
\end{itemize}

\section{Proof of Proposition 2.1} \label{app:prp1}

Equation~(3) in the main text provides the mixture representation of the normalizing constant as 
\begin{equation*}
h(x) = \int _0^{\infty} \!\!\! \int _0^{\infty}  N\!\left( x \left| 0, \tau _1+\tau _2   \right. \right) p_1(\tau _1 ) \ p_2(\tau _2 ) \ d\tau _1 d\tau _2,
\end{equation*}
where $\tau _1 \sim \mathrm{Ex} (\alpha ^2/2)$ and $\tau_2\sim \mathrm{Ex} (\beta ^2/2)$ independently. Set $z = \tau _1 + \tau _2$ and $w = \tau _1 / (\tau _1 + \tau _2)$. Then, by change of variable, we have 
\begin{equation*}
\begin{split}
p(w,z) &= p_1(wz) \ p_2((1-w)z) \ z \\
&= \frac{\alpha ^2\beta ^2}{4} z \exp \left\{ -\frac{\alpha^2}{2} z - \left( \frac{\beta^2}{2} - \frac{\alpha^2}{2} \right) (1-w)z \right\} .
\end{split}
\end{equation*}
Then, the marginal of $z$ is 
\begin{equation*}
\begin{split}
p(z) &= \int _0^1 p(w,z) dw = \frac{\alpha ^2\beta ^2}{4} z e^{ - (\alpha ^2/2) z } \int _0^1 \exp \left\{ - \left( \frac{\beta^2}{2} - \frac{\alpha^2}{2} \right) (1-w)z \right\} dw \\
&= \frac{\alpha ^2\beta ^2}{4} z e^{ - (\alpha ^2/2) z } \frac{1 - e^{ -(\beta^2/2 - \alpha ^2/2)z }}{(\beta ^2 - \alpha ^2)z/2} \\
&= \frac{\alpha ^2\beta ^2}{2(\beta ^2 - \alpha^2)} \left\{ e^{ - (\alpha ^2/2) z } - e^{ - (\beta ^2/2) z }   \right\} \\
&= \frac{1}{(\beta ^2 - \alpha^2)} \left\{ \beta ^2\mathrm{Ex}(z|\alpha ^2/2) - \alpha ^2\mathrm{Ex}(z|\beta ^2/2)  \right\} 
\end{split}
\end{equation*}
The normalizing constant is the mixture of normals with scale $z$. That is, 
\begin{equation*}
\begin{split}
h(x) &= \int _0^{\infty} \!\!\! \int _0^{\infty}  N\!\left( x \left| 0, \tau _1+\tau _2   \right. \right) p_1(\tau _1 ) \ p_2(\tau _2 ) \ d\tau _1 d\tau _2 \\
&= \int _0^{\infty} N\!\left( x \left| 0, z \right. \right) p( z ) dz \\
&= \frac{1}{(\beta ^2 - \alpha^2)} \left\{ \beta ^2\mathrm{DE}(x|\alpha ) - \alpha ^2\mathrm{DE}(x|\beta ) \right\} \\
&= \frac{\alpha \beta ^2}{2(\beta ^2 - \alpha^2)} e^{-\alpha |x|} \left\{ 1 - \left( \frac{\alpha}{\beta} \right) e^{-(\beta -\alpha)|x|}\right\} ,
\end{split}
\end{equation*}
which is the expression of the proposition.

\section{Normalizing constant for $\alpha = \beta$} \label{app:nc}

Throughout our research, the inequality $\alpha < \beta$ is assumed to represent our prior belief on the structure of sparseness in the dynamic coefficients. However, the same argument can be applied to the other cases: $\alpha = \beta$ and $\alpha > \beta$. Here, we provide the functional form of normalizing constant when $\alpha = \beta$, which is used in drawing the density function of the DFL prior in Figure~1 in the main text. 

\begin{prp}
	\label{prp:nc2}
	If $\alpha =\beta$, the normalizing function of the DFL prior is  
	\begin{equation*} 
	h(x) = \frac{\alpha}{4} e^{-\alpha |x|} (1+\alpha |x|).
	\end{equation*}
\end{prp}

{\it Proof.} \ The same scale mixture representation of Proposition~2.1 applies to this one. When $\alpha =\beta$, the marginal of $z=\tau _1+\tau_2$ is obtained by 
\begin{equation*}
p(z) = \frac{\alpha ^4}{4} z e^{ - (\alpha ^2/2) z }. 
\end{equation*}
Then, we have 
\begin{equation*}
\begin{split}
h(x) &= \int _0^{\infty} N\!\left( x \left| 0, z \right. \right) p(z) \\
&= \frac{\alpha^4}{4\sqrt{2\pi}} \int_0^{\infty} z^{3/2 - 1} \exp \left\{ -\frac{1}{2} \left\{ \alpha ^2 z + \frac{x^2}{z} \right\} \right\} dz \\
&= \frac{\alpha^4}{4\sqrt{2\pi}} \frac{ 2K_{3/2}(\alpha |x|) }{(\alpha / |x|)^{3/2}} \\
&= \frac{\alpha^4}{4\sqrt{2\pi}} \frac{2}{(\alpha / |x|)^{3/2}} e^{-\alpha |x|} \sqrt{ \frac{\pi}{2\alpha |x|} } \left( 1 + \frac{1}{\alpha |x|}  \right) \\
&= \frac{\alpha}{4} e^{-\alpha |x|} (1+\alpha |x|),
\end{split}
\end{equation*}
where we read-off the density kernel of $GIG(1/2, \alpha ^2 , x^2)$ in the integral in the second line, and $K_p(\cdot )$ is the modified Bessel function of the second kind with order $p$. The expression of $K_{3/2}(\cdot )$ can be found, for example, \cite{abramowitz1964handbook}, Section~10.2.17. \qed

\section{Proof of Proposition 2.3} \label{app:prp3}

Propositions 2.1 and 2.2 
prove the geometric series representation of the transition density as  
\begin{equation*}
p(x_t|x_{t-1}) = \frac{f(x_t) g(x_t,x_{t-1})}{f(x_{t-1})} \frac{2(C_0+C_+)(\alpha +\beta )}{\beta ^2} q(x_{t-1}|\alpha ,\beta ) 
\end{equation*}
where $q(x_{t-1}|\alpha ,\beta )$ is the discrete mixture given in Proposition 2.2 with the running index of the series denoted by $n_{t-1}$. 
By interpreting $n_{t-1}$ as the latent variable in the mixture representation, the transition density is understood as the marginal of $(x_t,n_{t-1})$. As the marginal distribution of $n_{t-1}$, we have $n_{t-1}=0$ with probability $C_0/(C_0+C_+)$, and $n_{t-1}>0$ with probability $C_+/(C_0+C_+)$. Jointly, for $n_{t-1}=0$, 
\begin{equation*}
p(x_t,n_{t-1}=0|x_{t-1})= \frac{f(x_t) g(x_t,x_{t-1})}{f(x_{t-1})} \frac{\beta ^2-\alpha ^2}{\beta ^2} 
\end{equation*}
and, for $n_t>0$,
\begin{equation*}
p(x_t,n_{t-1}|x_{t-1})= \frac{f(x_t) g(x_t,x_{t-1})}{f(x_{t-1})} \frac{2(\alpha +\beta )}{\beta ^2} \frac{1}{n_{t-1}} \left( \frac{\alpha}{\beta} \right) ^{n_{t-1}} DE(x_{t-1}|n_{t-1}(\beta -\alpha )) 
\end{equation*}
For notational convenience, we define $w(n)$ by 
\begin{equation*}
w(n) = \frac{2(\alpha +\beta )}{\beta ^2} \left\{ \frac{\beta -\alpha }{2} 1[n=0] + \frac{1}{n} \left( \frac{\alpha}{\beta} \right) ^n 1[n>0] \right\} 
\end{equation*}
In the joint density of state $x_{1:T}$ and auxiliary variable $n_{2:(T-1)}$, observe that $f(x_t)$ appears both in the numerator and denominator for $t=2,\dots ,T-1$ and cancels out one another as 
\begin{equation*}
\begin{split}
p(x_{1:T},n_{2:(T-1)}) &= \pi (x_1) p(x_2|x_1) \prod _{t=3}^T p(x_t|x_{t-1}) \\
&= \pi(x_1) \frac{f(x_2)g(x_2,x_1)}{h(x_1)} \prod _{t=3}^T \frac{f(x_t)g(x_t,x_{t-1})}{f(x_{t-1})} w(n_{t-1}) \prod _{\substack{t=3:T \\ n_{t-1} > 0}} DE(x_{t-1}|n_{t-1}(\beta -\alpha )) \\
&= f(x_1) f(x_T) \prod _{t=2}^T g(x_t,x_{t-1}) \prod _{t=2}^{T-1} w(n_{t}) \prod _{t:n_t>0} DE(x_t|n_t(\beta -\alpha )),
\end{split}
\end{equation*}
where we write $\{ t: n_t>0 \}$ to mean that the product is taken only for $t=2,\dots ,T{-}1$ that satisfies $n_t>0$. Using the scale mixture representation of double exponential distributions, we can read off the conditional distribution in the joint density above, as 
\begin{equation*}
p(x_{1:T} | n_{2:(T-1)},\Lambda_T ) = \left\{ N(z_T|x_T,\lambda _{\alpha,T}) \prod _{t:n_t>0} N(z_t|x_t,\lambda _{n,t}) \right\} \left\{ N(x_1|0,\lambda _{\alpha,1}) \prod _{t=2}^T N(x_t|x_{t-1},\lambda _{\beta,t}) \right\} 
\end{equation*}
where $z_t=0$ for all $t$, and $\Lambda_T$ is the set of all the latent variables up to time $T$, i.e., $\Lambda _T = \{ \lambda _{\alpha,1},\lambda _{\alpha,T},\lambda _{\beta,2:T}, \lambda _{n,2:(T-1)} \}$. The densities in the first parenthesis imply the linear and Gaussian likelihood, and those in the second parenthesis are the Gaussian AR(1) prior. This is exactly a dynamic linear model conditional on all the latent variables, the joint density of which is given by 
\begin{equation*}
p(n_{2:(T-1)},\Lambda _T ) = \prod _{t=1,T} Ga(\lambda _{\alpha,t}| 1,\alpha ^2/2) \prod _{t=2}^{T} Ga(\lambda _{\beta ,t}| 1,\beta ^2/2) \prod _{t=2}^{T-1} w(n_t) \prod _{t:n_t>0}Ga(\lambda _{2t}| n_t^2(\alpha -\beta )^2/2) 
\end{equation*}
As a whole, this is exactly the CDLM of the proposition. 

\section{Subjective elicitation of weight parameters} \label{app:prior}

In this subsection, the conditional expectation $E[x|x']$ and prior density of $(z,w)$, where $z=\tau _1+\tau _2$ and $w = \tau _1/(\tau_1+\tau_2)$, are investigated for the better understanding of weight parameters $(\alpha ,\beta )$. The details of derivation are given in the next subsection. 

Recall the expression of the conditional density given in Equation (4) in the main text, which is the location-scale mixture of the Gaussian AR(1) process, 
\begin{equation} \label{eq:cond}
p(x|x') = \int _0^{\infty} \!\!\! \int _0^1 N(x|wx',w(1-w)z ) q(z ,w | x') dz dw,
\end{equation}
where the mixing distribution is 
\begin{equation} \label{eq:zw}
q(z ,w | x') = \frac{N(x'|0,z )}{h(x')} \frac{\alpha ^2\beta ^2}{4} z  e^{-\frac{\beta ^2}{2} z} e^{-\frac{(\beta ^2-\alpha ^2)}{2}wz },
\end{equation}
which is obtained by change of variables with $z =\tau _1+\tau _2$ and $w=\tau _1/(\tau _1+\tau _2)$. The conditional expectation $E[x|x']=E[w|x']x'$ is one of the useful information that statisticians can interpret as ``conditional shrinkage effect'' and elicit their prior belief based on this quantity. The distribution $q(z ,w|x')$ itself is analytically available as well and provides further information on one's prior belief. 
\begin{prp}
	\label{prp:4}
	For DFL prior with $\beta >\alpha$, the conditional transition density in \eq{eq:cond} has the mean $E[w|x']x'$, where 
	\begin{equation} \label{eq:Ew}
	E[w|x'] = \frac{1 - \frac{2\alpha }{(\beta ^2-\alpha ^2)|x'|} (1-e^{-(\beta -\alpha )|x'|}) }{1-\frac{\alpha}{\beta}e^{-(\beta -\alpha )|x'|}}
	\end{equation}
\end{prp}

\begin{prp}
	\label{prp:42}
	The joint distribution in \eq{eq:zw} is decomposed into the compositional form. The conditional distribution of $(z|w)$ is $GIG(3/2,w\alpha ^2+(1-w)\beta ^2,(x')^2)$ and the marginal density of $w$ is 
	\begin{equation} \label{eq:w}
	q(w|x') = \frac{\alpha ^2\beta ^2}{4h(x')} e^{-|x'|\sqrt{ w\alpha ^2 + (1-w)\beta ^2 } } \left\{ \frac{|x'|}{w\alpha ^2+(1-w)\beta ^2} + \frac{1}{(w\alpha ^2+(1-w)\beta ^2)^{3/2}} \right\}
	\end{equation}
\end{prp}

\

\noindent 
See the Section~\ref{app:prp4} and \ref{app:sim} for the computation of equations~(\ref{eq:Ew}) and (\ref{eq:w}), respectively. Here, we focus on the implication of this form on the prior specification. 

\begin{figure}[!ht]
	\centering
	\begin{subfigure}{.45\textwidth}
		\centering
		\includegraphics[width = 1\textwidth]{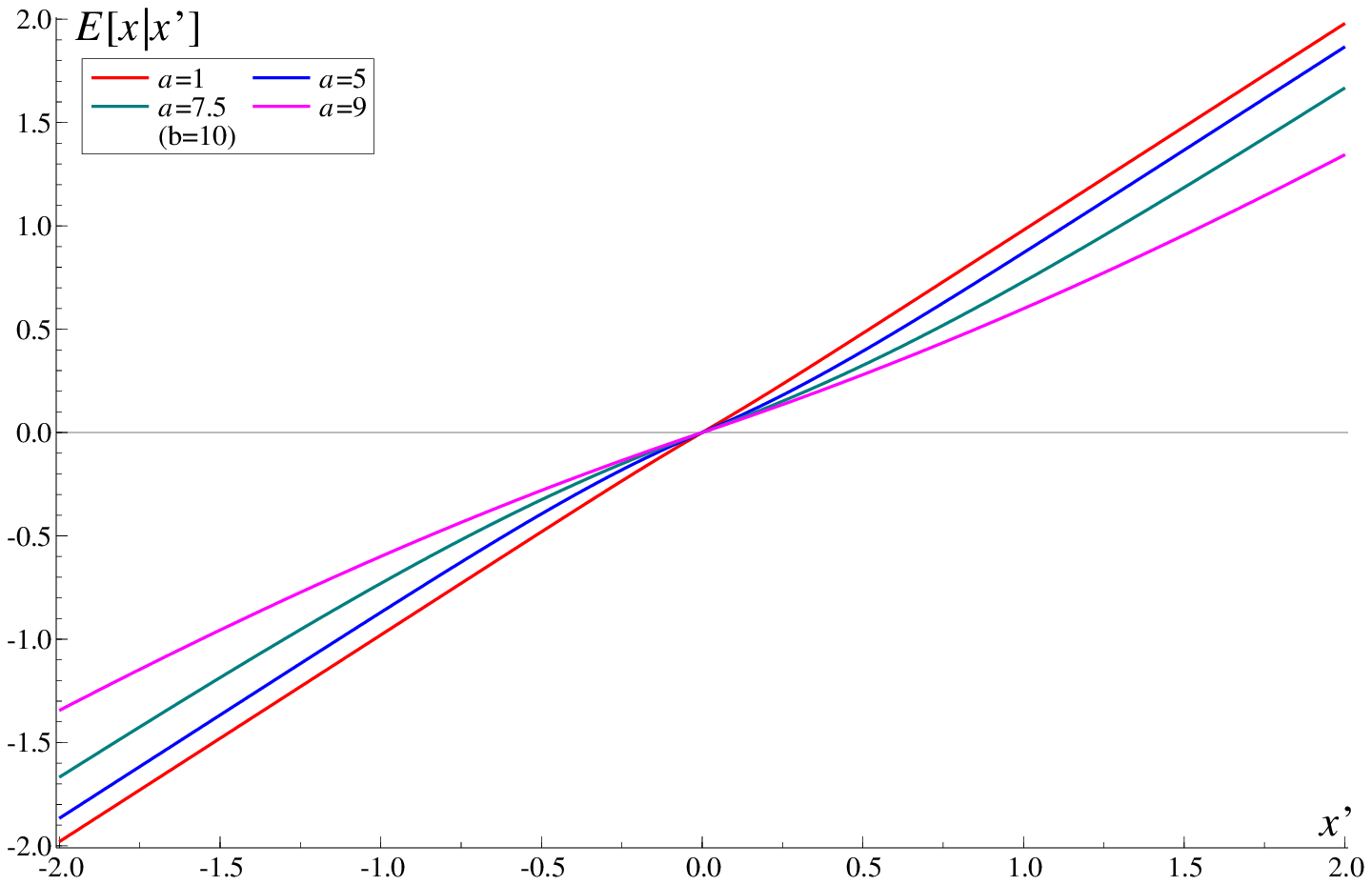}
		\caption{Conditional prior mean of state $E[w|x'] x$}
		\label{subfig:mean}
	\end{subfigure}
	\begin{subfigure}{.45\textwidth}
		\centering
		\includegraphics[width = 1\textwidth]{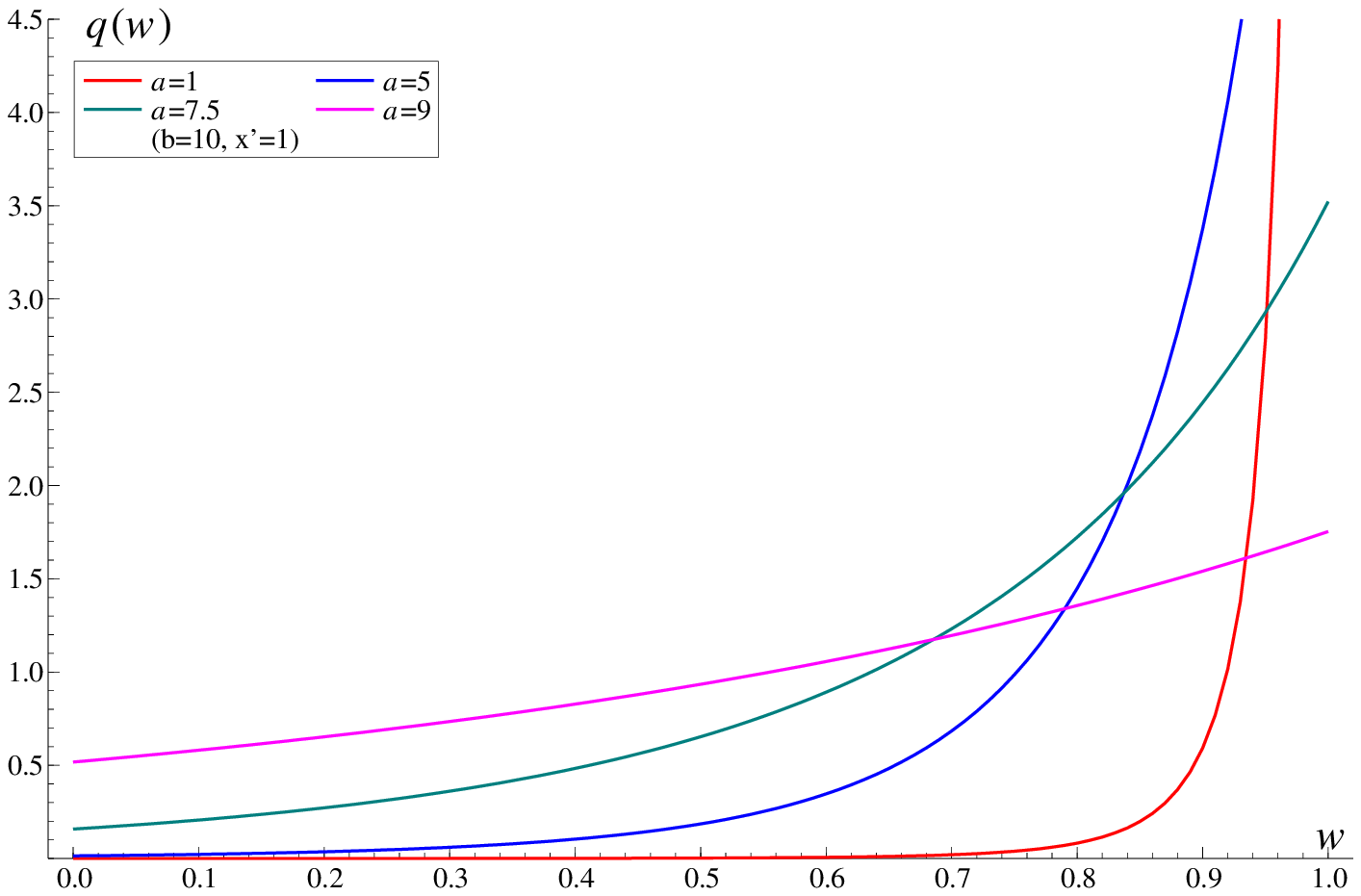}
		\caption{Density of conditional shrinkage $w$}
		\label{subfig:coef}
	\end{subfigure}
	\caption{Left: Conditional prior mean $E[x|x'] = E[w|x']x'$ for $b=10$ and various $a$ as the function of $x'$. The larger $a$ is, the more shrinkage to zero is observed. Conversely, if $a$ is small enough, then $E[x|x']\approx x'$. The shrinkage effect is also dependent on $x'$ in non-linear way. Right: Prior density of $(w|x')$ for $b=10$ and various $a$ with $z$ marginalized out. Small $a$ implies $w\approx 1$ with high probability, having little shrinkage to zero. For large $a$, it is still likely that $w$ is close to unity, but more probability mass is on smaller values of $w$, resulting in the additional shrinkage to zero. }
\end{figure}

The conditional mean $E[x|x']$ in \eq{eq:cond} is plotted against $x'$ in Figure~\ref{subfig:mean}. When $\alpha$ is small, the conditional means are almost on the diagonal line, showing more shrinkage effect to $x'$ as in the random walk models. In contrast, for large $\alpha$, the shrinkage effect to zero becomes strong, making the conditional means off the diagonal line in the figure. The conditional mean is the non-linear function of $x'$ as shown in \eq{eq:Ew}, and this non-linearity is also visually confirmed in the figure. Figure~\ref{subfig:coef} depicts the density of shrinkage effect $q(w|x')$ in \eq{eq:w}, conditional on $x'=1$, with $z$ marginalized out. For all the values of $a$ examined here, the prior mass concentrates around $w\approx 1$, implying the dominating conditional shrinkage effect is directed to $x'$, not to zero. For smaller $\alpha$, however, the more probability mass is placed on the smaller values of $w$ as well.

These two figures are just one aspect of the prior structure; for different choices of $(\alpha ,\beta ,x')$, the conditional prior mean and density look completely different, and it is difficult to summarize those differences in a simple manner. This might be the potential difficulty in subjectively specifying the prior, which motivates the estimation of hyperparameters. The automatic adjustment of hyperparameters via posterior analysis is discussed later in Section~\ref{app:weight}, but the choice of hyperparameters based on the discussion here still remains important.

\section{Computation for Proposition~\ref{prp:4}} \label{app:prp4}

In deriving the expectation $E[w|x']$, the other way of decomposition of the joint density of $(z,w)$, i.e., $q(w|z,x')q(z|x')$, is useful. By reading off the conditional density of $(w|z,x')$ in \eq{eq:zw}, observe that $q(w|z,x') \propto e^{(\beta ^2-\alpha ^2)zw/2}$, whose normalizing constant is 
\begin{equation*}
\int _0^1 e^{(\beta ^2-\alpha ^2)zw/2} dw = \frac{2}{(\beta ^2-\alpha ^2)z} \left\{ e^{(\beta ^2-\alpha ^2)z/2} -1 \right\} 
\end{equation*}
Next, the marginal of $z$ is 
\begin{equation*}
q(z|x') = \frac{\alpha ^2\beta ^2}{2\sqrt{2\pi}(\beta ^2-\alpha ^2) h(x')} z^{-1/2} \exp \left\{ - \frac{(x')^2}{2z} \right\}  \left\{ e^{-\alpha ^2z/2} - e^{-\beta ^2z/2} \right\} 
\end{equation*}
This is the {\it difference} of two GIG densities, not the mixture, hence not suitable for the random number generation that we will discuss in the next Section. This form is rather useful in computing the moments. 

{\it Proof of Proposition~\ref{prp:4}.} 
Compute the following integral, by using the integral by parts, as
\begin{equation*}
\begin{split}
\int _0^1 w e^{(\beta ^2-\alpha ^2)zw/2} dw &= \left[ w \frac{2}{(\beta ^2-\alpha ^2)z} e^{(\beta ^2-\alpha ^2)zw/2} \right] _0^1 - \frac{2}{(\beta ^2-\alpha ^2)z} \int _0^1 e^{(\beta ^2-\alpha ^2)zw/2} dw \\
&= \frac{2}{(\beta ^2-\alpha ^2)z} \left[ e^{(\beta ^2-\alpha ^2)z/2} - \frac{2}{(\beta ^2-\alpha ^2)z} \left\{ e^{(\beta ^2-\alpha ^2)z/2} - 1 \right\} \right] 
\end{split}
\end{equation*}
Denote the normalizing constant of $GIG(p,a,b)$ by $G(p,a,b)$. Then, 
\begin{equation*}
\begin{split}
E[w|x'] &= \int _0^{\infty} \int _0^1 wq(z,w) dwdz \\
&= \frac{\alpha ^2\beta ^2}{2\sqrt{2\pi}(\beta ^2-\alpha ^2)h(x')} \left[ G(1/2,\alpha ^2,(x')^2) - \frac{2}{(\beta ^2-\alpha ^2)} \left\{ G(-1/2,\alpha ^2,(x')^2) - G(1/2,\beta ^2,(x')^2) \right\} \right] \\
&= \frac{\alpha ^2\beta ^2}{2(\beta ^2-\alpha ^2)h(x')} \left[ \frac{1}{\alpha} e^{-\alpha |x'|} - \frac{2}{(\beta ^2-\alpha ^2)|x'|} \left\{ e^{-\alpha |x'|} - e^{-\beta |x'|} \right\} \right] 
\end{split}
\end{equation*}
which is further simplified by substituting the expression of $h(x')$ in Proposition~2.1 in the main text and shows the expression of Proposition~\ref{prp:4}. \qed

\section{Simulation from the prior} \label{app:sim}

Simulating the random variable $x$ from the conditional distribution $p(x|x')$ is an important step at the one-step and multi-step ahead predictions. The location-scale mixture representation in \eq{eq:cond} enables the simulation from the normal distribution $N(wx',w(1-w)z )$, given that $(z,w)$ is sampled from $q(z,w|x')$. Proposition~\ref{prp:4} provides the compositional form of this joint density, in which $q(z|w,x')$ is the generalized inverse Gaussian distribution and relatively easy to sample from. 

The problem remains in the sampling of $w$ from marginal $q(w|x')$ in \eq{eq:w}. We first verify this expression.

{\it Proof of Proposition~\ref{prp:42}.} \ The joint density of $(w,z) = (\tau _1/(\tau _1+\tau _2),\tau _1+\tau _2)$ is 
\begin{equation*}
\begin{split}
q(w,z) &= \frac{N(x'|0,z)}{h(x')} Ga(wz|1,\alpha ^2/2)Ga((1-w)z|1,\beta ^2/2)z \\
&= \frac{\alpha ^2\beta ^2}{4\sqrt{2\pi}h(x')} z^{3/2-1} \exp \left\{ - \frac{wa^2+(1-w)b^2}{2}z - \frac{(x')^2}{2z} \right\}
\end{split}
\end{equation*}
where we read off the conditional/marginal densities, $q(z|w)q(w)$ or $q(w|z)q(z)$. In the former, given $w$, we see $z|w \sim GIG(3/2,w\alpha ^2+(1-w)\beta ^2, (x')^2)$. Marginalizing $z$ out, we have 
\begin{equation*}
\begin{split}
q(w) &= \frac{\alpha ^2\beta ^2}{4\sqrt{2\pi}h(x')} \frac{2K_{3/2}(\sqrt{(w\alpha ^2+(1-w)\beta ^2)(x')^2})}{(w\alpha ^2+(1-w)\beta ^2 /(x')^2 )^{3/4}} \\
&= \frac{\alpha ^2\beta ^2}{4h(x')} e^{-|x'|\sqrt{w\alpha ^2+(1-w)\beta ^2}} \frac{|x'|}{w\alpha ^2+(1-w)\beta ^2} \left( 1+\frac{1}{|x'| \sqrt{w\alpha ^2+(1-w)\beta ^2}} \right)
\end{split}
\end{equation*}
which proves Proposition~\ref{prp:42}.\qed 

It is immediate from this expression that the density of $w$ is transformed into the relatively simple form by re-scaling.  

\begin{prp}
	\label{prp:5}
	The marginal density of $w$ in the scale of $y=|x'|\sqrt{w\alpha ^2+(1-w)\beta ^2}$ is
	\begin{equation} \label{eq:y}
	q(y) = \frac{\alpha \beta |x'|}{\beta e^{-\alpha |x'|} - \alpha e^{-\beta |x'|}} \left\{ y^{-1}e^{-y} + y^{-2}e^{-y} \right\} ,
	\end{equation}
	where $y \in (\alpha |x'|,\beta |x'|)$. The distribution function is 
	\begin{equation} \label{eq:dist}
	Q(y) = \int _{\alpha |x'|}^y q(t)dt = \frac{\beta e^{-\alpha |x'|} - (\alpha \beta |x'|)y^{-1}e^{-y}}{\beta e^{-\alpha |x'|} - \alpha e^{-\beta |x'|}} 
	\end{equation}
\end{prp}

\noindent 
{\it Proof.} \ Apply the change of variable by $dw = (-ydy) / (\beta ^2-\alpha ^2)$ to obtain equation~(\ref{eq:y}). To see equation~(\ref{eq:dist}), compute the integral of (\ref{eq:y}) based on the equality 
\begin{equation*}
\int _{\alpha |x'|}^y t^{-1}e^{-t}dt = \Big[ -t^{-1}e^{-t} \Big] _{\alpha |x'|}^y - \int _{\alpha |x'|}^y t^{-2}e^{-t}dt,
\end{equation*}
that is obtained by integral by parts. \qed 

The explicit formula of distribution function helps computing the inverse distribution function numerically, thus allows the inverse sampling. \\ \hrulefill \\Sampling from $p(x|x')$.
\begin{enumerate}
	\item Sample $u\sim U(0,1)$. 
	
	\item Given $u$, set $y = Q^{-1} ( u ) $. 
	
	Recover $w$ by $w = (\beta ^2-y^2/|x'|^2) / (\beta ^2-\alpha ^2)$.
	
	\item Given $w$, sample $z\sim GIG(3/2,w\alpha ^2+(1-w)\beta ^2,(x')^2)$.
	
	\item Given $(w,z)$, sample $x\sim N(wx',w(1-w)z)$. 
\end{enumerate}
\hrulefill

For the computation of the inverse of the distribution function $Q^{-1}$, the simple Newton method can be applied and satisfactorily efficient. To compute $y=Q^{-1}(u)$ for given $u$, one must solve the equality
\begin{equation*}
u = \frac{\beta e^{-\alpha |x'|} - \alpha \beta |x'| y^{-1}e^{-y} }{\beta e^{-\alpha |x'|} - \alpha e^{-\beta |x'|}},
\end{equation*}
or, equivalently, 
\begin{equation*}
y^{-1} e^{-y} = y_0 = \frac{\beta e^{-\alpha |x'|} - u \{ \beta e^{-\alpha |x'|} - \alpha e^{-\beta |x'|} \} }{ \alpha \beta |x'| }
\end{equation*}
in $y$. 
Further, by taking $z=\log (y) \in ( \log (\alpha ) + \log |x'|, \log (\beta ) + \log |x'|$ and $z_0= \log (y_0)$, we have 
\begin{equation*}
g(z) = z + e^z + z_0 = 0.
\end{equation*}
The Newton method for solving this equation is defined by the update rule of the sequence $\{ z_n \}$ as 
\begin{equation*}
z_{n+1} = z_n - \frac{g(z_n)}{g'(z_n)} = z_n - \frac{z_n+e^{z_n}+z_0}{1+e^{z_n}} 
\end{equation*}
until the increment is less than the tolerance level.

\section{Estimation of hyperparameters} \label{app:weight}

\subsection{Likelihood and conjugate priors}

To consider the hierarchical version of the DFL priors, where we assume the prior for weights $\alpha$ and $\beta$, we need to consider the ``likelihood,'' or the joint distribution of state variables $x_{1:T}$ as the function of $(\alpha ,\beta )$. As stated in the main text, we re-parametrize the weights by $\alpha = \rho \beta $ with $\rho \in (0,1)$ to guarantee the restriction $\alpha < \beta$. The joint density of state variables as the function of $(\alpha ,\beta)$ is
\begin{equation*}
\begin{split}
&p(x_{1:T}|n_{2:(T-1)},\alpha ,\beta ) \propto f(x_1) f(x_T) \prod _{t=2}^T g(x_t,x_{t-1}) \left( \frac{\beta ^2-\alpha ^2}{\beta ^2} \right) ^{T-2} \prod _{t:n_t>0} \left( \frac{\alpha}{\beta} \right) ^{n_t} e^{-n_t(\beta - \alpha)|x_t|} \\
&\propto \rho ^{2+\sum n_t} (1-\rho ^2)^{T-2} \beta ^{T+1} \exp \left\{ - \rho \beta (|x_1|+|x_T|) - \beta \sum _{t=2}^T |x_t-x_{t-1}| - (1-\rho)\beta \sum _{t:n_t>0} n_t|x_t| \right\} ,
\end{split}
\end{equation*}	
where, by writing ${t:n_t>0}$, we take the product/summation of those that satisfies $n_t>0$ for $t=2,\dots ,T{-}1$. For this ``likelihood,'' the conditionally conjugate prior for $\beta$ is the gamma distribution. If $\beta \sim Ga(r^b_0,c^b_0)$, then the conditional posterior of $\beta$ is $Ga(r^b,c^b)$, where
\begin{equation*}
r^b = r^b_0+T+2 \ \ \ \mathrm{and} \ \ \ c^b = c^b_0 + \rho (|x_1|+|x_T|) + \sum _{t=2}^T |x_t-x_{t-1}| + (1-\rho) \sum _{t:n_t>0} n_t|x_t|
\end{equation*}
The full conditional posterior density of $\rho$ is given by, with the prior density $\pi (\rho )$, 
\begin{equation*}
p(\rho | \mathrm{-} ) \propto \pi (\rho ) \rho ^{2+\sum n_t} (1-\rho ^2)^{T-2} \exp \left\{ - \rho \left( \beta (|x_1|+|x_T|) - \beta \sum _{t:n_t>0} n_t|x_t| \right) \right\}
\end{equation*}
In this research, we pursue the simplicity of computation by using the discrete prior on the interval $(0,1)$. For some positive integer $N$ and $M$ ($M<N$), the grid is defined by $d = 1/N$ and the prior support of $\rho$ is $\{ d,2d,\cdots Md \}$.
In practice, it is advised to avoid the value $Md$ not too close to 1, e.g., $Md=0.9$, for the numerical issue. 
In our analysis, $d=0.001$ ($N=1000$), $M=900$ and $\rho$ follows the discrete uniform distribution on $\{ d,2d,\dots, Md \}$. The prior probabilities on those points are proportional to the beta density $Be(a_0^r,b_0^r)$ evaluated on the grids. 

\begin{figure}[!ht]
	\centering
	\begin{subfigure}{.45\textwidth}
		\centering
		\includegraphics[width = 1\textwidth]{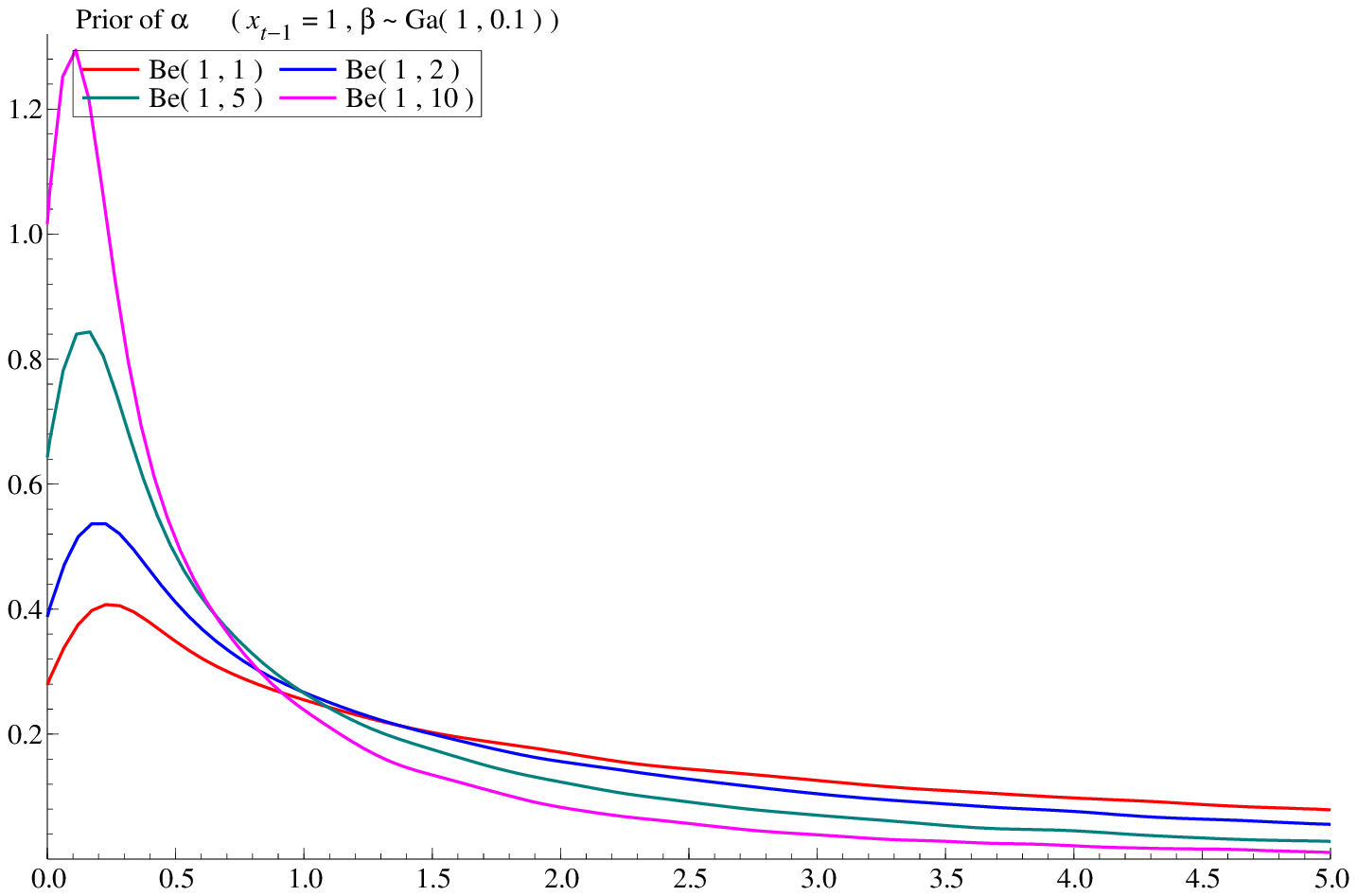}
		\caption{Prior density of $\alpha$}
		\label{subfig:ran_alpha}
	\end{subfigure}
	\begin{subfigure}{.45\textwidth}
		\centering
		\includegraphics[width = 1\textwidth]{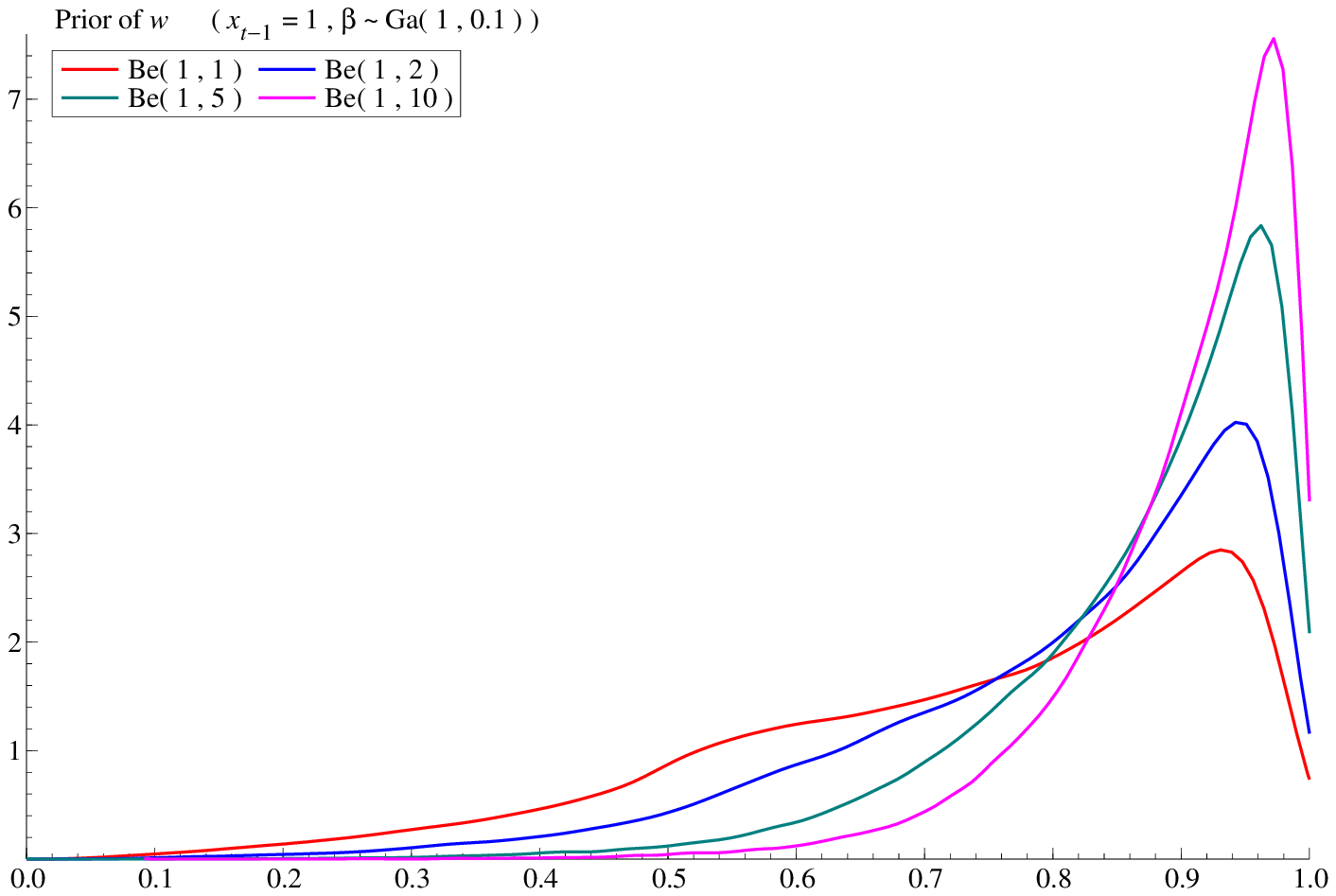}
		\caption{Prior density of $w$}
		\label{subfig:ran_coef}
	\end{subfigure}
	\caption{Left: the density of $\alpha = \rho \beta$ when $\rho \sim Be(1,b_0^r)$ and $\beta \sim Ga(1,0.1)$, Right: the density of $w=\tau _1/(\tau _1+\tau _2)$ conditional on $x_{t-1}=1$. The larger $b_0^r$ is, the more  prior mass concentrates around smaller $\rho$ (and $\alpha$) as seen in the left figure. This property is inherited to the densities of $w$ in the right figure; the large $b_0^r$ (i.e., small $\rho$ and $\alpha$) makes the density of $w$ skewed toward unity, implying the stronger shrinkage of $x_t$ to $x_{t-1}$, not to zero. Note that $x_{t-1}=1$ in these figures; if this value of the state variable is regarded an significant signal, then choosing small $b_0^r$ (e.g., $b_0^r=1$) and shrinking the next state $x_t$ to zero are not appropriate. The densities are computed by simulation based on 10,000 samples of $(\rho , \beta )$ in both figures. 
	}
\end{figure}

The careful, subjective choice of hyperparameters in the prior of $(\beta , \rho)$ is still unavoidable for the appropriate representation of one's prior belief. To see this point, the marginal prior densities of $\alpha$ and $w$ are drawn in Figure~\ref{subfig:ran_alpha} and \ref{subfig:ran_coef}, respectively, for different choices of hyperparameters $b_0^r=1,2,5,10$, while the other parameters are fixed as $a_0^r=1$, $(r_0^b,c_0^b) = (1,0.1)$ and $x_{t-1}=1$. As clearly seen in the prior of $\alpha$ in Figure~\ref{subfig:ran_alpha}, the larger $b_r$ is, the less weight is placed on the shrinkage effect toward zero. The difference of those hyperparameters is emphasized in the density of $w$-- the conditional shrinkage effect to $x_{t-1}$ defined in the location-scale mixture representation in (\ref{eq:cond}). In Figure~\ref{subfig:ran_coef}, the density with $b_r=1$ shows the heavier tail in smaller values of $w$, implying the excess shrinkage effect to zero, even though the strong signal is assumed by $x_{t-1}=1$. From this viewpoint, although this choice means $\rho$ follows the uniform distribution and is seemingly ``less informative'' prior, it is in fact regarded an extreme choice for its strong shrinkage effect applied to even large value of $x_{t-1}$.

\subsection{Marginal distribution implied by hyperprior}

\begin{figure}[!htbp]
	\centering
	\includegraphics[width=5in]{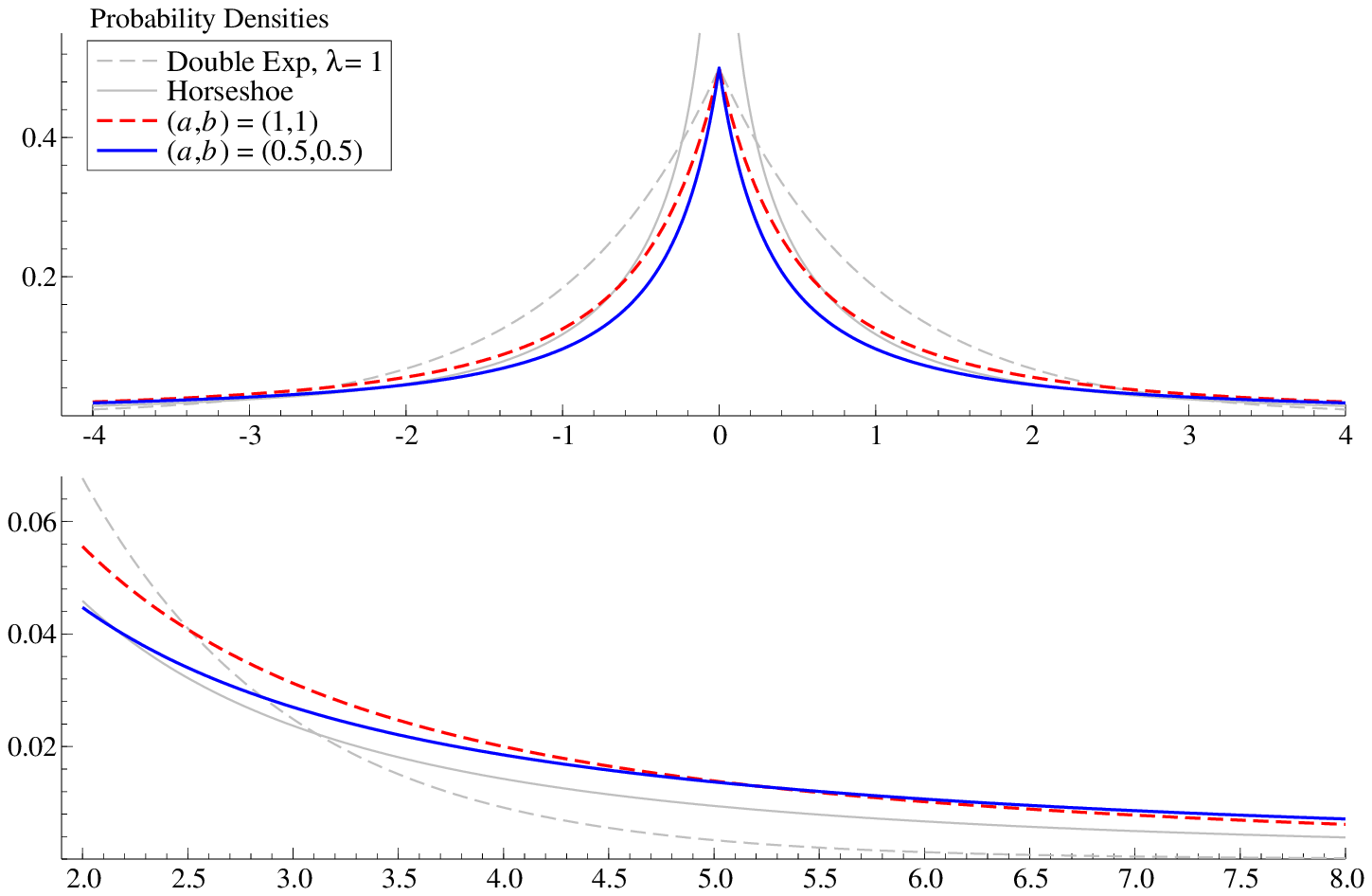}
	\caption{\small Top: the densities of hierarchical double-exponential distribution, i.e., the marginal of $x|\beta \sim DE(\beta)$ and $\beta \sim Ga(r,c)$. The red dashed line and blue solid lines are the densities of $(r,c)=(1,1)$ and $(r,c)=(0.5,0.5)$, respectively. For comparison, the densities of double-exponential distribution $DE(1)$ and horseshoe prior are drawn by gray dashed and solid lines. Bottom: the tails of the densities in the top panel. The hierarchical prior on weight $\beta$ realizes the heavier tails, while the shrinkage effect to zero is at most as strong as the original double-exponential model. 
	} \label{fig:mde} 
\end{figure}

The implied marginal distribution of state variable $x$ under the DFL prior with the conjugate prior on $\beta$ is illustrated in the top panel of Figure~3 in the main text by simulation. Here, we confirm the property of the marginal density analytically by studying the simpler model with $\rho = 0$. That is, consider $x|\beta \sim DE(\beta )$ and $\beta \sim Ga(r_0^b , c_0^b)$. Then, the marginal of $x$ is 
\begin{equation*}
p(x) = \frac{r_0^b (c_0^b)^{r_0^b}}{2} (c_0^b + |x| )^{-(r_0+1)}.
\end{equation*}
In this form of the density function, we observe the heavier tails of the polynomial order that accommodate the sudden change of state variables. On the other hand, the density evaluated at $x=0$ is finite, which contrasts the density of the horseshoe prior whose density is divergent at the origin. Figure~\ref{fig:mde} shows the two examples of the marginal densities of $x$. The hyperprior can realize the heavier tails than those of the horseshoe priors, but little change can be seen in the behavior around the original from that of the double-exponential distribution. From this figure, in addition to the observation in Figure~3 in the main text, the hierarchical DFL prior is expected to react to the dynamics of state variables and shift the posterior locations accordingly, but lacks the strong shrinkage effect toward the previous state.

\subsection{Prior of half-Cauchy type}

As verified in the next section, the DFL prior can yield the horseshoe distribution as the marginal distribution of state variable if the hyperprior of $\beta$ is modified as 
\begin{equation*}
\beta^2 | \delta \sim Ga( 1/2, \delta /2 ), \ \ \ \ \ \delta \sim Be(1/2,1/2).
\end{equation*}
The conditional posteriors of $\beta$ is, after the reparametrization by $\gamma = w\beta ^2/2$, the extended gamma distribution, whose density is given by 
\begin{equation*}
\pi (\gamma | - ) \propto \gamma ^{ r_{\gamma} - 1 } \exp \left\{ -\gamma - 2 c_{\gamma} \sqrt{\gamma}  \right\},
\end{equation*}
where $r_{\gamma} = (T+1)/2$ and 
\begin{equation*}
c_{\gamma} = \sqrt{ \frac{1}{2w} } \left\{ \rho (|x_1|+|x_T|) + \sum _{t=2}^T |x_t-x_{t-1}| + (1-\rho) \sum _{t:n_t>0} n_t|x_t| \right\}.
\end{equation*}
The acceptance-rejection algorithm for sampling from the general class of extended gamma distributions has been discussed in the literature \citep{finegold2011robust,liu2012rejection}. We utilize the fact that the ``rate'' parameter $c_{\gamma}$ is always positive in our model to devise the sampling algorithm based on Algorithm 3 in \cite{liu2012rejection}. 
\\ \hrulefill \\Sampling $\gamma$ from the extended gamma distribution with positive parameter $r_{\gamma}$ and $c_{\gamma}$. 
\begin{enumerate}
	\item Generate $\sqrt{\gamma } \sim Ga( 2r_{\gamma}, d_{\gamma} )$, where $d_{\gamma} = c_{\gamma} + \sqrt{ c_{\gamma }^2 + 4r_{\gamma} }$. 
	
	\item Accept $\gamma$ with probability $\min \{ 1,\ \exp\{ -(\sqrt{\gamma} - (d_{\gamma}/2) + c_{\gamma} )^2 \} \}$. Otherwise, reject $\gamma$ and return to Step 1.
\end{enumerate}
\hrulefill

The full conditional of $\delta$ is 
\begin{equation*}
\pi (\delta | - ) \propto (1-\delta )^{-1/2} \exp \left\{ -\frac{\beta^2}{2} \delta \right\},
\end{equation*}
which is the special case of Kummer-beta distributions. Despite the amount of works on this class of distributions (e.g., \citealt{gordy1998generalization}), the methodology for random number generation has not been fully developed. For this problem, as discussed in the main text, we simply replace the beta prior of $\delta$ by the discrete distribution on grids $\{ d,2d,\dots , Md \}$ for $d=1/N$ and $M<N$. The probability on each grid is proportional to the original beta density, i.e., 
\begin{equation*}
Pr[ \delta = md ] \propto m^{-1/2} (N-m)^{-1/2},
\end{equation*}
for $m=1:M$. The posterior of $\delta$ is the discrete distribution defined by 
\begin{equation*}
Pr[ \delta = md | - ] \propto (N-m)^{-1/2} \exp \left\{ -\frac{\beta ^2}{2} (md) \right\}.
\end{equation*}

\subsection{Estimation of baselines}

The baseline is the location parameter of the synthetic likelihoods and the prior of initial state $x_1$. The normal prior for $\mu$ is conditionally conjugate; for $\mu \sim N(m_0,s_0^2)$, the conditional posterior of $\mu$ is $N(m_1,s^2_1)$, where
\begin{equation*}
s_1^{-2} = \frac{ 1 }{ s_0^2 } + \frac{ 1 }{ \lambda _{\alpha ,1} } + \frac{ 1 }{ \lambda _{\alpha ,T} } + \sum _{t:n_t>0} \frac{ 1 }{\lambda _{n,t}} \ \ \ \ \ \mathrm{and} \ \ \ \ \ 
\frac{m_1}{s_1^2} = \frac{ m_0 }{ \sigma _0^2 } + \frac{ x_1 }{ \lambda _{\alpha ,1} } + \frac{ x_T }{ \lambda _{\alpha ,T} } + \sum _{t:n_t>0} \frac{ x_t }{\lambda _{n,t}} 
\end{equation*}
Shrinkage on this baseline can be introduced by another hierarchical prior on $\sigma ^2$.

\section{Hierarchical LASSO and its marginal} \label{app:hs}

Consider the univariate $x$ following the scale mixture of normals, i.e., $x|y \sim N(0,y)$. If $y\sim Ga(1,z/2)$, then $x\sim DE(\sqrt{z})$. In this subsection, we prove the hierarchical double-exponential distribution can be expressed as the three-parameter-beta distribution, 
\begin{equation*}
\pi _{TPB} (y) = \frac{\phi ^b}{Be(a,b)} y^{a-1} (y+\phi )^{-(a+b)},
\end{equation*}
where $a,b$ and $\phi$ are all positive, in addition that we assume $0<a<1$. The special case of $a=b=0.5$ is the half-Cauchy prior that induces the horseshoe distribution as the marginal of $x$. In our research, we also assume $\phi=1$. 

\begin{prp}
	\label{prp:hs}
	The three-parameter-beta distribution is expressed as the marginal of the following hierarchical model:
	\begin{equation*}
	y|z,w \sim \mathrm{Ex}(z/2),\qquad z|w \sim Ga(b,\phi w /2), \qquad w\sim Be(a,1-a)
	\end{equation*}
\end{prp}

Importantly, the conditional distribution of $y$ is the exponential distribution which implies the double-exponential marginal as $x|z,w \sim DE(\sqrt{z})$. This means that the introduction of the appropriate prior distribution on weight $\sqrt{z}$ leads to the strong shrinkage and robustness of the horseshoe-type. 

\

\noindent 
{\it Proof.} \ For any positive $r$ and $c$, we have
\begin{equation*}
c^{-r} = \int _0^{\infty} \frac{t^{r-1}e^{-ct}}{\Gamma (r)}dt,
\end{equation*}
as the integral of density function of $Ga(r,c)$. By applying this augmentation to the three-parameter-beta distribution, we obtain 
\begin{equation*}
\begin{split}
\pi _{TPB}(y) &= \frac{\phi ^b}{Be(a,b)} \int _0^{\infty}\!\!\! \int _0^{\infty} \frac{\tilde{t}^{(1-a)-1}e^{-y\tilde{t}}}{\Gamma (1-a)} \frac{\tilde{s}^{(a+b)-1}e^{-(y+\phi)\tilde{s}}}{\Gamma (a+b)} d\tilde{t}d\tilde{s} \\
&= \frac{\phi ^b2^{-(b+1)}}{ \Gamma (a) \Gamma (b) \Gamma (1-a) } \int _0^{\infty}\!\!\! \int _0^{\infty} t^{(1-a)-1}s^{(a+b)-1} \exp \left\{ -\frac{y}{2} t - \frac{y+\phi}{2} s  \right\} dtds \\ 
& \hspace{320pt} (\tilde{t}=t/2,\ \tilde{s}=s/2) \\
&= \frac{\phi ^b2^{-(b+1)}}{\Gamma (a) \Gamma (b) \Gamma (1-a) } \int _0^{\infty}\!\!\! \int _0^1 z^b  w^{(a+b)-1}(1-w)^{(1-a)-1} \exp \left\{ -\frac{y}{2} z - \frac{\phi}{2} wz  \right\} dzdw \\
& \hspace{280pt} (z=t+s,\ w=s/(t+s)) 
\end{split}
\end{equation*}
The integrand is further computed as 
\begin{equation*}
\begin{split}
& z^b  w^{(a+b)-1}(1-w)^{(1-a)-1} \exp \left\{ -\frac{y}{2} z - \frac{\phi}{2} wz  \right\} \\ 
&= 2\Big\{ \frac{z}{2} e^{-(z/2)y} \Big\} z^{b-1}  w^{(a+b)-1}(1-w)^{(1-a)-1} \exp \left\{ -\frac{\phi}{2} wz  \right\} \\ 
&= 2 \mathrm{Ex}( y | z/2) \left\{ \frac{( \phi w/2)^b}{\Gamma (b)} z^{b-1} e^{ -(\phi w/2)z } \right\} \frac{\Gamma (b)}{( \phi w/2)^b}  w^{(a+b)-1}(1-w)^{(1-a)-1} \\ 
&= 2^{b+1} \phi ^{-b} \Gamma (b) \ \mathrm{Ex}( y | z/2 ) \ Ga( z | b,\phi w/2 ) \ w^{a-1}(1-w)^{(1-a)-1} \\ 
&= \frac{ \Gamma (b)\Gamma (a)\Gamma (1-a) }{\phi ^b 2^{-(b+1)}} \ \mathrm{Ex}( y | z/2 ) \ Ga( z | b,\phi w/2 ) \ Be(w|a,1-a). 
\end{split}
\end{equation*}
Plug-in the obtained expression in the integral and observe that the desired mixture representation is obtained. \qed 

The special case of half-Cauchy distribution, where $a=b=1/2$ and $\phi = 1$, is stated as the following corollary. 

\begin{cor}
	\label{cor:hs}
	The half-Cauchy distribution, $\pi _{HC}(y) = y^{-1/2}(1+y)^{-1} / \pi$, is expressed as the marginal of the following hierarchical model:
	\begin{equation*}
	y|z,w \sim Ga(1,z/2),\qquad z|w \sim Ga(1/2,w /2), \qquad w\sim Be(1/2,1/2)
	\end{equation*}
\end{cor}

\begin{figure}[!htbp]
	\centering
	\includegraphics[width=4.5in]{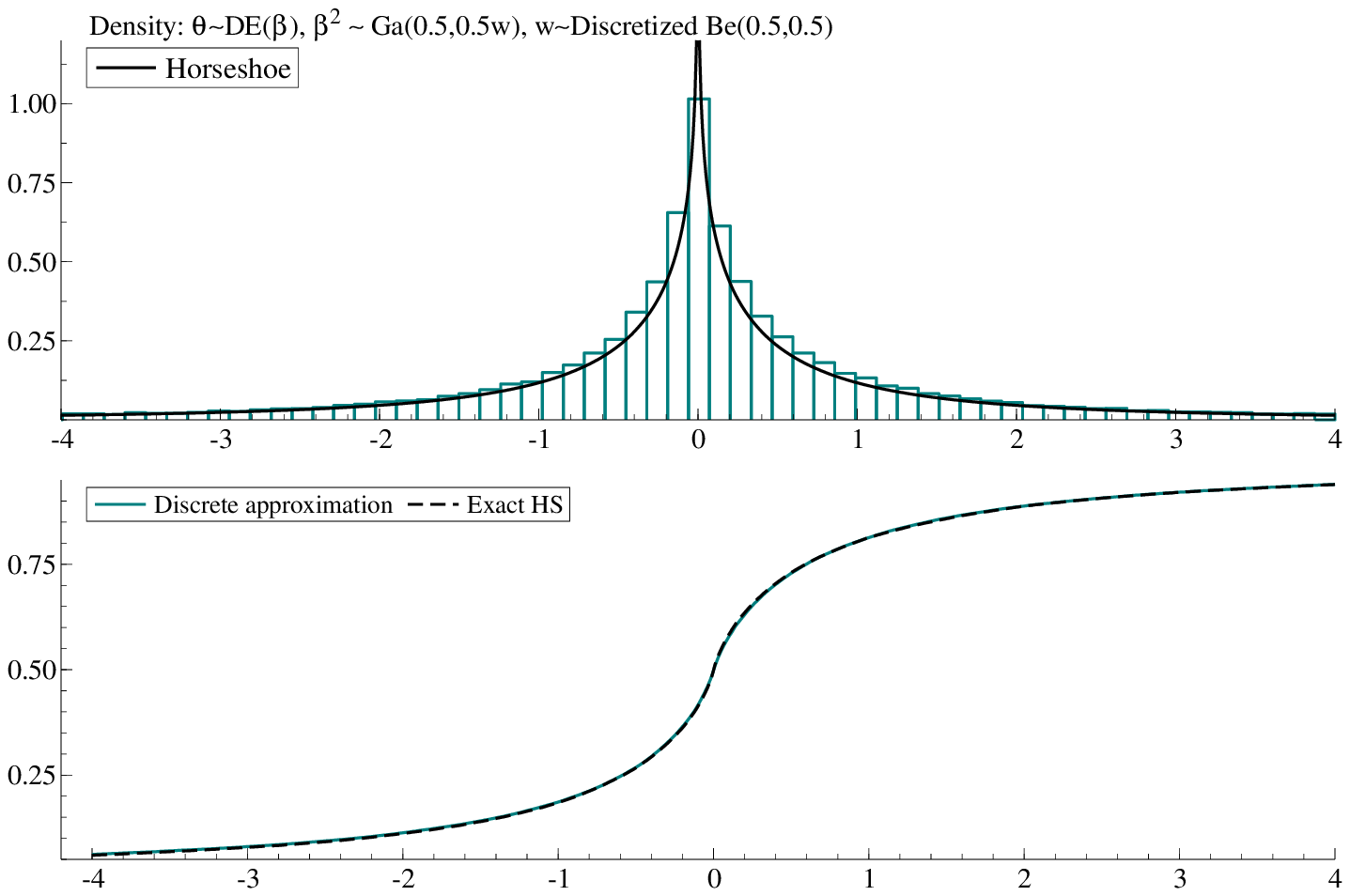}
	\caption{\small Top: the densities of horseshoe prior (black solid line) and its expression by hierarchical double-exponential distribution with discrete approximation of beta distribution (green bins). To draw the histogram, 100,000 random samples are generated from the model $x|\beta \sim DE(\beta)$, $\beta \sim Ga(0.5,0.5)$ and $w$ from the discrete distribution on $\{ d,2d,\dots , (N-1)d \}$ where $N=1000$ and $d=1/N$, whose probability function is proportional to the density of $Be(0.5,0.5)$. Bottom: the comparison by the empirical distribution functions. In both panels, the difference between two distributions is hard to recognize visually, implying that little bias is caused by approximating the original beta distribution by discretization. 
	} \label{fig:mhs} 
\end{figure}

\noindent 
In the main text, for the convenience in posterior computation, the beta distribution of $w$ is replaced by the discrete distribution on the girds of $(0,1)$ with probability proportional to the density of the beta distribution. This modification potentially affects the marginal distribution of $x$, which might deviate from the original half-Cauchy distribution. We examine this potential discrepancy by generating the random variables from the hierarchical model in Corollary~\ref{cor:hs} with the discrete distribution on $w$ in Figure~\ref{fig:mhs}. Both in the density form and empirical distribution function, the difference from the target horseshoe distribution is negligible.

\section{Modeling of observational variance} \label{app:sv}

\subsection{Constant variance for scale-free DLMs} \label{app:constv}

The scale-free DLM with the DFL prior has the CDLM form as follows; 
\begin{equation} \label{synV}
\begin{split}
y_t | \theta _t, V &\sim N(F_t'\theta _t,V) \\
z_{it} | \theta _t, V &\sim N(\theta _{it},V\lambda _{n,it}), \ \ \ \ \ \mathrm{if} \ n_{it}>0 \\
z_{iT} | \theta _T, V &\sim N(\theta _{iT},V\lambda _{\alpha ,it}) \\
\theta _{i1} | V &\sim N(\mu _i,V\lambda _{\alpha,i1}) \\
\theta _{it} | \theta _{i,t-1},V &\sim N(\theta _{i,t-1},V\lambda _{\beta ,it}) 
\end{split}
\end{equation}
where $z_{it}=\mu _i$ for all $t$ and $i$. The MCMC algorithm is modified by replacing the sampling of state variables as follows; 

\noindent \hrulefill \\ Gibbs sampler for scale-free models: replace Step 1 of the algorithm in Section~4 by the following. 
\begin{enumerate}	
	\item Sampling $\theta _{1:p,1:T}$ and $V$.
	
	\item[(i)] Forward filtering. 
	
	For the DLM in \eq{synV}, implement the forward filtering to compute the one-step ahead predictive density for $y_t^{\ast} = (y_t,z_t)'$, where $z_t$ is the collection of observed $z_{1:p,t}$ ($z_{it} \in z_t$ if $n_{it}>0$). Specifically, compute the one-step ahead predictive mean and variance, $f_t$ and $Q_t^{\ast}$, defined by 
	\begin{equation*}
	p(y_t^{\ast}|y^{\ast}_{1:{t-1}}) = N(y_t^{\ast}|f_t,VQ_t^{\ast})
	\end{equation*}
	
	\item[(ii)] Sampling of $V$.
	
	Generate $V$ from its posterior $Ga(n_T/2,n_TS_T/2)$, where the sufficient statistics can be computed by 
	\begin{equation*}
	n_t = n_{t-1}+1, \ \ \ \ \ S_t = n_{t-1} S_{t-1} + (y_t^{\ast} - f_t)' (Q_t^{\ast})^{-1} (y_t^{\ast} - f_t) 
	\end{equation*}
	
	\item[(iii)] Sampling of $(\theta _{1:p,1:T} | V)$.
	
	Given $V$, implement the forward filtering and backward sampling for the model in \eq{synV} to generate $\theta _{1:p,1:T}$. 
\end{enumerate}
\hrulefill 

The sampling of the other parameters remain the same for the scaled state variable $\theta _{it}/\sqrt{V}$.

\subsection{Stochastic volatility of log-Gaussian type} \label{app:gauss}

In the macroeconomic application in Section~5.2, we consider the model in \cite{NakajimaWest2013JBES} where the observational variance, or stochastic volatility, is the time-varying parameter. The log-volatility, $h_t \equiv \log V_t$ is modeled by the Gaussian AR(1) process, 
\begin{equation*}
h_t - \mu _h = \phi _h(h_{t-1}-\mu _h) + \eta _t, \ \ \ \ \ \ \eta _t \sim N(0,\sigma _h^2),
\end{equation*}
with parameters $(\mu _h, \phi _h , \sigma _h^2)$ to be estimated with the specific priors \cite{jacquier1994bayesian}. The priors for the set of  $(\mu _h, \phi _h , \sigma _h^2)$'s in the time-varying VAR models are taken from \cite{NakajimaWest2013JBES}. Likewise, the posterior sampling of $h_{1:T}$ is based on the multi-move sampler (\citealt{shephard1997likelihood}; \citealt{watanabe2004multi}). While the efficiency of this sampling procedure is sufficient for our purpose as reported in \cite{NakajimaWest2013JBES}, there has been many methodological updates on the posterior computation of the stochastic volatility models; readers who seek for the better practice should refer to, for example, \cite{kastner2014ancillarity}.

\section{Supplemental results for the simulation study} \label{app:simstudy}

\subsection{Posterior udpate of latent count}

It is of interest here how much one can learn about the latent count, $n_t$, from data. We evaluate the change from prior to posterior of $n_t$ in the simulation study. 

The prior of $n_t$ is given in Proposition~2.3 and 2.3 as the mixture of the point mass on zero and the log-geometric distribution. The prior probability of having $n_t>0$ is 
\begin{equation*}
P[n_t > 0] = \frac{C_+}{C_+ + C_0},
\end{equation*}
where $C_0 = (\beta - \alpha )/2$ and $C_+ = \log ( \beta / (\beta - \alpha) )$. By parameterizing $\alpha = \rho \beta$ with $\rho \in (0,1)$, we have $C_0 = \beta (1-\rho )/2$ and $C_+ = - \log (1-\rho )$. The prior probability of positive $n_t$ is 
\begin{equation*}
P[n_t > 0] = \frac{\log (1-\rho)}{\log (1-\rho) - (1-\rho )\beta /2}. 
\end{equation*}
In practice, we also place priors on $\beta$ and $\rho$. The prior on $\rho$ is the discrete distribution, so it is easy to integrate $\rho$ out in the above expression. To compute the expectation with respective to $\beta$, we simply use the Monte Carlo integration. 

The prior mean of $n_t$ is, 
\begin{equation*}
E[n_t] = \frac{C_+}{C_+ + C_0} E[ n_t | n_t >0 ],
\end{equation*}
where the conditional expectation in the right is the mean of log-geometric distribution with parameter $\alpha /\beta$. It is shown that 
\begin{equation*}
E[n_t|n_t>0] = \frac{1}{C_+} \frac{\alpha}{\beta - \alpha}.
\end{equation*}
Hence, in terms of $\rho$ and $\beta$, the target expectation is computed as 
\begin{equation*}
E[n_t] = \frac{1}{ (1-\rho )\beta /2 - \log (1-\rho) } \frac{\rho}{1-\rho}. 
\end{equation*}

\begin{figure}[!htbp]
	\centering
	\includegraphics[width=4.5in]{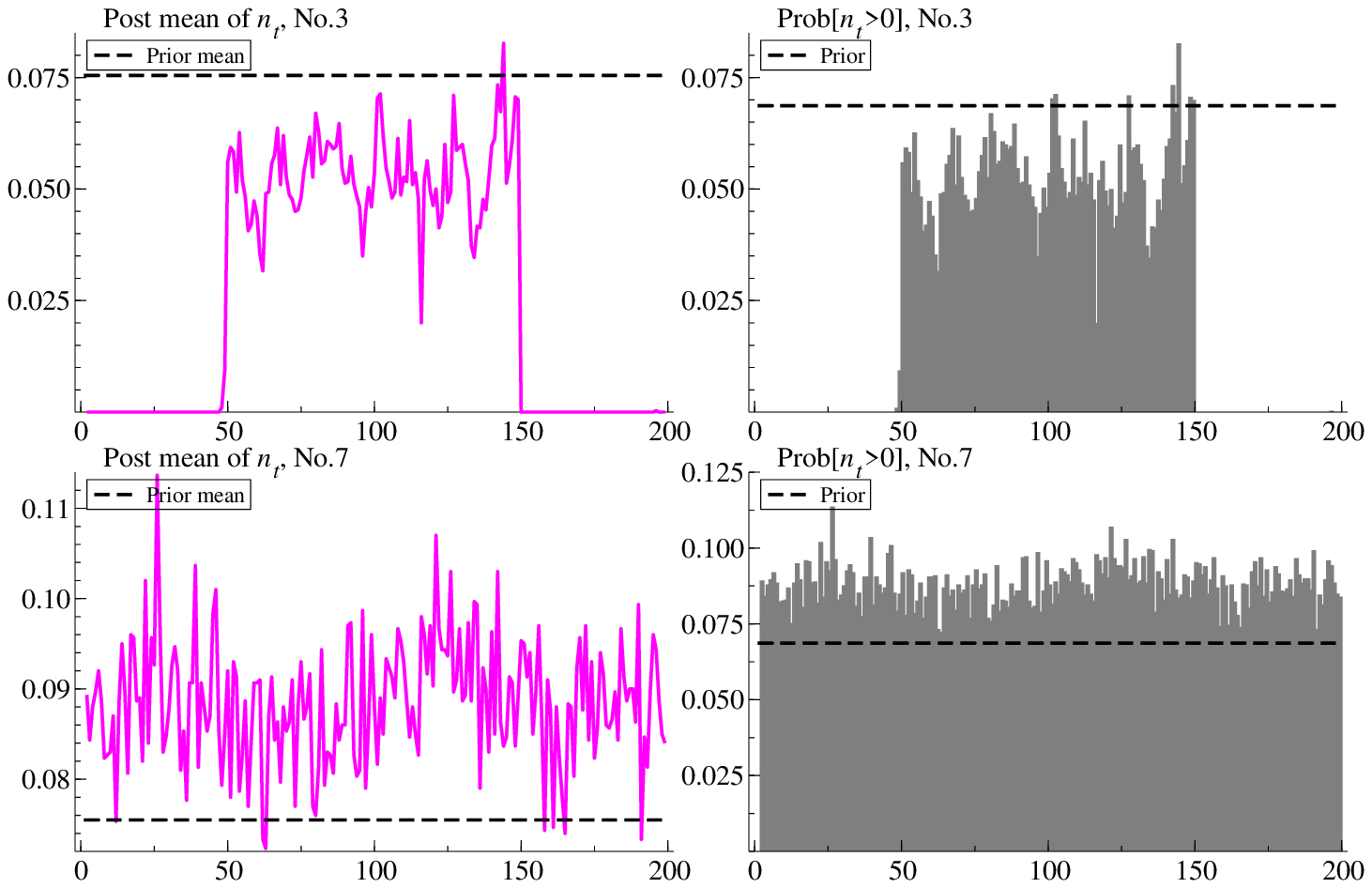}
	\caption{\small Top-Left: posterior mean of $n_{3t}$. Top-Right: posterior probability of having $n_{3t}>0$. Bottom-Left: posterior mean of $n_{7t}$. Bottom-Right: posterior probability of having $n_{7t}>0$. All of those plots are overlayed by their prior counterparts, $E[n_{it}]$ an $P[n_{it}>0]$, for $i=3,7$. The process of posterior learning is evident in the difference of the prior and posterior quantities. } \label{fig:count_prior} 
\end{figure}

Figure~\ref{fig:count_prior} shows the prior and posterior mean and probability of positive $n_{it}$ for $i=3$ (dynamically significant predictor) and $i=7$ (noise predictor) under the model $\mathcal{M}_3$. The posterior probability of $n_{3t} > 0$ here is identical to the one shown in Figure~5 of the main text. For the active predictor $i=3$, the probability of positive $n_t$ is lowered from that of the prior. This change in posterior distribution is local; the probability of positive $n_t$ is not completely zero when the coefficient is non-zero. For noise predictor $i=7$, the posterior probability of positive $n_{7t}$ and the posterior mean of $n_{7t}$ are slightly larger than those of the prior, reflecting the belief on the ``insignificance'' of the predictor enhanced by the observed data. 

\subsection{Posterior plots for baselines}

\begin{figure}[!htbp]
	\centering
	\includegraphics[width=4in]{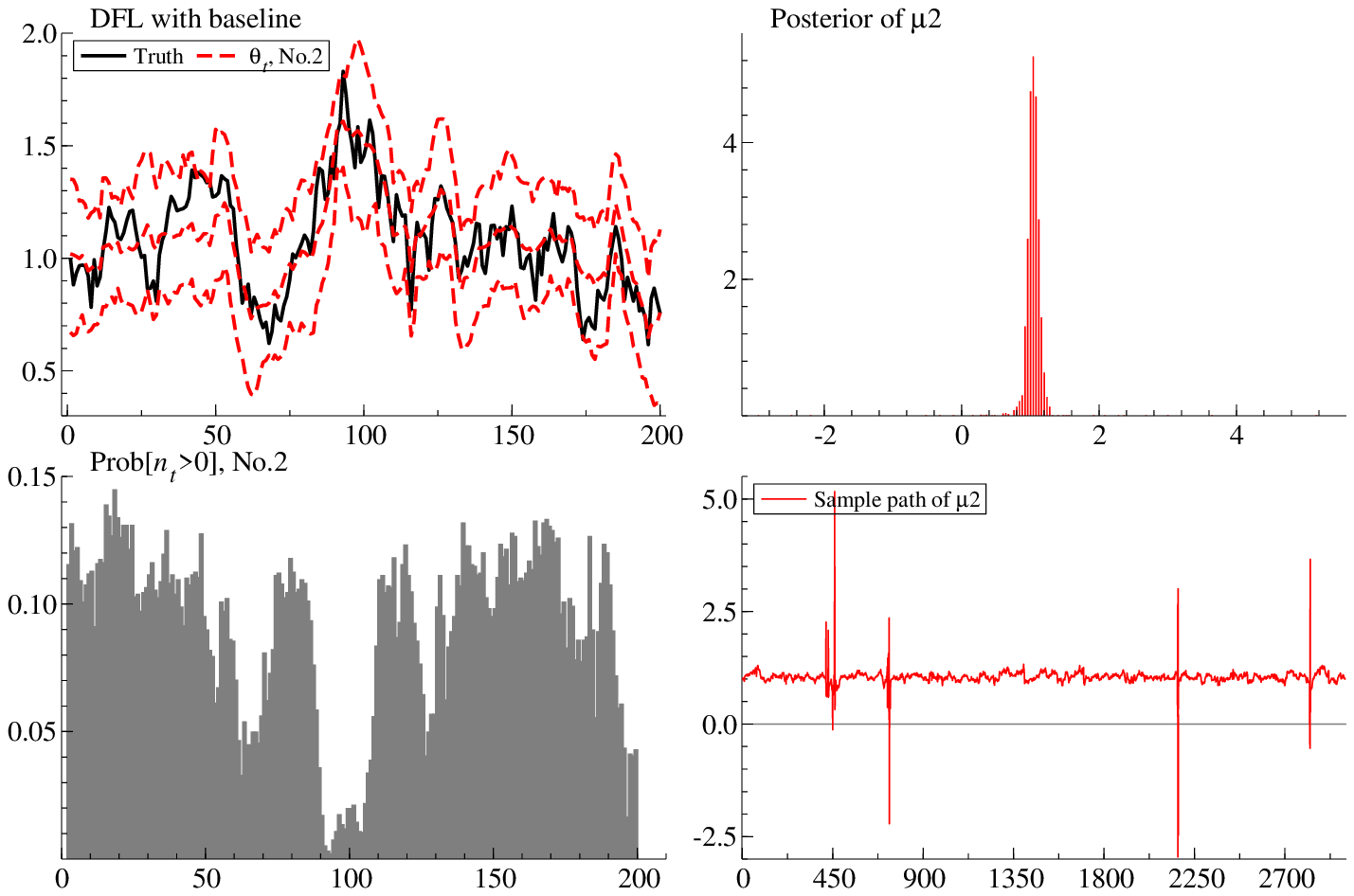}
	\caption{\small The posterior summaries about $\theta _{2t}$ and $\mu_2$ under the DFL prior with baseline (Model $\mathcal{M}_6$). Top-Left: the posterior of state variables of $\theta _{2t}$. Top-Right: the histogram of posterior samples of $\mu_2$. Bottom-Left: posterior probabilities of having $n_t > 0$. Bottom-Right: sample path of $\mu_2$. This is an example where the true values of state variable are stable around unity, hence the use and estimation of baseline is successful. } \label{fig:base2} 
\end{figure}

Figure~\ref{fig:base2} collects four posterior plots on baseline $\mu_2$ in model $\mathcal{M}_6$: the histogram of posterior samples of $\mu_2$, sample path, the posterior mean and 95\% credible intervals of state variable $\theta _{2t}$ and latent count $n_{2t}$. This is an example where the use of baseline is successful, for the true values of $\theta _{3t}$ in Figure~4 are stably varying around unity, resembling the realization of a stationary process. In this case, it is efficient to have a single baseline to capture the stability of state variables. 

\begin{figure}[!htbp]
	\centering
	\includegraphics[width=4in]{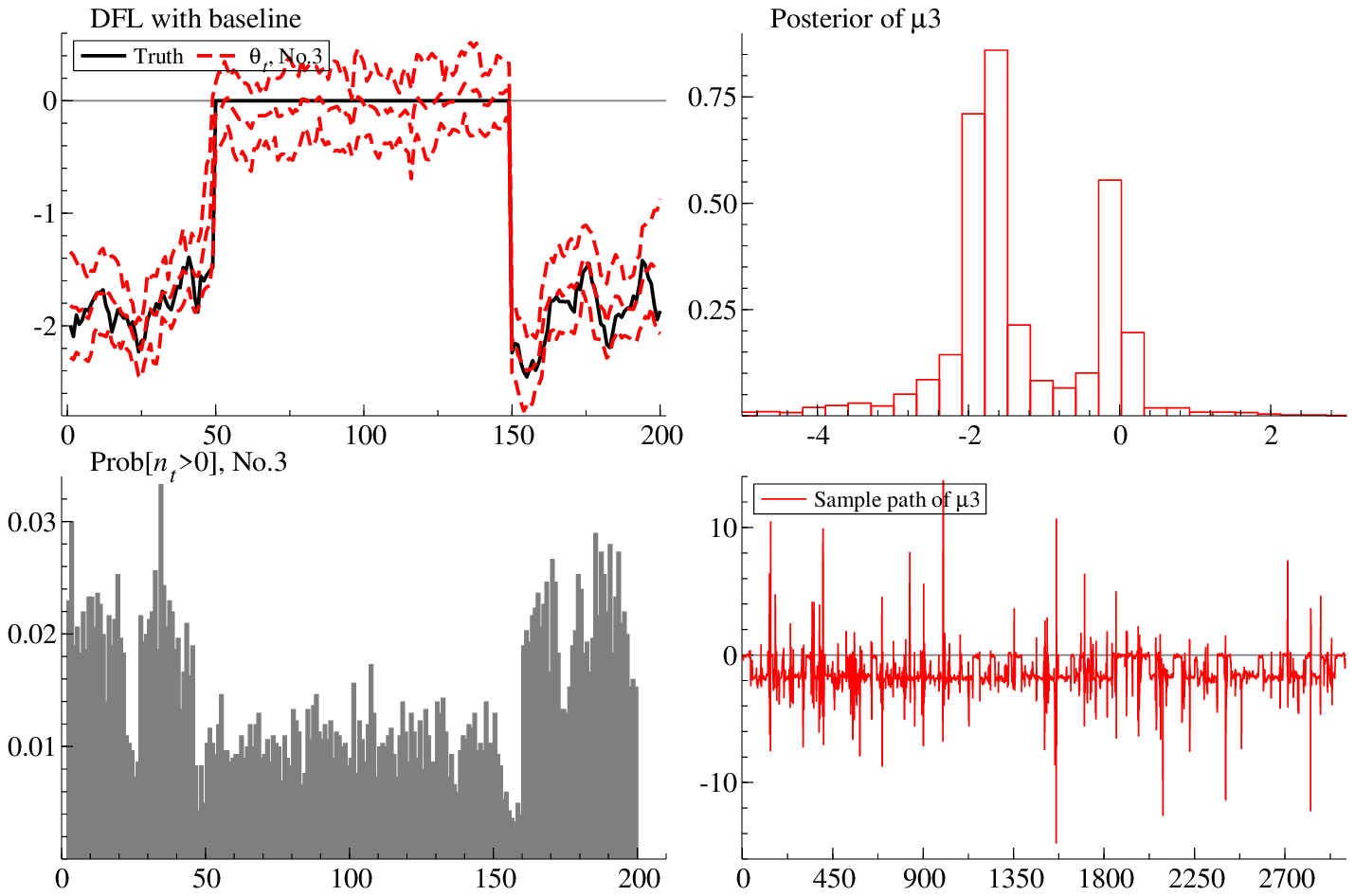}
	\caption{\small The same posterior summaries as in Figure~\ref{fig:base2} about $\theta _{3t}$ and $\mu_3$. Because the true values of state variable switches between zero and non-zero values, the posterior of baseline $\mu_2$ becomes bimodal. Under this circumstance, the posterior of $n_{3t}$ is difficult to interpret.} \label{fig:base3} 
\end{figure}

However, the baseline could complicate the description of models and result in the less efficient inference and prediction. Figure~\ref{fig:base3} summarizes the same information on $\mu _3$. In this example, where the true value of state variable $\theta _{3t}$ switches between zero and non-zero values, the posterior of baseline $\mu_3$ becomes bimodal to capture the two main locations that the state variable stays at. For this reason, the latent count $n_{3t}$ is suspected to be the redundant parameter for the model, in addition to losing its interpretation as the indicator of shrinkage to the baseline. This could possibly explain why we had little gain in predictive performances from the additional baselines, as summarized in Table~1.

\subsection{Length of 95\% credible intervals}

\begin{figure}[!htbp]
	\centering
	\includegraphics[width=4.5in]{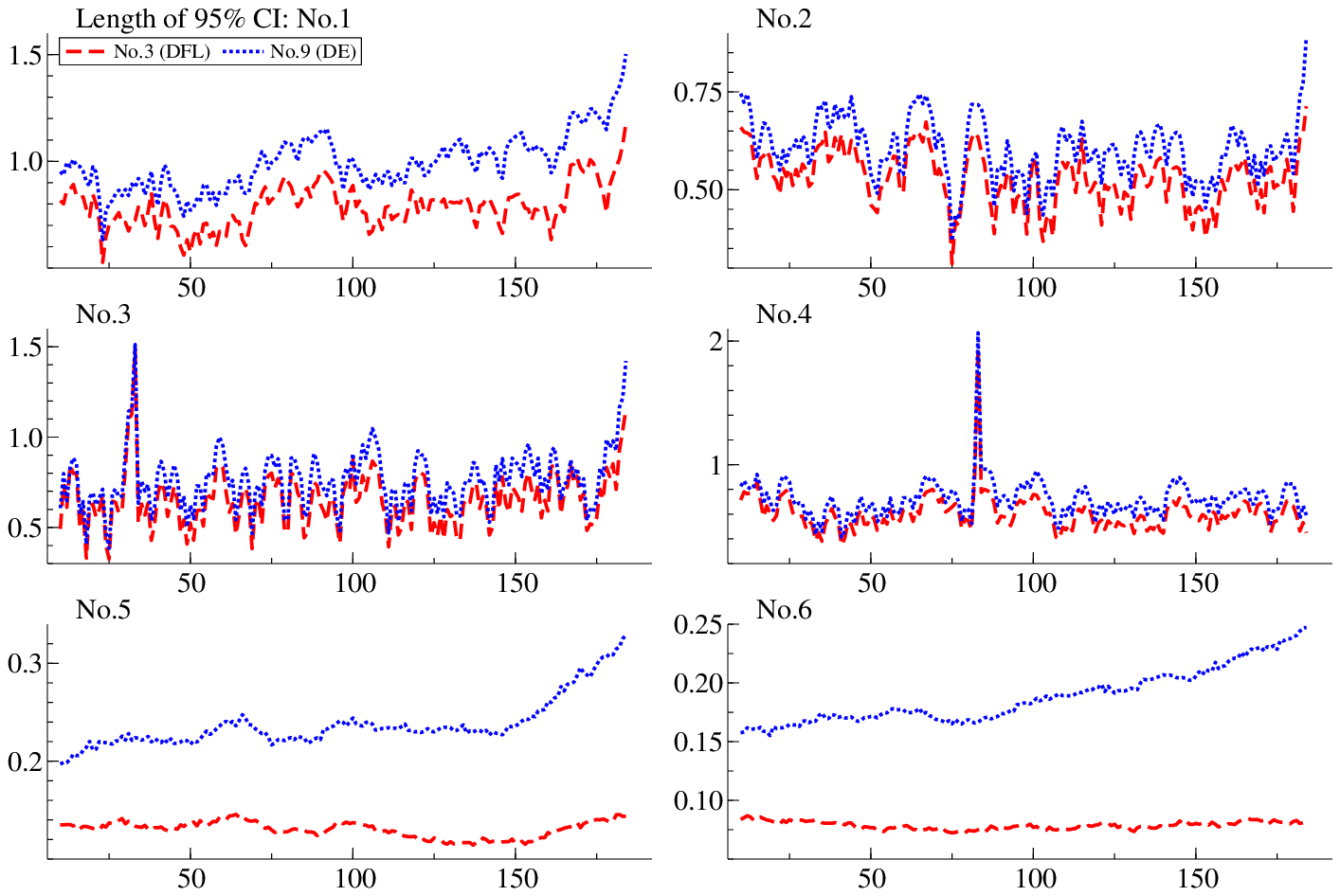}
	\caption{\small Lengths of 95\% posterior credible intervals of $\theta _{it}$ for $i=1,\dots,6$ under the DFL $\cM_3$ and DE $\cM_9$.} \label{fig:CIstate} 
\end{figure}

The lengths of 95\% posterior credible intervals of state variables $\theta _{it}$ for $i=1,\dots ,6$ under $\mathcal{M}_3$ and $\mathcal{M}_9$ are shown in Figure~\ref{fig:CIstate}. For each $i$, the average of those lengths over time is summarized in Table~1 in the main text. The reduced posterior uncertainty under the DFL-DLM $\mathcal{M}_3$ is clear in the zero coefficients. The credible intervals of $\mathcal{M}_3$ is narrower than those of $\mathcal{M}_9$ for the active predictors as well, but the difference is relatively slight.  

\section{Supplemental results for the application} \label{app:app}

\subsection{Posteriors of log volatilities}

\begin{figure}[!htbp]
	\centering
	\includegraphics[width=4.5in]{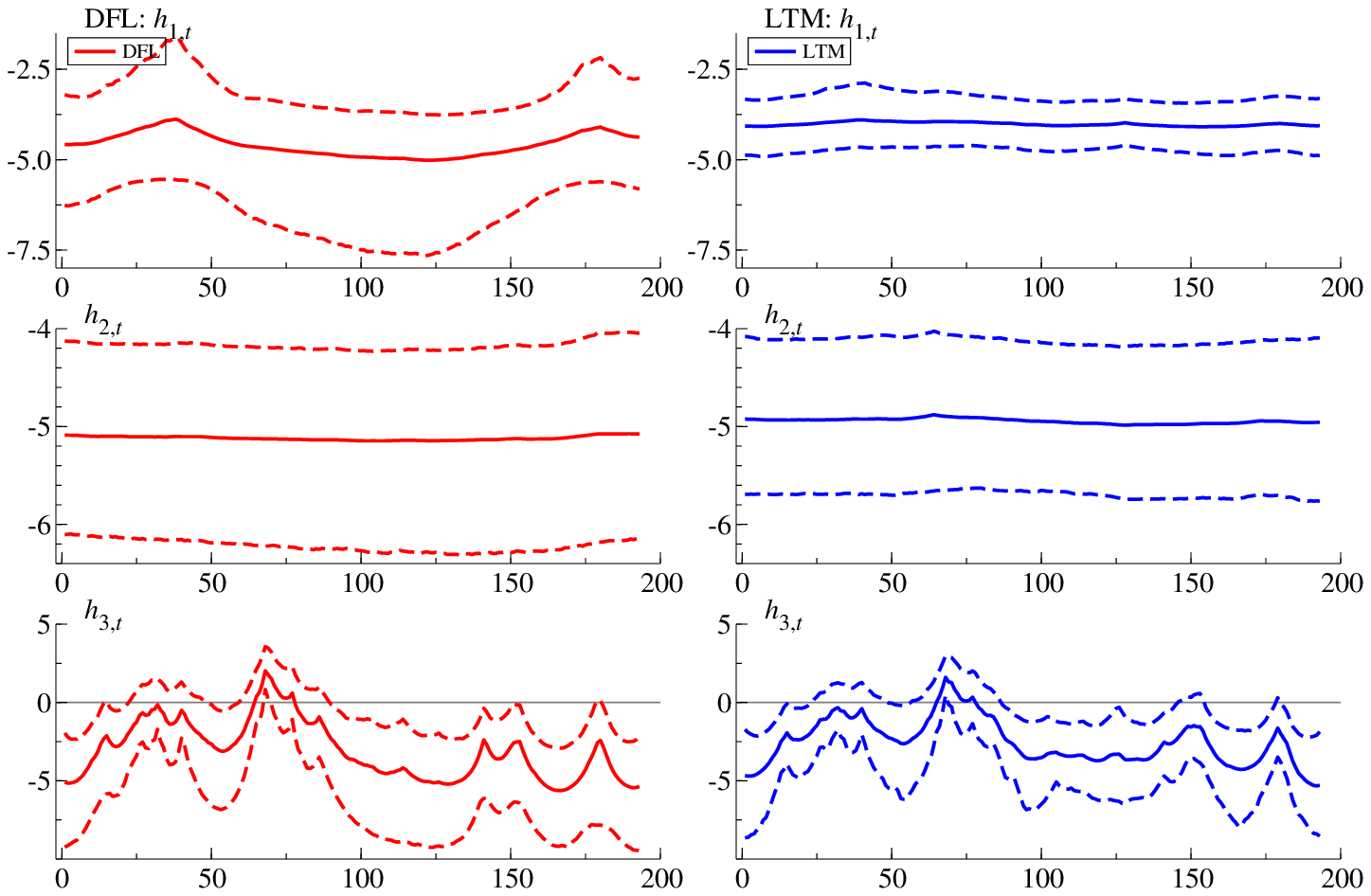}
	\caption{\small Posterior means and 95\% posterior credible intervals of log volatilities, $h_{it}$, for $i=1,\dots,6$ under the DFL-DLM (left) and LTM (right).} \label{fig:svh} 
\end{figure}

Figure~\ref{fig:svh} shows the three posteriors of log volatilities, $h_{it}$, for both models. The posteriors of volatilities of the DFL-DLM are similar to those of the LTM. Because of this similarity of the variance part, the difference of the two models, if any, must be summarized in the estimation of regression coefficients in the main text. In fact, the LTM is very sensitive in this example to the choice of hyperparameters, and its slight change affects the posteriors of both variance and regression coefficients. To make the posterior of the LTM (and DFL-DLM) interpretable, we chose the hyperparameters of the DFL prior so that both models have little difference in the estimation of volatilities.

\subsection{Posterior means of latent counts}

\begin{figure}[!htbp]
	\centering
	\includegraphics[width=4.5in]{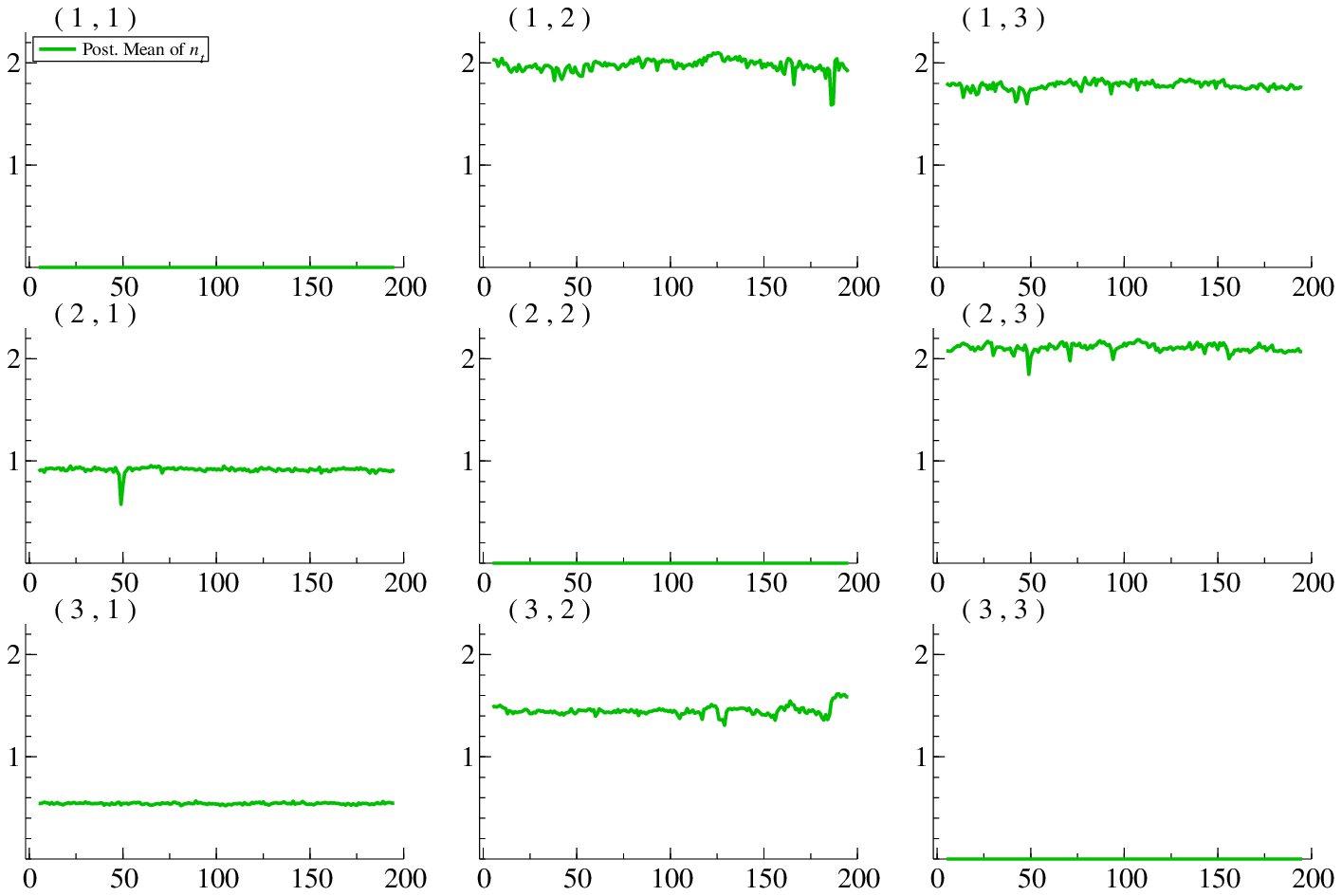}
	\caption{\small Posterior means of latent counts $n_t$ for coefficient matrix of lag 1.} \label{fig:ltmn} 
\end{figure}

In Figure~\ref{fig:ltmn}, the posterior means of latent count $n_t$ for the regression coefficients of lag 1 are plotted. The patterns of posteriors are almost identical to those in Figure~9 in the main text, where we computed the posterior probabilities of sampling positive counts $n_t$. The difference of the amount of shrinkage applied to each entry of the coefficient matrix is seen in the expected values of latent count. For example, the upper diagonal entries is shrunk to zero more strongly than those of off diagonal ones. The dynamics of shrinkage are less apparent in this example, compared with the simulation study. Although the slight variations of the posterior expectations of $n_y$ can be seen in some entries, they are likely to be indistinguishable from the Monte Carlo errors.

\end{document}